\documentclass[11pt,a4paper]{article}

\usepackage{a4wide}
\usepackage{amsmath}
\usepackage{amssymb}
\usepackage{graphicx}
\usepackage{url}
\usepackage{hyperref}
\usepackage{cite}
\usepackage{booktabs}
\usepackage{multirow}

\newcommand{\GeV}{\,\mathrm{GeV}}
\newcommand{\TeV}{\,\mathrm{TeV}}
\newcommand{\Erf}{\mathrm{Erf}}
\newcommand{\ie}{i.e.\ }
\newcommand{\av}{\mathrm{Av}}
\newcommand{\Av}{\mathrm{Av}}
\newcommand{\med}{\mathrm{med}}
\newcommand{\trans}{\mathrm{Trans}}

\newcommand{\Min}{\mathrm{Min}}
\newcommand{\Max}{\mathrm{Max}}
\newcommand{\crit}{\mathrm{crit}}
\newcommand{\tile}{\mathrm{tile}}
\newcommand{\soft}{\mathrm{soft}}
\newcommand{\hard}{\mathrm{hard}}
\newcommand{\jet}{\mathrm{jet}}
\newcommand{\tot}{\mathrm{tot}}
\newcommand{\ext}{\mathrm{ext}}
\newcommand{\as}{\alpha_s}
\newcommand{\order}[1]{{\cal O}\left(#1\right)}

\newcommand{\JA}{\text{JA}}
\newcommand{\Sd}{S_d}
\newcommand{\cP}{\mathcal{P}}
\newcommand{\la}{\langle}
\newcommand{\ra}{\rangle}
\newcommand{\mean}[1]{\left\la\smash{#1}\right\ra}

\title{On the characterisation of the underlying event }
\author{Matteo Cacciari,\!$^{1,2}$ Gavin P.~Salam$^1$ and Sebastian Sapeta$^1$\\
\\
{\it  \normalsize $^1$LPTHE, UPMC Univ.~Paris 6 and CNRS UMR 7589, Paris, France}\\
{\it  \normalsize $^2$Universit\'e Paris Diderot, Paris, France}\\
{\it  \normalsize E-mail:}
{\tt \normalsize cacciari@lpthe.jussieu.fr, salam@lpthe.jussieu.fr,}\\
{\tt \normalsize sapeta@lpthe.jussieu.fr }
}

\date{}

\begin{document}

\maketitle
\thispagestyle{empty}

\begin{abstract}
  The measurement of the underlying event (UE) and its separation from
  hard interactions in hadron-collider events is a conceptually and
  practically challenging task.
  We develop a simple, mostly analytical toy model for the UE in order
  to understand how different UE measurement approaches fare on the
  practical aspects of this problem,
  comparing the traditional approach used so far at Tevatron with a
  recently proposed ``jet-area/median'' approach.
  Both are found to perform comparably well in measuring average
  properties of the UE, such as the mean transverse momentum flow, but
  the jet-area/median approach has distinct advantages in determining
  its fluctuations.
  We then use the jet-area/median method to investigate a
  range of UE properties in existing Monte Carlo event-generator
  tunes, validating the main results of the toy-model and
  highlighting so-far unmeasured characteristics of the UE such as its
  rapidity dependence, as well as its intra- and inter-event
  fluctuations and correlations.
%   %
\end{abstract}

\clearpage

\setcounter{page}{1}

\tableofcontents

%======================================================================
\section{Introduction}
\label{sec:introduction}

The ``underlying event'' (UE) in high-energy hadron-hadron collisions
can be thought of as the low transverse momentum ($p_t$) part of the
event activity that is not naturally associated with the hard
interaction.

Despite being a low-$p_t$ phenomenon, the underlying event has a large
impact on high-$p_t$ physics at hadron colliders. For example in
measurements of the inclusive jet spectrum at $p_t \sim 50 \GeV$ it
can affect the result by up to $50\%$ \cite{Abulencia:2007ez}.
It also biases kinematic reconstructions (for example in top mass
measurements), degrades their resolution and affects the efficiency of
isolation criteria that enter into the experimental identification of
particles like photons and electrons.
It has even been suggested that certain new physics scenarios might
show up in ``anomalous'' characteristics of the underlying event
\cite{Harnik:2008ax}.
For these reasons it is important to have a good understanding of
its properties.

The purpose of this article is to investigate some different ways in which
the underlying event can be measured and/or constrained
experimentally. Such measurements enter into tunes of Monte Carlo
models of the UE
\cite{Sjostrand:1987su,Butterworth:1996zw,Sjostrand:2004ef,Sjostrand:2007gs,Bahr:2008tf,Bahr:2008dy,Gleisberg:2008ta,Corke:2009tk}. They
can also serve as an input to analytical methods of accounting for the
average UE correction to a jet's transverse momentum
\cite{Dasgupta:2007wa,Cacciari:2008gn} and to approaches that correct for
the UE on a jet-by-jet and event-by-event
basis~\cite{Cacciari:2007fd}, as well as for related work that seeks to
optimise jet definitions.

\begin{figure}
  \centering
  \includegraphics[width=\textwidth]{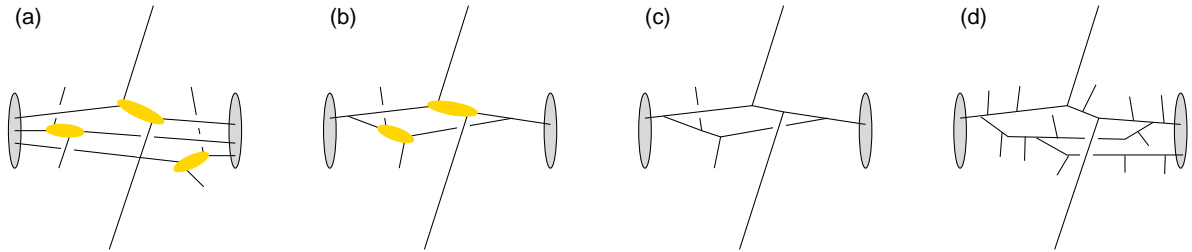}
  \caption{Views of a hard scattering plus UE; (a) shows 
    multiple $2\to2$ interactions, as incorporated in the most
    successful models; (b) illustrates how a collinear splitting in
    the initial-state can lead to correlation between the partons
    involved in a double $2\to2$ scattering; from (c) one sees the
    similarity with a perturbative 1-loop $2\to4$ diagram; and (d)
    represents a BFKL-inspired picture for the UE.}
  \label{fig:UE-model-pics}
\end{figure}

%% %%%%%%%%%%%%%%%%%%%%%%%%%%%%%%%%%%%%%%%%%%%%%%%%%%%%%%%%%%%%%%%%%%%%%
%% %
%% %
%% %
%% %
%% %
%% %
%% %
%% %

A difficulty in discussing the measurement of the UE is that
there exists no good definition of what the UE actually is, or how to
distinguish it, in a conceptually unambiguous way, from the hard interaction.
For instance, the most successful phenomenological models of the UE
involve multiple $2\to2$ scattering as in
fig.~\ref{fig:UE-model-pics}a (``multiple parton interactions'' ---
MPI). Very simply, they supplement the one
hard interaction in the event with multiple other lower-$p_t$
interactions, whose multiplicity is determined by the $2\to2$ cross
section, regulated with a low-$p_t$ cutoff of the order of a couple of
GeV.
Fig.~\ref{fig:UE-model-pics}a can only be part of the picture because
some of the partons entering multiple $2\to2$ scatterings are
necessarily correlated, e.g.\ due to energy conservation
(cf. ref.~\cite{Gaunt:2009re}), or because they can have a common
origin from an initial-state collinear splitting, as in
fig.~\ref{fig:UE-model-pics}b.
However the contribution of fig.~\ref{fig:UE-model-pics}b is itself
also part of the $2\to 4$ 1-loop scattering diagram
fig.~\ref{fig:UE-model-pics}c (as discussed for example in
\cite{Nagy:2006xy}), which is relevant as a N$^4$LO correction to the
dijet cross section or at N$^3$LO in its interference with tree-level
$2\to 4$ scattering.
This means there are non-trivial questions of double-counting between
multiple parton interaction and perturbative higher orders.
In addition, the radiation that fills the event is not bound to come
just from $2\to 2$ scatterings, but may also arise from BFKL type
configurations which can involve (multiple) chains with low-$p_t$
emissions spread in rapidity, as in fig.~\ref{fig:UE-model-pics}d
(some work towards modelling this while retaining consistency with the
total cross section is to be found in \cite{Avsar:2006jy}).
Though these ways of viewing the UE represent just a subset of the
diversity of physics considerations that are of potential relevance
for its modelling (more detailed reviews are to be found in
Refs.~\cite{Sjostrand:2004pf,Corke:2009tk}), they do illustrate the
difficulties that arise in ascribing unambiguous physical meaning to
it.

Given this complexity in discussing what the UE might be, how are we
to go about measuring it?
A feature present in most models is that, on some low $p_t$ scale, UE
activity fills the whole event.
One way then of characterising the UE is to say that it \emph{is}
whatever physical effect fills most of the event with radiation.
To help understand what implications this picture has for UE
measurements, we shall develop
(section~\ref{sec:systematics}) a semi-analytical two-component toy
model: one component will be purely soft and dispersed across the
event, corresponding to the UE, while the second component will
involve the hard scattering and a simple approximation for the
perturbative radiation with which it is associated.

Though our toy model is undoubtedly too simple to fully reflect
reality, the fact that we know exactly what goes into it will make it
quite powerful: it will, for example, allow us to take different
UE-measurement approaches and examine to what extent their results are
affected both by the radiation associated with the hard scattering, and
by the techniques used to limit that hard contamination.
This can be investigated both for averaged quantities, and for
event-by-event extractions of information about the UE.
A number of the results will be given in analytical form, in terms of
the characteristics of the UE and the hard scattering and of the
parameters of the measurement methods.
This will give insight into the compromises that arise
when measuring the UE, especially when extending existing methods to
the greater phase space that is available at LHC relative to Tevatron.

The two UE measurement methods that we shall investigate are both
reviewed in section~\ref{sec:overview-of-approaches}.
One, which we call the ``traditional'' approach (see for example
\cite{Albrow:2006rt,KarFieldDY}), is currently the default approach for most UE
studies.
Another, the jet-area/median based approach of
\cite{Cacciari:2007fd,Cacciari:2008gn}, was originally developed for
evaluating $pp$ pileup (and backgrounds in heavy-ion collisions, as used
for example in \cite{Salur:2009hw}), but may also have benefits for UE
studies. 
The basic results for the two approaches will be derived in
section~\ref{sec:systematics}. 
Some readers may prefer to skip most of this section and read just the
final summary of these results, as presented in
section~\ref{sec:toy-model-summary}.

For the purpose of determining the quantity that we call $\rho$, the
UE's mean transverse momentum ($p_t$) flow per unit rapidity-azimuth
area, both methods will turn out to have systematics that are under
control and of similar magnitude, at about the $20\%$ level (except
at high $p_t$ for the traditional method).
However it is also important to have knowledge of fluctuations of the
underlying event (both intra and inter-event).
Since it is the jet-area/median method that will prove to be the more
robust tool for measuring them,
it is  this method that we will use when, in section
\ref{sec:area-based-stuff}, we look at a range of possibly interesting
measurements of the UE.
They will be carried out on Monte Carlo events simulated with Pythia
6.4\cite{Sjostrand:2006za} and with Herwig 6.5~\cite{Herwig} with
Jimmy~4.3~\cite{Butterworth:1996zw}.
This part of
the study will  help validate the
understanding developed with the toy-model, and illustrate determinations
of the UE average $p_t$ flow as a function of rapidity, its event-to-event
fluctuations, its intra-event fluctuations, and the degree of
intra-event correlation.
These observables go beyond the kinds of measurements that are commonly discussed
for Tevatron or envisaged so far for the LHC and, as we shall see,
there will be substantial differences between Jimmy and Pythia UE
tunes on a number of them.

% 
% 
% %
% %
% 
% %
% %
% 
% 
% 
% 
% %
% %
% %
% 
% % 
% %
% 

%======================================================================
\section{Overview of measurement approaches}
\label{sec:overview-of-approaches}

%----------------------------------------------------------------------
\subsection{Traditional approach}
\label{sec:trad-appraoch}

\begin{figure}
  \centering
  \includegraphics[width=0.25\textwidth, trim=  0 -150 0 0]{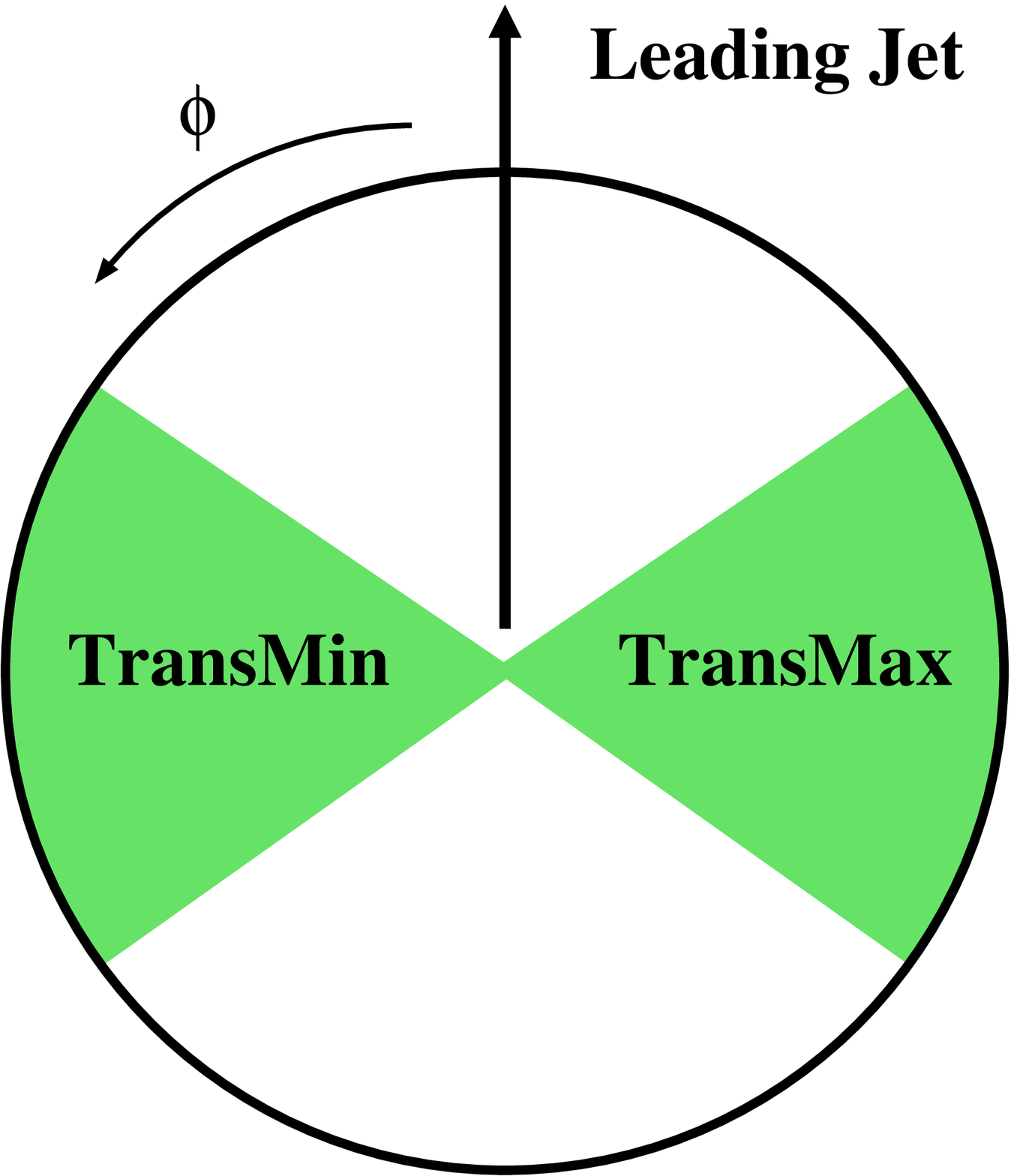}
  \hspace{5pt}
  \includegraphics[width=0.20\textwidth, trim = 0 -50 0 0]{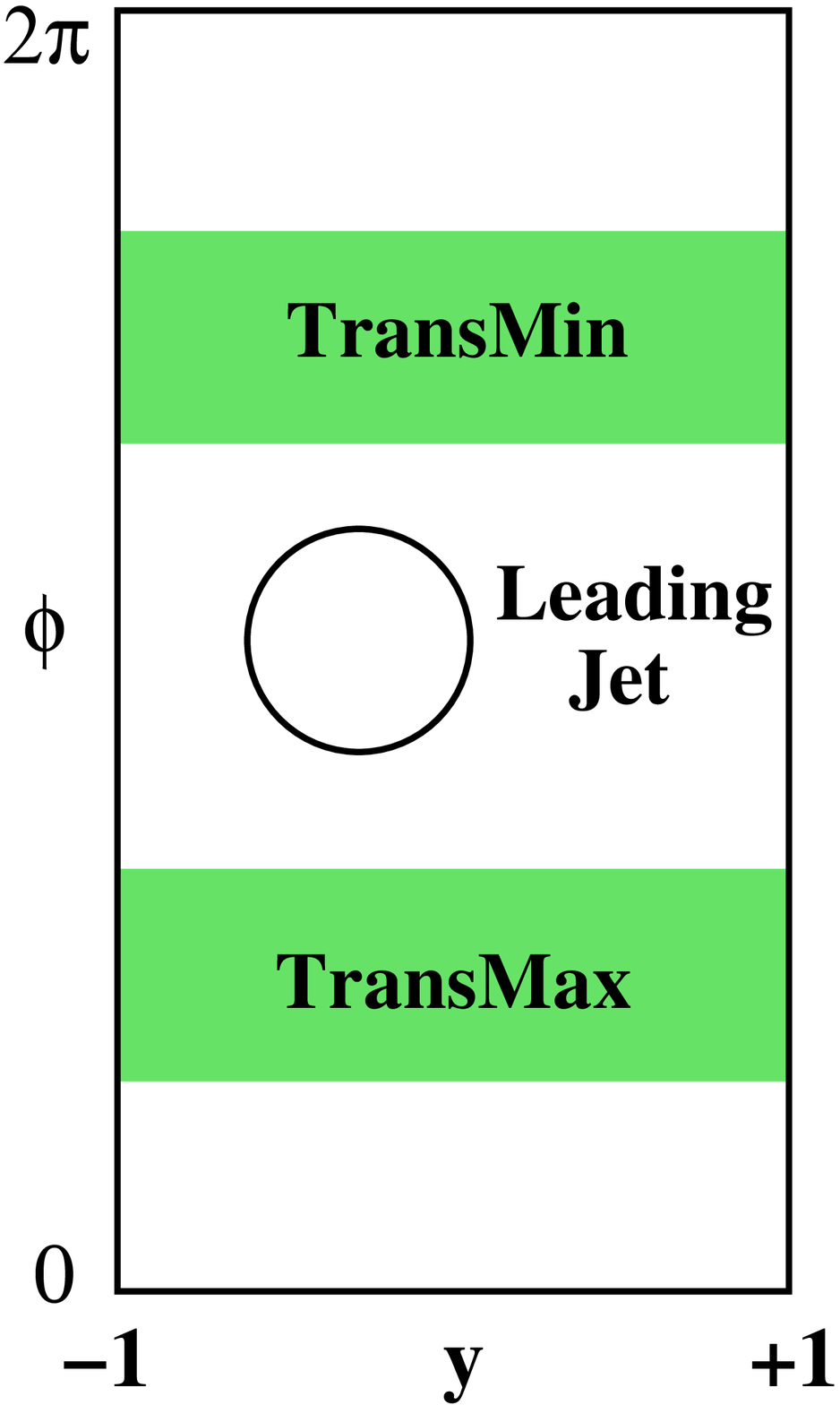}
  \hspace{20pt}
  \includegraphics[width=0.4\textwidth]{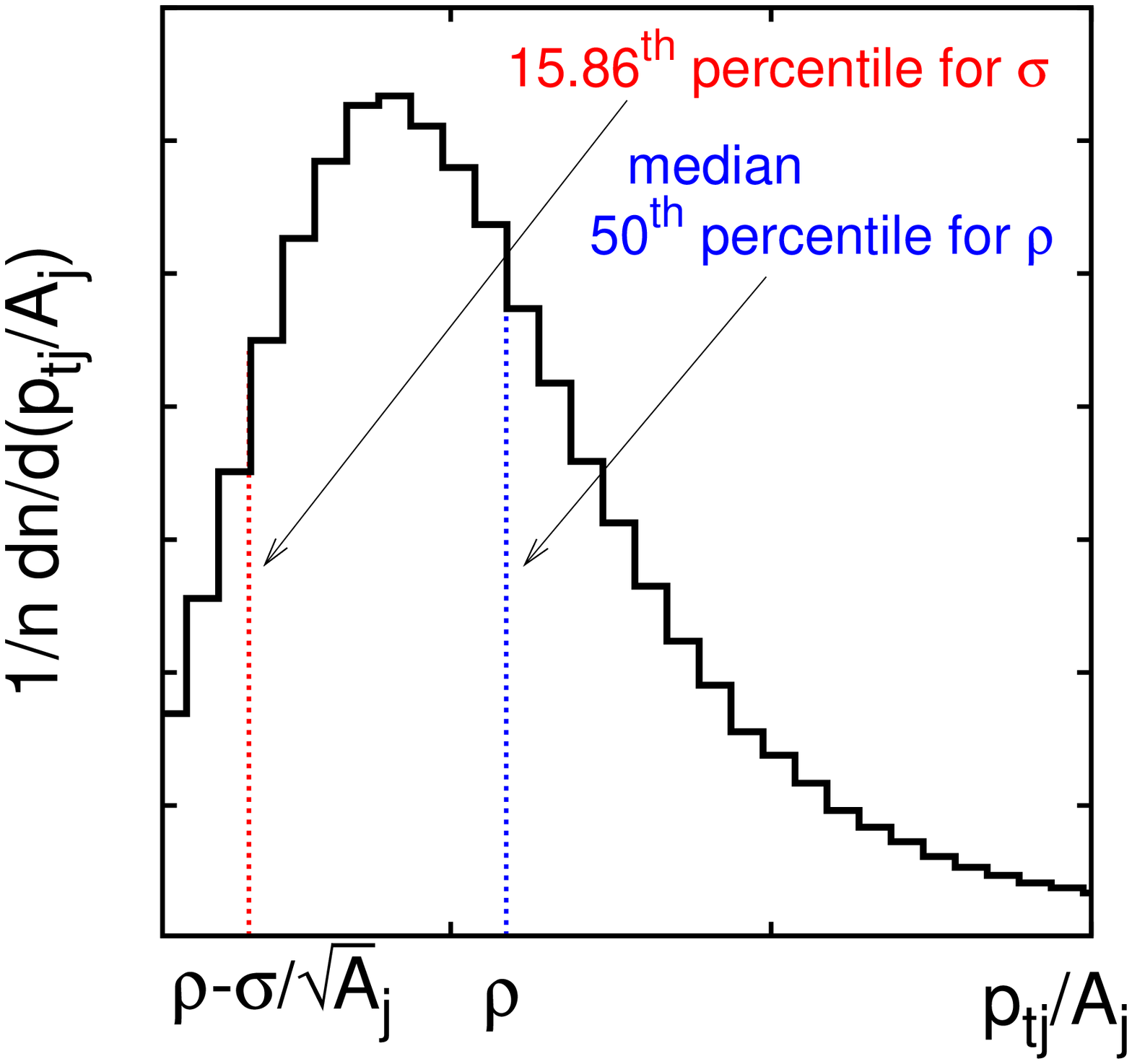}
  \caption{%
    Illustration of underlying-event measurement methods. 
    Left: representation of TransMin and TransMax regions of the
    rapidity azimuth plane in the traditional approach.
    Right: representative distribution of $p_{tj}/A_j$ for the set of
    (real and ghost) jets in a single event, as used to determine
    $\rho$ and $\sigma$ in the area/median approach.
    %
    % ~(left) the level of transverse momentum for UE is determined as
    % an average $p_t$ of particles in the regions TransMin, TransMax
    % (or both) of the transverse plane. In the area/median based
    % approach (right) median of the distribution of $p_{tj}/A_j$ for
    % a set of (real and ghost) jets is used to find a value of $p_t$
    % per unit area typical for UE.  
  }
  \label{fig:approaches}
\end{figure}

The main current approach \cite{Albrow:2006rt} to measuring the
properties of the underlying event involves considering the central
part of the detector, say $|\eta| < 1$, where the pseudorapidity
$\eta$ is defined as $\eta = -\ln\tan\frac\theta2$. One tags events based on the
presence of a jet (whose direction defines an azimuthal angle
$\phi=0$), and divides the central part of the event into four blocks
in azimuth: the ``towards'' region, typically $|\phi| < \pi/3$, 
an away region $2\pi/3 < |\phi| < \pi$, and two transverse
regions, covering $\pi/3 < |\phi| < 2\pi/3$.
This is illustrated in fig.~\ref{fig:approaches} (left).
Since the trigger and recoil jets will usually occupy the towards and
away regions, one then restricts one's attention to the two transverse
regions.
There one measures the multiplicity of charged tracks above some
transverse-momentum threshold as well as the total transverse momentum
contained in the charged tracks (sometimes normalised per unit
area, $d\eta d\phi$).
The results for the charged track multiplicity and charged $p_t$ flow
are usually presented as averages across many events, as a function of
the $p_t$ of the leading jet. One also sees measurements of the
charged momentum flow as a function of the multiplicity.

Since there is a probability of order $\as$ that at least one of the
transverse regions is contaminated by perturbative radiation from
the dijet event, which substantially affect the extracted
information about the UE's $p_t$ flow.
To work around this, it is usual to label the two transverse regions
as TransMin and TransMax, respectively the less and more active of the
two.
The largest component of perturbative contamination should be
restricted to the TransMax region, while TransMin should be less
affected.

In the earliest variants of this ``two-region'' method
\cite{Marchesini:1988hj,UA1-UE}, the two regions used for sampling the
UE were actually placed at non-central rapidities, rather than central
rapidities and transverse azimuth, and it was the total transverse
energy flow that was considered rather than just its charged
component. 
Another variant measured charged momentum flow in cones
\cite{Acosta:2004wqa}. 
These differences reflect the freedom inherent to this method: the
question of where to place the ``transverse'' regions, and the choice
of their shape and size.
In the rest of this article we will always assume that the transverse
regions are well separated from the dijet system in an event, as shown
in fig.~\ref{fig:approaches} (left), and we will quote our results as
a function of the \emph{area} $A_\trans$ of each of the transverse
regions.

%----------------------------------------------------------------------
\subsection{Jet-area/median approach}
\label{sec:area-median-approach}

In the jet-area/median approach, one first clusters the event with a
Cambridge/Aachen (C/A) \cite{Cam,Aachen} or inclusive $k_t$
\cite{KtHH,Kt-EllisSoper} type jet algorithm.
To each jet, $j$, one attributes an ``active'' jet-area, $A_j$, as
described in more detail in \cite{Cacciari:2008gn}.
This is calculated by adding a large number of ``ghost'' particles to
the event (each with negligible $p_t\sim 10^{-100}\GeV$) and including
them in the clustering.
The area of a jet is then proportional to number of ghosts it
contains. Some jets contain just ghost particles (``pure ghost
jets'') and are considered to have $p_t=0$.

% 
% %

A proposal~\cite{Cacciari:2007fd} for a way to measure the level
$\rho$ of uniform background 
noise in an event is then to take it to be the median of the
distribution of the $p_{tj}/A_{j}$ for the ensemble of jets in that
event
\begin{equation}
  \label{eq:median}
  \rho = \mathop{\mathrm{median}}_{j \in \mathrm{jets}} 
  \left[ \left\{ \frac{p_{tj}}{A_j}\right\}\right]\,,
\end{equation}
as shown schematically in fig.~\ref{fig:approaches} (right).
The logic of the use of the median is that it is much less susceptible to
contamination from outliers (i.e.\ hard perturbative jets) than the
mean.
In addition to measuring $\rho$ one can also determine the
intra-event fluctuations of the UE. 
We introduce a quantity $\sigma$, defined such that a fraction
$X/2$ of jets satisfy
\begin{equation}
  \label{eq:sigma-def}
  \rho - \sigma/\sqrt{\langle A_j \rangle} < p_{tj}/A_j < \rho\,,
\end{equation}
where $X=\Erf(1/\sqrt{2}) \simeq0.6827$ is the fraction of a Gaussian
distribution within one standard deviation of the mean.
\footnote{This definition is such that, if the noise distribution is Gaussian
and uncorrelated from point to point within the event, then a patch of
area $A$ will have an average noise $p_t$ contamination of $\rho A$ and
standard deviation $\sigma \sqrt{A}$.}
The approach to measuring $\sigma$ is analogous in spirit to the use
of the median for determining $\rho$. As shown in
fig.~\ref{fig:approaches} (right), it is one sided (i.e.\ just
considering jets with $p_{tj}/A_j < \rho$). This choice has been made
so as to limit contamination of $\sigma$ from hard jets when the total
number of jets is small.

This method was originally suggested in \cite{Cacciari:2007fd} as a
way of measuring average pileup noise across an entire event (e.g. for
$|y|<5$).
One difficulty when using it for the UE will come from the fact that
the UE is significantly softer, and has relative fluctuations that are
larger and less Gaussian than those in (say) 20 pileup minimum-bias
events, both consequences of the lower density of particles.

Another point is that in measuring the UE it is important to obtain
differential information on the UE's rapidity ($y=\ln
\frac{E+p_z}{E-p_z}$) dependence.
This leads us to consider
\begin{equation}
  \label{eq:median-y}
  \rho(y) = \mathop{\mathrm{median}}_{j \in \mathrm{jets},\;
    |y_j - y|< \delta y} \left[ \left\{ \frac{p_{tj}}{A_j}\right\}\right]\,,
\end{equation}
where $2\delta y$ is the width of a rapidity window in which one
carries out the measurement.
The fact that one has a relatively limited number of jets in a given
rapidity window will be one further challenge that we will face in
using this method for studying the UE, because the relative impact of
the presence of a hard jet in the region of interest is amplified by
the small total number of jets.
A modification of the method that helps address these difficulties is
``hard jet removal,'' first employed in the STAR collaboration's
\cite{Salur:2008hs} use of the techniques of \cite{Cacciari:2007fd}
for estimating the (very large) UE for jet measurements in heavy-ion
collisions.
We will investigate the impact of this modification here for events
with dijet topologies, in which case we will simply remove the two
hardest jets in the event from the overall list of jets.

A choice that is present in both pileup and UE measurements with the
area/median method is that of the jet algorithm and jet radius
$R$. Both the $k_t$ \cite{KtHH,Kt-EllisSoper} and Cambridge/Aachen
\cite{Cam,Aachen} algorithms are suitable options, because they
produce jets whose area distribution is quite regular. 
In contrast, algorithms that give mostly conical jets (like
anti-$k_t$~\cite{Cacciari:2008gp} and, to a lesser extent,
SISCone~\cite{Salam:2007xv}) 
tend not to be, because they fill in the ``holes'' between the cones
with jets with very small areas, which can have unrepresentative
$p_t/A$ values.
The question of what $R$ value to use is one of the freedoms of the
method and will be discussed in the coming section.

% 
% 
% 

%% 
%% 

%======================================================================
%\section{A toy model for investigating systematic biases}
\section{A toy model}
\label{sec:systematics}

To understand the strengths and weaknesses of different UE-measurement
approaches, it is helpful to consider events as consisting of two
components: a low-$p_t$ (``soft'') noise component, \emph{defined} to
be the UE, supplemented with hard jets from a perturbative
scattering and associated higher-order corrections.
We will introduce models for each of these two components and
investigate how the methods behave when either of the two components
is present alone and when both are present together.
Our guiding principle in designing these models has been to keep them
sufficiently simple as to be treatable analytically, while also
maintaining a reasonable degree of realism.
While our combined hard and soft models will not quite have the
continuous transition between hard and soft components that is present
in Monte Carlo MPI models, we will see that they nevertheless lead to
certain signature behaviours in UE measurement methods that 
correspond nicely to what is observed in Monte Carlo simulations.

%

%----------------------------------------------------------------------
\subsection[Low-$p_t$ component]{Low-$\boldsymbol{p_t}$ component}
\label{sec:toy-low-pt-part}

As a simple model for the underlying event let us imagine that on
average in a patch of unit of area there are $\nu$ particles, that the
probability distribution for the number of particles $n$ in a specific
patch of area $A$ follows a Poisson distribution, $P_n = (\nu A)^n
e^{-\nu A}/n!$, and that the single-particle transverse-momentum
probability distribution is given by
\begin{equation}
  \label{eq:pt-dist-P1}
  %% GPS: changed normalisation so that we include 1/P_1 (since
  %%      we have now explicitly introduced P_1). Did this everywhere
  %%      else too
  \frac{1}{P_1}\frac{dP_1}{dp_t} = \frac1\mu e^{-p_t/\mu}\,,
\end{equation}
where $\mu$ is the mean transverse momentum per particle.
This particular form has been chosen mainly because it will allow us
to carry out analytical calculations.\footnote{Another simple variant
  will be considered in appendix~\ref{sec:more-toy-model}. It would
  also be interesting to consider a distribution suppressed as
  $1/p_t^n$ at large $p_t$, more in line with the scaling that is to
  be found with multiple parton interactions. However, we have not
  found a form with this property that can be handled analytically
  throughout the calculation. Given that the results from the toy
  model will reproduce many features that we see in Monte Carlo
  simulation, we believe that the form we are using is adequate for
  our purposes.}
If a patch contains $n$ particles then the probability distribution for
its $p_t$ is given by
\begin{equation}
  \label{eq:pt-dist-Pn}
  \frac{1}{P_n} \frac{dP_n}{dp_t} = \frac{1}{(n-1)!}\frac{p_t^{n-1}}{\mu^n}  e^{-p_t/\mu}\,.
\end{equation}
For a patch of area $A$, summing over the Poisson distribution for the
number of particles in the patch ($\langle n \rangle = \nu A$) gives
us the overall probability distribution for the transverse momentum in
the patch as 
\begin{subequations}
    \label{eq:pt-dist-PA}
  \begin{align}
    \label{eq:pt-dist-PA-sum-ingredients}
    \frac{dP}{dp_t}(A) &= 
    \sum_{k=0}^\infty \frac{dP_k}{dp_t} =
    \delta(p_t)e^{-\nu A} + \sum_{n=1}^\infty
    \frac{(\nu A)^n}{n!} e^{-\nu A}
    \frac{1}{(n-1)!}\frac{p_t^{n-1}}{\mu^n}  e^{-p_t/\mu}\\
    \label{eq:pt-dist-PA-sum-results}
    &= \delta(p_t)e^{-\nu A} + e^{-\nu A - p_t/\mu} 
       \sqrt{\frac{A\nu}{\mu p_t}}\; I_1\left(2\sqrt{\frac{A \nu p_t}{\mu}}\right)\,,
  \end{align}
\end{subequations}
where $I_1$ is the (first order) modified Bessel function of the first kind.
The mean and standard deviation of the distribution are given by $\nu
A \mu$ and $\sqrt{2\nu A} \mu$. It is convenient to express this as
saying that the transverse momentum in a patch of area $A$ is
\begin{equation}
  \label{eq:ptA}
  p_t(A) = \rho A \pm \sigma \sqrt{A}\,, 
\end{equation}
where, in the model discussed here,  $\rho$ and $\sigma$ are given by
\begin{subequations}
\label{eq:rho-sigma-toy}
  \begin{align}
  \rho & =  \nu \mu\,,  \\
  \sigma & = \sqrt{2\nu} \mu = \rho\, \sqrt{\frac{2}{\nu}}\,. 
  %\rho & =  \nu \mu$ and $\sigma = \sqrt{2\nu} \mu = \rho \sqrt{2/\nu} \\
  \end{align}
\end{subequations}
In the limit in which $\nu A \gg 1$, the distribution in
eq.~(\ref{eq:pt-dist-PA}) tends to a Gaussian with mean and standard
deviation as given in eqs.~(\ref{eq:ptA}) and (\ref{eq:rho-sigma-toy}). This is a consequence of the
central-limit theorem.

\paragraph{Traditional approach.} %
Taking the area of each of the transverse regions to be $A_\trans$, the
traditional approach will extract the following results for $\rho$ in
the transverse Average, Min and Max regions
\begin{subequations}
\label{eq:pt-trad-Gauss}
  \begin{align}
    \label{eq:pt-trad-Gauss-av}
    \mean{\rho_\mathrm{\ext,Av}} &= \rho\,, \\
    \label{eq:pt-trad-Gauss-min}
    \mean{\rho_\mathrm{\ext,Min}} &= \rho  - 
    \frac{\sigma }{ \sqrt{\pi A_\trans}}
    %\frac{\sigma}{\sqrt{\pi A}}
    \,,  \\
    \label{eq:pt-trad-Gauss-max}
    \mean{\rho_\mathrm{\ext,Max}} &= \rho  + 
    \frac{\sigma}{\sqrt{\pi A_\trans}}
    %\frac{\sigma}{\sqrt{\pi A}}
    \,,
  \end{align}
\end{subequations}
where the Min and Max results have been derived using the Gaussian
limit of eq.~(\ref{eq:pt-dist-PA}).
The only one of the above results that correctly estimates $\rho$ is
the average. The Min and Max results tend slowly to the correct answer in
the limit in which $A_\trans$ is large.

In the literature (e.g.\ \cite{Albrow:2006rt}), $A_\trans$ has usually been
taken equal to $2\pi/3$.  As we shall see in section
\ref{sec:area-based-stuff}, a typical value for $\sigma$ at the LHC
($\sqrt{s}=10\TeV$) is $\sigma\simeq 0.5\rho -0.75 \rho$, which implies
that the $\sigma / \sqrt{\pi A_\trans}$ term is about $20-30\%$ of $\rho$.

\paragraph{Area/median-based approach.} 
To help understand the behaviour of the area/median-based approach, let
us replace the jets (which have a range of areas) with uniform
rectangular tiles, each of which has a fixed area $A_{\tile}$.
It is important to use the full distribution $dP/dp_t(A_{\tile})$ as
given by eq.~(\ref{eq:pt-dist-PA}) rather than the Gaussian
distribution, because a physically interesting domain is that in which
$\nu A_{\tile}$ is of order $1$.
The extracted value $\rho_\ext$ of the UE $p_t$ density in the tiled approximation is given by
the median value of $p_{t,\tile}/A_{\tile}$ across the many tiles in a single
event. 
It can be determined from the solution of the equation
\begin{equation}
  \label{eq:rho-ext-tiles}
  \int_0^{A_{\tile}\, \rho_{\ext}} dp_t \frac{dP}{dp_t}(A_{\tile}) = \frac12\,,
\end{equation}
This result has been obtained in the limit of there being a large
number of tiles, i.e.\ large $A_\tot$, which allows us to approximate
the distribution of tile transverse momenta in a specific event with
the average probability distribution $\frac{dP}{dp_t}(A_{\tile})$ (see
also appendix~\ref{sec:pure-soft-case-fluct}).
The integral in eq.~(\ref{eq:rho-ext-tiles}) is non-trivial to evaluate
analytically, however an
approximation to the solution for  $\rho_{\ext}$ that is accurate
to a couple of percent and has the correct asymptotic behaviours is
given by
\begin{equation}
  \label{eq:rho-ext-tiles-approx}
  {\rho_{\ext}} \simeq \rho\, \frac{\nu A_{\tile} - \ln 2}{\nu A_{\tile}
    - \ln 2 + \frac12} \,\Theta(\nu A_{\tile} - \ln 2)\,.
\end{equation}
The result is non-zero only for $\nu A_{\tile} > \ln 2$, which stems
from the requirement that tiles with no particles, i.e.\ the
$\delta(p_t)$ contribution in eq.~(\ref{eq:pt-dist-PA}), should not
account for more than half of the total number of tiles.
This property of a sudden turn-on, as well as the fact that at large
$\nu A_\tile $ the offset from the correct $\rho$ goes as $1/(\nu
A_\tile )$,
\begin{equation}
  \label{eq:rho-tile-offset}
  \frac{{\rho_\ext} - \rho}{\rho} = - \frac{1}{2 \nu A_\tile } +
  \order{(\nu A_\tile )^{-2}} \,,
\end{equation}
are features that we have found to hold for certain other analytic forms
of $dP_1/dp_t$, notably all those with a structure $p_t^m
e^{-(m+1) p_t/\mu}$ (for arbitrary positive $m$).
Other characteristics, such as the particular coefficient of the
$1/(\nu A_\tile)$ offset in eq.~(\ref{eq:rho-tile-offset}), or the
analytic structure close to the turn-on, do depend on the form taken
for $dP_1/dp_t$.

The determination of $\rho$ from the median of jets' $p_t$ densities
differs from the above ``tiled'' model in that jets do not have a
fixed area.
There is no simple way of extending the analytical model so as to
account for this, however one can study the impact of the question
numerically, by examining toy events in which many soft particles have
been generated according to eq.~(\ref{eq:pt-dist-P1}), with random
positions in the $y-\phi$ plane. 
On each event one runs the jet-based procedure to determine
$\rho_\ext$ and compares it to $\rho_\ext$ for the tile-based
procedure and to the analytic approximation
eq.~(\ref{eq:rho-ext-tiles-approx}).
We assume that tiles of $A_{\tile}$ are comparable to jets of the same
average area, $A_{\tile} = \langle A_\mathrm{jet} \rangle$.
One might think that  the average jet area should be given by the
result for ghost jets in Ref.\cite{Cacciari:2008gn}, $\langle
A_\mathrm{ghost-jet} \rangle \simeq 0.55 \pi R^2$. However, this only
holds in the limit of very dense UE; for the typical kinds of
configuration that are of interest to us, it will be more appropriate
to use 
\begin{equation}
  \label{eq:ghostjet-area-c_N}
    \langle A_\mathrm{jet} \rangle \simeq c_J R^2 \,, 
    \qquad 
    c_J \simeq 0.65\pi \simeq 2.04\,,
\end{equation}
where we have defined a constant $c_J$ that will reappear in several
places below.

Given this relation between the typical jet area and radius $R$, we
can deduce the critical radius, $R_{\crit}$, below which $\rho$ is
zero,
\begin{equation}
  \label{eq:Rcrit}
  R_\crit \,\simeq\, \frac{\sigma}{\rho} \cdot \sqrt{\frac{\ln 2}{2c_J}}
  \,\simeq\, 0.41 \,\frac{\sigma}{\rho}\,.
\end{equation}
We can also rewrite eq.~(\ref{eq:rho-tile-offset}) in terms of $R$ and
$\sigma/\rho$ or $R_{\crit}$:
\begin{equation}
  \label{eq:rho-jet-offset}
  \frac{{\rho_\ext}-\rho}{\rho} = -\frac{\sigma^2}{4\rho^2 c_J R^2} +
  \order{\frac{\sigma^4} {\rho^4 R^4}}
  = - \frac{1}{2 \ln 2} \frac{R_\crit^2}{R^2} + \order{\frac{R_\crit^4}{R^4}}\,.
\end{equation}
The above results depend on the specific form of toy model that one
chooses. 
To estimate the importance of this model dependence, one can replace
$dP_1/dp_t$ as given in eq.~(\ref{eq:pt-dist-P1}) with the alternative form
\begin{equation}
  \label{eq:pt-dist-P1-variant}
  \frac{1}{P_1}\frac{dP_1}{dp_t} = 4 p_t \mu^{-2} e^{-2p_t/\mu}\,,
\end{equation}
as discussed in more detail in appendix~\ref{sec:more-toy-model}. The
essential relations for this model are $\sigma/\rho =
\sqrt{3/(2\nu)}$, $R_{\crit} = \frac\sigma\rho \sqrt{(2\ln 2)/(3
  c_J)}\simeq 0.48 \frac\sigma\rho$ and
\begin{equation}
  \label{eq:rho-jet-offset-variant}
  \frac{{\rho_\ext}-\rho}{\rho} = -\frac{2\sigma^2}{9 \rho^2 c_J R^2} +
  \order{\frac{\sigma^4} {\rho^4 R^4}}
  = - \frac{1}{3 \ln 2} \frac{R_\crit^2}{R^2} + \order{\frac{R_\crit^4}{R^4}}\,.
\end{equation}
These results involve the same analytic structures as for the original
form of the toy model, with numerical coefficients that imply slightly
smaller corrections for finite $R$.
\begin{figure}[t]
  \centering
  \includegraphics[width=0.6\textwidth]{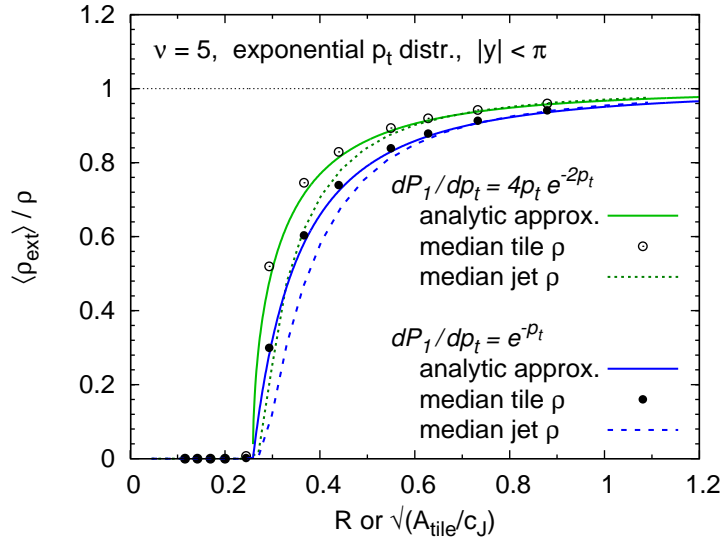}
  \caption{%
    Comparison of jet and tile median-based determinations of $\rho$
    for two toy soft-event models (as described in the text), together
    with the corresponding analytical approximations. 
    The mean number of particles per unit area, $\nu$, was $5$
    (commensurate with typical expectations for the UE at LHC) and
    the jets that were used had $|y|<\pi$ (this choice stems from our
    use of square tiles). 
    To achieve sub-percent-level agreement between the analytical and
    jet-based results at larger $R$ values, it was necessary to carry out
    the determination of $\rho_\ext$ using $p_t$-scheme recombination
    of $4$-momenta in the jets.}
  \label{fig:analytical-soft-UE}
\end{figure}

In fig.~\ref{fig:analytical-soft-UE}, the approximate analytical toy
model results as a function of $R$ (or the equivalent $A_{\tile}$) are
compared to the average
results obtained by applying the ``tiled'' median approach as well as
the jet-area median approach to toy-model configurations. 
% In fig.~\ref{fig:analytical-soft-UE}, the approximate analytical toy
% model results as a function of $R$ (or the equivalent $A_{\tile}$) are
% compared to the exact 
% results for the ``tiled'' median approach as well as to those obtained
% by applying the jet-area median approach to toy-model configurations.
%
This is done for both variants of the toy model.
%
% %
% 
%
There is near perfect agreement between the analytical approximations
and the average median tile-based results. 
This is indicative both of the quality of the analytical approximation
and of the limited impact of the practical use of finite $A_\tot$
(here $4\pi^2$), as opposed to the large $A_\tot$ limit that went into
the analytical results.
The median jet-based results are  rather close to the tile-based results for larger
$R$ values,\footnote{To get this level of agreement for $R\sim 1$, it
  turned out to be necessary to use the $p_t$ recombination scheme in
  determining the jet momenta.}
though the precise shape in the turn-on region
differs a little with respect to the tile-based expectation.
Moderate differences exist between the results with the two choices for
$dP_1/dp_t$, and these can be taken as indicative of the magnitude of the
model-dependence in the above analysis.

One message to take from fig.~\ref{fig:analytical-soft-UE} is that for
an $R$ value that is twice that where the turn-on occurs,
$\rho_{\ext}$ underestimates $\rho$ by about $10-20\%$.
This can be kept in mind as a ballpark value for the accuracies that
we will be able to achieve and can be compared to the $20-30\%$ effect
discussed above for the TransMin $\rho$ determination in the
traditional approach.

% 
% 
% 
% %
% %
% %
% 

%----------------------------------------------------------------------
\subsection{Purely perturbative events}
\label{sec:toy-pure-pert}

In our definition of purely perturbative events there is no
underlying event and the \emph{only} regions of the event
that have non-zero momentum are those that contain a perturbative
emission. 
Nevertheless UE-determination methods may still give a non-zero result
for their estimate of the UE energy density, because of the way they
are affected by those extra jets.

We work here with the assumption that we have selected events with at
least two hard jets (or with a $W$ or $Z$ boson), and that extra jets
may be present at higher perturbative orders.
A crude, but illustrative approximation for those higher orders will
be obtained as follows. We will take the dominant source of
emissions to be radiation from the initial-state partons that enter the hard
reaction.\footnote{Ignoring radiation from any outgoing (Born) partons
  is not too poor an approximation, because a significant part of that
  radiation will be contained within the corresponding jets. }
Furthermore we will assume that those emissions are soft, distributed
uniformly in rapidity and azimuth, and with a $p_t$ spectrum given by
$\as(p_t)/p_t$:
\begin{equation}
  \label{eq:dndptdy}
  \frac{dn}{dp_t dy d\phi}\simeq \frac{C_i}{\pi^2} \frac{\as(p_t)}{p_t}\,,
\end{equation}
where $C_i$ is the colour factor associated with the incoming partons
($C_i=C_A\equiv3$ for gluons, $C_i=C_F\equiv\frac43$ for quarks).
For a dijet event whose two hard jets have transverse momenta
$p_{t,\hard}$, we will take eq.~(\ref{eq:dndptdy}) to be valid
independently of rapidity for $Q_0 < p_t < Q$, where $Q\sim \frac12
p_{t,\hard}$ and $Q_0 \sim 1 \GeV$.\footnote{Based on collinear
  factorisation, one would expect that the upper limit on the $p_t$ of
  emissions to have significant rapidity dependence. For example, if
  the hard process takes place at central rapidities, then one might
  write $p_t \lesssim p_{t,\hard} e^{-|y|}$.
  %(there would also be some
  %enhancement of central emissions due to radiation from the dijet
  %system itself).
  %
  The rapidity-independent approximation is instead inspired by a
  high-energy factorisation picture, relevant when $p_t \ll \sqrt{s}
  e^{-|y|}$.
  Studies with Herwig at parton-level (based on collinear
  factorisation) give a distribution for the upper $p_t$ limit on
  extra jets that is intermediate between these two expressions.}
We will also assume that emissions are independent. Thus,
the probability distribution for the number of emissions will be a
Poisson distribution whose mean is obtained by the integral of
eq.~(\ref{eq:dndptdy}) over the relevant phase space.

\paragraph{Traditional approach.} 
The average $p_t$ densities in the Average and the Max regions will
both receive contributions from the emission of one gluon (relative to
the Born diagram for the process). In contrast, the Min region only
receives a contribution when at least two gluons have been
emitted.
One can obtain the $\rho_{\ext,\av}$ value just by integrating
eq.~(\ref{eq:dndptdy}) up to $Q$, which we do in a fixed coupling
approximation, $\as =\as(Q)$, since the integral is dominated by
values of $p_t \sim Q$:
\begin{equation}
  \label{eq:pt-trad-pert-av}
  \langle \rho_{\ext,\av} \rangle = 
  \frac{1}{2A_\trans} \int^Q_0  p_t dp_t 
      \int_{2A_\trans}\hspace{-2.0em} dy d\phi\; \frac{dn}{dp_t dy d\phi} = 
  \frac{C_i\as}{\pi^2} Q\,.
\end{equation}
Here we have neglected the (small) impact of $Q_0$ and, in the
fixed-coupling approximation, the result is complete to all orders in
$\as$.
One feature to note about the result is that it scales with $Q$. 

To determine $\langle \rho_{\ext,\Min}\rangle$ to $\order{\as^2}$, we
assume that the left and right transverse regions each contain one
gluon, and that the left-hand gluon ($L$) is harder than the
right-hand one ($R$);
$\mean{\rho_{\ext,\Min}}$ is then given by the average transverse momentum in the
right-hand region,
\begin{equation}
  \label{eq:pt-trad-pert-min}
  \langle \rho_{\ext,\Min}\rangle = 
      2 \frac{1}{A_\trans} \left(\frac{C_i\as A_\trans}{\pi^2}\right)^2
  \int_0^Q \frac{dp_{t,L}}{p_{t,L}} \int_0^{p_{t,L}}
  \frac{dp_{t,R}}{p_{t,R}} p_{t,R}
  = 2\left(\frac{C_i\as}{\pi^2}\right)^2 A_\trans\, Q\,,
\end{equation}
with an additional factor of $2$ to account for the case where the
right-hand gluon is the harder one.
The result is proportional to $\as^2$, i.e.\ suppressed by an extra
factor of $\as$ compared to $ \rho_{\ext,\av}$, however it is enhanced
by a factor of $A_\trans$.

Finally we can estimate $\langle \rho_{\ext,\Max}\rangle$ using the
relation $\rho_{\ext,\Min} + \rho_{\ext,\Max} = 2\rho_{\ext,\av}$,
giving us 
\begin{equation}
  \label{eq:pt-trad-pert-max}
  \langle \rho_{\ext,\Max}\rangle 
  = 2\frac{C_i\as}{\pi^2} Q - 2\left(\frac{C_i\as}{\pi^2}\right)^2 A_\trans\, Q\,.
\end{equation}
To appreciate the impact of the various terms, let us take $C_i \equiv
C_A=3$, $Q=50\GeV$, $\as(Q)=0.13$, and $A_\trans=2\pi/3$. Then we obtain
\begin{subequations}
\label{eq:rho-trad-pert-est}
  \begin{align}
    \langle \rho_{\ext,\av} \rangle &\simeq 2.0\GeV, \\
    \langle \rho_{\ext,\Min}\rangle &\simeq 0.3\GeV, \\
    \langle \rho_{\ext,\Max}\rangle &\simeq 3.6\GeV.
  \end{align}
\end{subequations}
These numbers scale roughly linearly with $Q$. The crudeness of our
approximations for the perturbative part of the event means that they
are not be trusted to better than within a factor of two (worse in the
case of $\mean{\rho_{\ext,\Min}}$). However the rough orders of magnitude are
still instructive and highlight the advantage of the Min region.
 
The above analytic estimates can be verified by using more realistic
events from a~Monte Carlo generator at parton level. For the case of
dijets from Herwig $pp$ collisions at $\sqrt{s}=10\TeV$, with $p_t$ of
the partons in the hard 2$\to$2 process required to be above $100
\GeV$ (consistent with $Q=50\GeV$) and the soft underlying event
turned off, one obtains 
\begin{subequations}
\label{eq:rho-trad-herwig}

  \begin{align}
    \langle \rho^{\rm MC}_{\ext,\av} \rangle &\simeq 2.1\GeV, \\
    \langle \rho^{\rm MC}_{\ext,\Min}\rangle &\simeq 0.5\GeV, \\
    \langle \rho^{\rm MC}_{\ext,\Max}\rangle &\simeq 3.8\GeV.
  \end{align}
\end{subequations}
These numbers are very close to those from the simple model of purely
perturbative underlying event described above for gluon jets (though
Herwig has an admixture of quark jets here).
We have verified that if we double the area over which the measurement
is carried out, the $\langle \rho^{\rm MC}_{\ext,\Min}\rangle$ result
roughly doubles, as expected from eq.~(\ref{eq:pt-trad-pert-min}).

One comment concerning the above results is that in the pure soft UE
case it was $\rho_{\ext,\av}$ that was the least biased estimate of
$\rho$. Here it is $\rho_{\ext,\Min}$ that is the least biased by hard
perturbative radiation.
If one restricts one's attention to $\rho_{\ext,\Min}$, then a further
property of interest is that in the soft UE case, the bias is reduced
by increasing the transverse region's area $A_\trans$, while for hard
perturbative contamination increasing $A_\trans$ increases the bias.
This trade-off between the two issues is characteristic of the
difficulty of accurately estimating~$\rho$.

\paragraph{Area/median-based approach}
Let us suppose that we extract $\rho$ based on jets contained in a
region of area $A_\tot$.
Assuming the typical area for jets as introduced in
eq.~(\ref{eq:ghostjet-area-c_N}), 
then the typical number
of jets $N$ in the region (including ghost jets)  should be given by 
\begin{equation}
  \label{eq:NofA}
    N \simeq \frac{A_\tot}{c_J R^2}\,.
\end{equation}
The exact value of $N$ in each given event will depend on that event's
detailed structure (and the exact set of ghosts), but
eq.~(\ref{eq:NofA}) should be adequate for our illustrative discussion
here.

Of the $N$ jets, we will assume that $n_h$ are ``hard'' jets, of which
$n_b$ correspond to the
final-state born particles ($n_b=2$ for dijet events, $n_b=0$ for
Drell-Yan events) and $n_p$ stem from perturbative radiation.
It is convenient, albeit somewhat simplistic, to model $n_p$ as being
given by the number of emitted gluons\footnote{Recall that our model
  includes only primary emissions from the incoming partons --- we
  neglect cases where two of these emissions end up in the
  same jet (which would decrease $n_p$), and also the fragmentation
  contribution from the Born and radiated particles (mostly contained
  within the respective jets, but which would increase $n_p$ in the cases where
  the fragmentation is outside those jets). }
\begin{equation}
  \label{eq:np-nb-pert}
  \langle n_h\rangle =  n_b +  \langle n_p\rangle  \simeq
  n_b + \frac{C_i}{\pi^2} \int_{A_\tot}\hspace{-1em} d\phi\, dy
  \int_{Q_0}^Q \frac{dp_t}{p_t} \as(p_t)
   = n_b  + A_\tot\, \frac{C_i}{\pi^2} \frac{1}{2 b_0} \ln
   \frac{\as(Q_0)}{\as(Q)} \,,
\end{equation}
where we consider the number of perturbative emissions between some
non-perturbative scale $Q_0$ and an upper limit $Q$ related to the
hard scale of the process (e.g.\ half the $p_t$ of the hardest jet, as
before) and we have used a 1-loop running approximation for $\as(p_t)$.

\begin{figure}[t]
  \centering
  \includegraphics[width=0.48\textwidth]{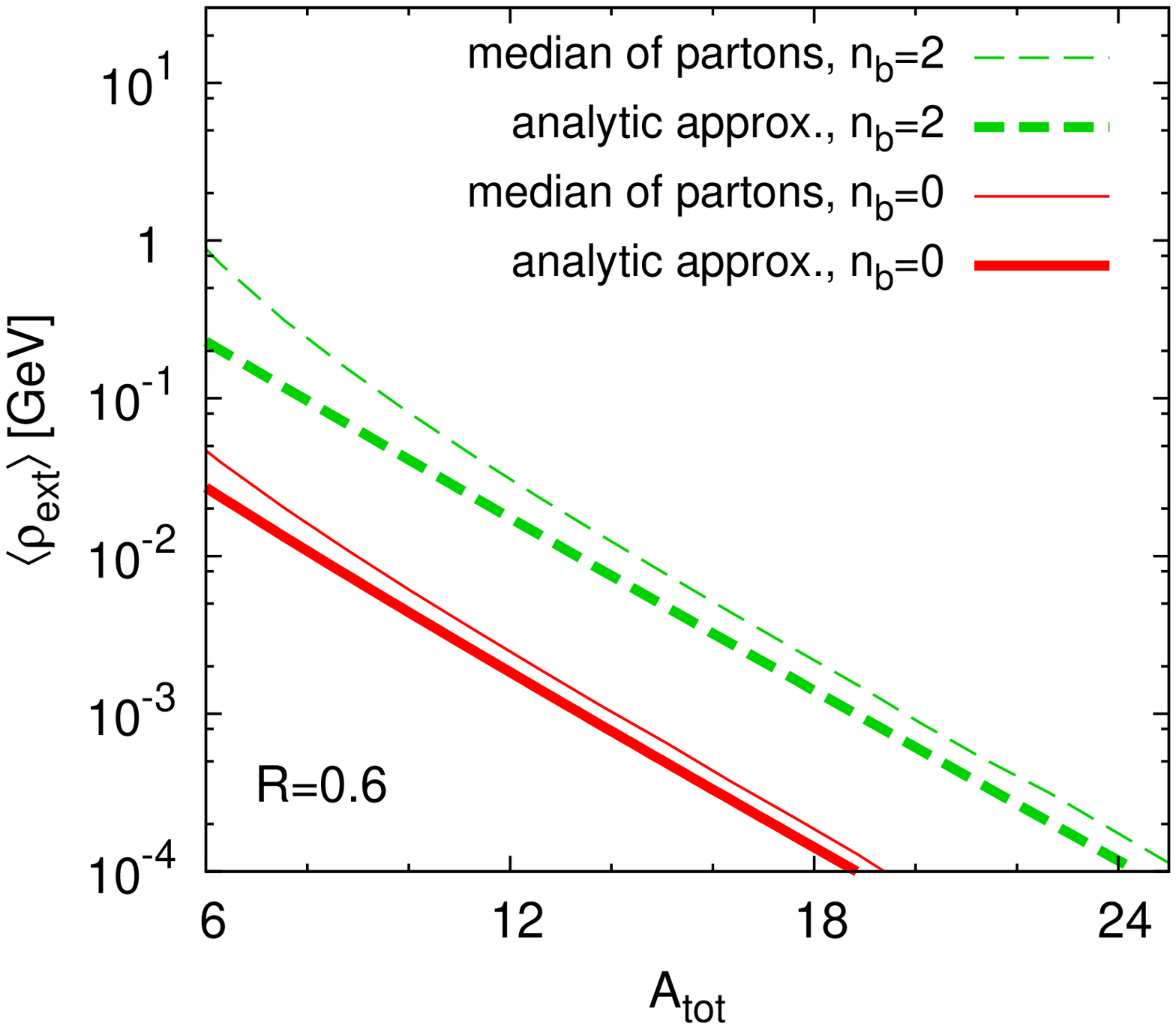}\hfill
  \includegraphics[width=0.48\textwidth]{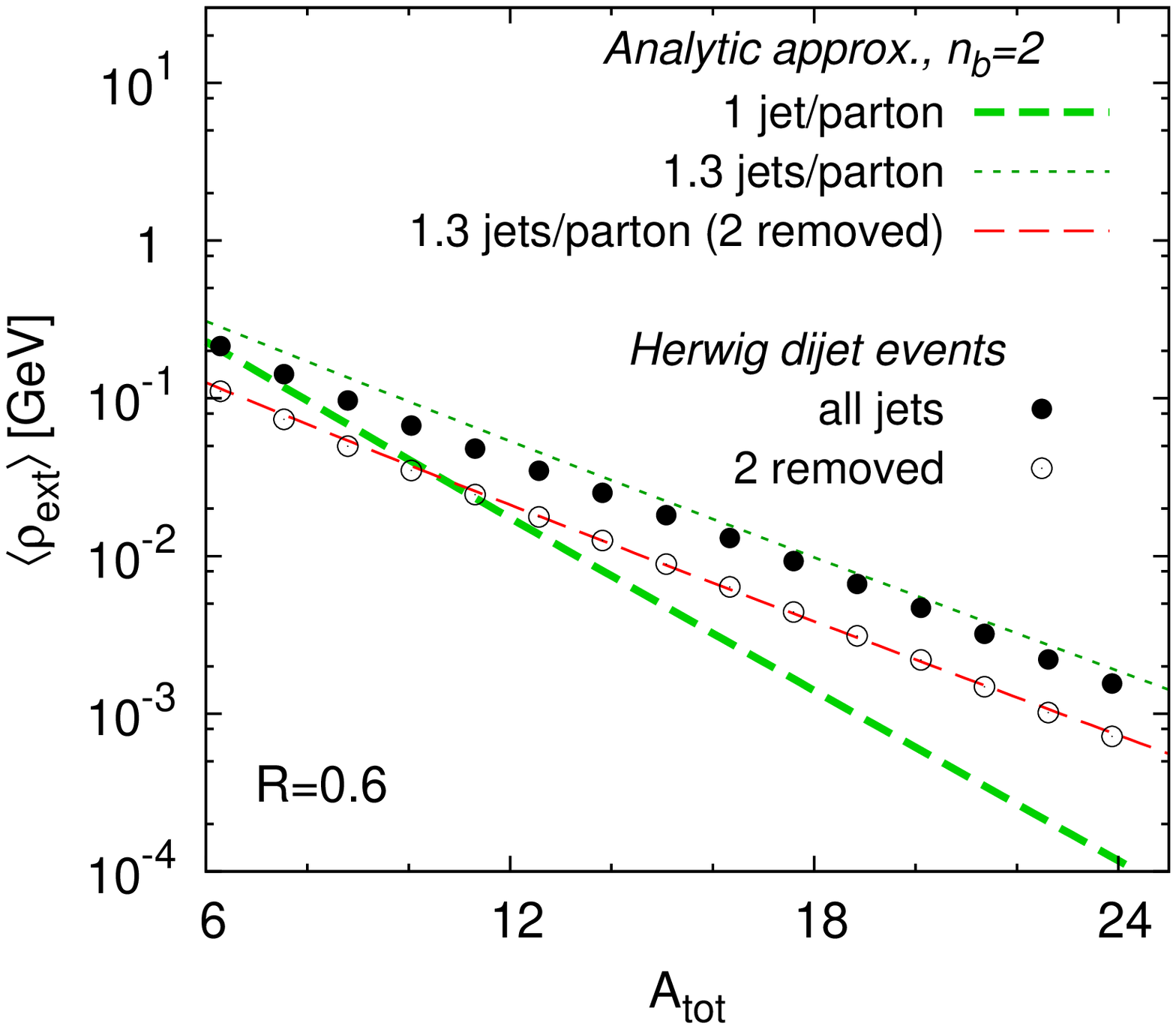} 
  \caption{ Left: pure perturbative underlying event extraction via the area/median
    method, as a function of the total area, for cases with zero and two Born
    particles.  %
    The curves correspond to the simple model of perturbative
    emissions described in the text. Thick lines show the analytic
    estimate of $\rho_\ext$ based on the RHS of
    eq.~(\ref{eq:prob-nonzero-pert-median}) and thin lines correspond to the
    use of the median on our model for the ensemble of parton emissions
    (see text for further details).  %
    Right: comparison of the analytical prediction for the case with 2
    Born particles with results from parton-level (no UE) events
    generated with Herwig, and clustered with C/A algorithm
    (points).  %
    Two analytical results are shown: one with $\mean{n_p}$ as given in
    eq.~(\ref{eq:np-nb-pert}) (1 jet/parton) and another in which $\mean{n_p}
    \to 1.3 \mean{n_p} + 0.3 n_b$, i.e.\ each parton generates an
    average of $1.3$ jets.  
    The figure also shows a curve (analytical) and points (Herwig) that
    illustrate the impact of removing the two hardest jets from the
    list used in determining the median.
    %
   %\comment{We have to do something about to many lines in the plot ue-simple-sebastian/install/plots/toy-pure-pert-paper.eps}
   %\comment{GPS: this plot in any case needs fixing}
    %Eq.~(\ref{eq:rho-ext-pure-pert}) shown as a function of the
    %total area for different numbers of Born partons  and assumptions
    %about the number of jets generated per parton; 
  }
  \label{fig:toy-pure-pert}
\end{figure}

For the median estimator of $\rho$ to be non-zero, at least half the
jets should contain perturbative radiation, \ie $n_p + n_b \ge N/2$.
Since the number of primary emissions follows a Poisson distribution,
we get the probability of non-zero $\rho$ from the following sum
\begin{equation}
  \label{eq:prob-nonzero-pert-median}
  P = \sum_{n = N/2-n_b}^N \frac{\langle n_p\rangle^{n} e^{-\langle n_p\rangle}}{n!}
  \simeq \frac{\langle n_p\rangle^{(N/2-n_b)} e^{-\langle n_p\rangle}}{(N/2-n_b)!}\,,
\end{equation}
where we have also made the approximation that the sum is dominated by
its first term, on the grounds that $\langle n_p \rangle/(N/2 -n_b)
\ll 1$.
Given $P$, one can estimate $\rho_\ext$ by observing  that
the $(N/2)^\mathrm{th}$ jet will be the softest of all the
perturbative jets, and therefore have $p_t \sim Q_0$, giving 
\begin{equation}
  \label{eq:rho-ext-pure-pert}
  \mean{\rho_\ext} \simeq 
  \frac{Q_0}{\langle A_\jet\rangle} P \simeq
  \frac{Q_0}{c_J R^2} P\,.
\end{equation}
This is plotted with thick lines in fig.~\ref{fig:toy-pure-pert}
(left) as a
function of $A_\tot$ for $R=0.6$, using a 1-loop 5-flavour coupling with
$\Lambda=0.1\GeV$ \footnote{This mimics a 2-loop coupling with
  $\as(M_Z)=0.120$ to within a couple of percent over a broad range of
  scales.}
and with $C_i=C_A=3$,
$Q=50$ GeV, $Q_0=1$ GeV and two
values for $n_b$, $0$ and $2$.
 
To estimate the uncertainties of our analytic formula introduced by
approximations~(\ref{eq:prob-nonzero-pert-median})
and~(\ref{eq:rho-ext-pure-pert}) we also plot with thin lines in
fig.~\ref{fig:toy-pure-pert} (left) the result of numerical studies of the same
simple set of perturbative emissions (equivalent to the full sum over
$n$).
As before, we associate each parton (\ie gluon in our case) with one
jet and assume that this jet has area $c_J R^2$. In the case $n_b=2$,
we make sure that the two Born particles are always present
in the region where the underlying event is measured. One sees in
fig.~\ref{fig:toy-pure-pert} that the exact numerical results
(equivalent to taking the full sum in
eq.~(\ref{eq:prob-nonzero-pert-median})) are moderately higher than the
corresponding analytic approximations, as would be expected.  In both
cases, however, the contribution to $\rho_\ext$ is negligible except
for small values of $A_\tot$ with $n_b=2$.

The approximation that we have used for the
distribution of the number of jets is rather simplistic. 
To estimate the order of magnitude of the uncertainties that are
present in our toy model of the perturbatively induced ``UE'', notably those
associated with secondary radiation from the partons, we have also
applied the area/median method to realistic  dijet events from
Herwig with soft UE turned off, taken at parton level. To make the
comparison, we required that the partons in the hard 2$\to$2 process
have $p_t>$ 100 GeV, as before (and consistent with our choice
$Q=50\GeV$). The result is shown in fig.~\ref{fig:toy-pure-pert} (right) with
black and open circles for the cases with and without removal of the
two hardest jets from the ensemble used for the median.
The $A_\tot$-dependence of the Herwig results differs somewhat from the
expectations in eqs.~(\ref{eq:prob-nonzero-pert-median})
and~(\ref{eq:rho-ext-pure-pert}). 
We attribute this to the fact that each parton (be it a Born parton or
a primary emission) can itself radiate extra gluons, some of which
will lead to additional jets. 
A simple way of accounting for this is to replace $\mean{n_p}$
in eq.~(\ref{eq:prob-nonzero-pert-median}) with $(1+X) \mean{n_p} + X
n_b$, where $X$ is the number of extra jets obtained per parton.%
\footnote{A more sophisticated approach would be to calculate $X$
  using soft-gluon resummation, which would lead to a
  single-logarithmic dependence of $X$ on the parton transverse
  momentum.}
Using a modest $X=0.3$ ($1.3$ jets per parton) brings the analytical result into accord with
the Herwig results.

In terms of the practical impact of perturbative emission on the
extracted $\rho$ values, one sees that for $A_\tot \ge 4\pi$ (the minimal
value that we shall use in section~\ref{sec:area-based-stuff}) it
remains a small effect, though the curves also highlight that at small
total area $A_\tot$, the area/median method can start to become sensitive
to perturbative radiation, especially when Born partons are present.

A final point to  comment on is the relation between these results and the
discussion in \cite{Cacciari:2007fd}, where it was argued that
$\rho_\ext$ would be non-zero starting only at
$\order{\smash{\as^{N/2-n_b}}}$.
Examining eq.~(\ref{eq:prob-nonzero-pert-median}), one sees that this
statement stems from the fact that $\langle n_p\rangle \sim \as$ (in a fixed-coupling
approximation).
However, each power of $\as$ is compensated in part by a power of $A
\ln Q/Q_0$ (which is large), and ultimately the small value of $P$ in
eq.~(\ref{eq:prob-nonzero-pert-median}) (and hence $\mean{\rho_\ext}$)
cannot be solely attributed to an $\as$ power-counting argument, but
rather involves a more subtle interplay of all the factors in
eq.~(\ref{eq:rho-ext-pure-pert}).

%  %
%  %
%}

%----------------------------------------------------------------------
\subsection{Two-component events}
\label{sec:toy-two-comp}

Realistic events are neither purely perturbative nor consist of pure
soft noise.
It is instructive to examine what happens if one considers events that
have both components together.

\paragraph{Traditional approach.} 

The transverse-momentum density extracted from the average of the two
transverse regions is straightforward to calculate in the two
component model: it is just given by the sum of the soft and
perturbative components,
\begin{equation}
  \label{eq:pt-trad-Gauss+pert-av}
  \langle \rho_{\ext,\av}\rangle = 
  \rho  + \frac{C_i \as}{\pi^2}Q\,.
\end{equation}
The results for the Min and Max regions are more complex: in the
pure soft-component case, it was the soft radiation that determined
which of the two regions was Min/Max; analogously, in the
perturbative case, it was the perturbative radiation that determined
this.
When both can be present one has to consider which of the two
fixes the Min/Max regions.
It is useful to define $\cP$ to be the fraction of events in which the
amount of perturbative radiation in each of the two transverse
regions is smaller than the size of soft fluctuations of those
regions, $p_{t,L}, p_{t,R} \ll \sigma \sqrt{A_\trans}$.
In this set of events, it is the soft component that defines which
region is Min/Max, and the bias in the extraction of $\rho$ is just
the soft bias, eqs.~(\ref{eq:pt-trad-Gauss}), with no perturbative
bias.
In the remaining events, in which one or more perturbative emissions
are much harder than $\sigma \sqrt{A_\trans}$, it is those perturbative
emissions that will determine which of the two regions is Min/Max.
For these events, there will be no bias in the contribution from the
soft component. 
This implies that the average bias in the Min/Max regions for the
soft component will be $\mp \cP\cdot \sigma/\sqrt{\pi A_\trans}$.

As concerns the perturbative contamination, the average results in
eqs.~(\ref{eq:pt-trad-pert-min}) and (\ref{eq:pt-trad-pert-max}) are
already dominated by the set of events in which there is at least
one hard emission, so these contributions remain unchanged in the
two-component case.
The final result for the Min/Max regions is therefore
\begin{subequations}
  \label{eq:pt-trad-Gauss+pert-min-max}
  \begin{align}
    \label{eq:pt-trad-Gauss+pert-min}
    \langle \rho_{\ext,\Min}\rangle &\simeq \rho 
    - \frac{\sigma \cP}{\sqrt{\pi A_\trans}}
    + 2\left(\frac{C_i\as}{\pi^2}\right)^2 A_\trans\, Q\,,
    \\  
    \label{eq:pt-trad-Gauss+pert-max}
    \langle \rho_{\ext,\Max}\rangle &\simeq \rho 
    + \frac{\sigma \cP}{\sqrt{\pi A_\trans}}
    + 2\frac{C_i\as}{\pi^2} Q - 2\left(\frac{C_i\as}{\pi^2}\right)^2 A_\trans\, Q\,,
  \end{align}
\end{subequations}
where $\cP$ is given by
\begin{equation}
  \label{eq:two-component-P-norad}
  \cP \simeq \exp\left( 
    -2 A_\trans\, \frac{C_i}{\pi^2} \frac{1}{2 b_0} \ln
    \frac{\as(\max(Q_0,\sigma\sqrt{A_\trans}))}{\as(Q)}
  \right)\,.
\end{equation}
Note that the choice of lower scale in the logarithm, $\sigma
\sqrt{A_\trans}$, or $Q_0$ if that is larger, is only controlled to within a
factor of $\order{1}$, just as is the choice of $Q$ for the upper
limit.

\paragraph{Area/median-based approach.}  
The combination of low-$p_t$ and perturbative components is
non-trivial also in the case of the area/median-based approach.
To treat it analytically, it will be convenient to work at $R$ values
that are sufficiently large that the distribution of $p_{t,\jet}/A_\jet$ for the
many jets can be considered approximately Gaussian --- i.e.\ we will
work away from the turn-on region $\nu A_\jet = \ln 2$ that was
discussed in section~\ref{sec:toy-low-pt-part}.

Of the $N$ jets that are used
in determining the median, some will contain hard perturbative
radiation with transverse momentum significantly above the scale of
the fluctuations of the UE.
Assuming that there are on average $\mean{n_h}$ hard partons, and that the
probability distribution of hard partons in a jet (or tile) is given
by a Poisson distribution with mean $\mean{n_h}/N$, then the average
number of jets not contaminated by the hard partons will be given
by $N e^{-\mean{n_h}/N}\simeq N - \mean{n_h} + \mean{n_h}^2/(2N)$.
These uncontaminated jets will have a distribution of values of $\rho_\jet
= p_{t,\jet}/A_\jet$ that is governed just by the soft component and
is roughly Gaussian
% 
% 
% %
% %
%
\begin{equation}
  \frac{dn^{(\soft)\!\!\!}}{d\rho_{\jet}} = \frac{1}{\sigma} \sqrt{\frac{\langle A_{\jet}\rangle}{2\pi}} 
  \exp\left(
    -\frac{\langle A_{\jet}\rangle  }{2\sigma^2 } (\rho_{\jet} - \rho)^2
    \right) \cdot N \exp\left(-\frac{\mean{n_h}}{N}\right)\,.
\end{equation}
%% %
%% 
%% %
%% 
Assuming $\mean{n_h} < N\ln 2$, the median procedure implies finding $\rho_\ext$
such that $N/2$ of the $Ne^{-\mean{n_h}/N}$ Gaussian-distribution jets have
$\rho_\jet < \rho_\ext$, \ie one must determine the value of
$\rho_\ext$ such that
\begin{equation}
  \int_{-\infty}^{\rho_{\ext}} d\rho_{\jet}   \frac{dn^{(\soft)\!\!\!}}{d\rho_{\jet}} = 
  \frac{N}{2}\,,
\end{equation}
(the unphysical negative lower limit of the integral, an artefact of
the Gaussian approximation, doesn't perturb the argument).
In the small $\mean{n_h}/N$ limit, this is easily solved and
gives%
\begin{equation}
  \label{eq:rho-AM-Gauss+pertN}
  \mean{\rho_{\ext}} \simeq \mean{\rho_{\ext}^{(\soft)}}+ 
  \sigma \sqrt{\frac{\pi}{2\langle A_{\jet} \rangle}}
  \left(\frac{\mean{n_h}}{N} 
     + \frac{\mean{n_h}^2}{2N^2} +\order{\frac{\mean{n_h}^3}{N^3}} 
  %\left(1+\frac{n_h}{N} +\order{\left(\frac{n_h}{N}\right)^2} 
  \right)\,,
\end{equation}
where $\mean{\rho_{\ext}^{(\soft)}}$ is the result obtained in the
pure soft case, eq.~(\ref{eq:rho-ext-tiles-approx}) of
section~\ref{sec:toy-low-pt-part}.\footnote{The additivity of soft and
  hard results is an approximation, justified only when
  $\mean{\rho_{\ext}^{(\soft)}}$ is close to $\rho$.
  An additional point is that when plotting soft+hard results for
  $\mean{\rho_{\ext}}$, we will eliminate the $\Theta$-function in
  eq.~(\ref{eq:rho-ext-tiles-approx}) and use the prescription that
  $\mean{\rho_{\ext}}$ is well-defined only when the sum of soft
  and hard contributions gives a positive result.
}
One can then use eqs.~(\ref{eq:ghostjet-area-c_N}) and (\ref{eq:NofA}) to
express $\langle A_\jet\rangle$ and $N$ in terms of $R$ and $A_\tot$. 
Keeping only the first two terms in the $R$ expansion gives
\begin{equation}
  \label{eq:rho-AM-Gauss+pert-v-R}
  \mean{\rho_{\ext}} \simeq \mean{\rho_{\ext}^{(\soft)}} + 
  \sqrt{\frac{\pi c_J}{2}}
   \sigma R
  \left(\frac{\mean{n_h}}{A_\tot} +c_J R^2 \frac{\mean{n_h}^2}{2A_\tot^2} \right)\,.
\end{equation}
Features to note here are that the discrepancy is proportional to
$\sigma R$, with the next correction going as $\sigma R^3$. 
Eq.~(\ref{eq:np-nb-pert}) provides a result for
the average number of hard emissions $\mean{n_h}$, and $Q_0$ there should be
replaced with $\max(Q_0,\order{\sigma \sqrt{A_\jet}})$, or
equivalently $\max(Q_0,\sqrt{c_J} \sigma R)$, because perturbative 
emissions whose transverse momentum is much smaller than the
scale of fluctuations of the underlying event will not bias the median.
This then gives us
\begin{equation}
  \label{eq:nh-over-Atot}
  \frac{\mean{n_h}}{A_\tot} \simeq
  \frac{n_b}{A_\tot}
  + \frac{C_i}{\pi^2} 
    \frac{L}{2b_0}\,, \qquad\quad L\equiv \ln \frac{\as(\max(Q_0, \sqrt{c_J} \sigma R))}{\as(Q)}\,.
\end{equation}
We see that the Born particles contribute when $A_\tot$ is not very large.
Perturbative emissions always contribute, essentially because the
number of emissions scales with the total area.
Substituting physically reasonable numbers into
eq.~(\ref{eq:nh-over-Atot}), \ie which corresponds to setting the logarithm $L$ equal to $1$,
gives
\begin{equation}
  \label{eq:nh-over-Atot-num}
  \frac{\mean{n_h}}{A_\tot} \simeq \frac{n_b}{A_\tot} + 0.25\frac{C_i}{C_A}\, ,
\end{equation}
which then gives the following numerical result for the bias,
\begin{equation}
  \label{eq:rho-AM-Gauss+pert-v-R-numbers}
  \mean{\rho_{\ext}} \simeq  \mean{\rho_{\ext}^{(\soft)}}  + 
  0.45 \,\sigma R \cdot \left(\frac{C_i}{C_A} + 4.0 \frac{n_b}{A_\tot} \right)
  + 0.11\, \sigma R^3 \cdot \left(\frac{C_i}{C_A} + 4.0 \frac{n_b}{A_\tot} \right)^2
\,.
\end{equation}
Ignoring the $n_b/A_\tot$ terms, for $R=0.6$ in gluon-initiated processes ($C_i=C_A$), the bias introduced
in $\rho$ is about $0.29\sigma$.
For values of $\sigma = 0.5\rho -0.7 \rho$, as we will obtain from the
Monte Carlo study in section~\ref{sec:area-based-stuff}, the positive
bias due to these perturbative effects is in the same $15-20\%$
ballpark as the negative bias due to finite-particle density that was
discussed in section~\ref{sec:toy-low-pt-part} for pure soft events.
There is an expectation that these two sources of bias should combine
linearly, at least when $R$ is sufficiently far above the critical
turn-on point that the Gaussian approximation used above is valid.
Since they have different $R$-dependences, $\sim +R$ and $\sim -1/R^2$
respectively, there exists only a limited  range of $R$ in which they compensate
each other.
In this respect the numbers given above tend to confirm the choice
$R\sim 0.5-0.6$ that had originally been recommended based on Monte
Carlo studies in \cite{Cacciari:2007fd}.%
\footnote{ An alternative method of extraction of $\rho$ comes to mind
  at this point, fitting a formula motivated by the results of this
  section and appendix~\ref{sec:more-toy-model}:
  \begin{equation}
    \label{eq:area-median-fit-formula}
    \rho_{\ext} = 
    \rho \left(\frac{R^2 - R_\crit^2}{R^2 + R_\crit^2\left[\frac{n}{(n+1)\ln 2} -1\right]}\right)^{\frac1n}
    \left(1 + c R \right)\,,
  \end{equation}
  with the fit parameters $\rho$, $R_\crit$, $n$, $c$. One might choose
  to forgo $n$, or try a finite number of choices, e.g. $n=1,2$. We
  have found that such a procedure eliminates a substantial part of the
  biases in the extraction of $\rho$ for some events, but in others
  statistical fluctuations lead to poor fits, with results for $\rho$ that
  are highly skewed towards low values. 
  For this reason,  we do not adopt this approach for the current study.}

\begin{figure}[t]
  \centering
  \includegraphics[width=0.6\textwidth]{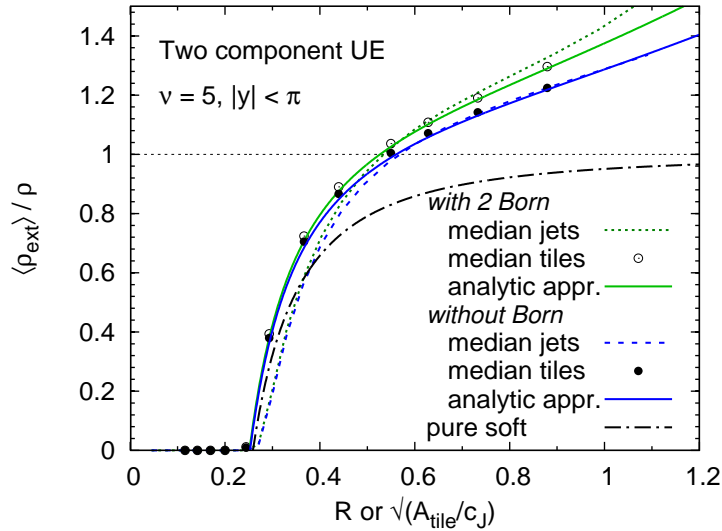}
  \caption{%
    Underlying event from the two component toy model analysed in the range
    of rapidity $|y|<\pi$ and the full span of the azimuthal angle. 
    The results for the average $\rho_\ext$ (normalised to $\rho$), extracted in 
    tile and jet median-based 
    approach (see text) are shown for two sets of events. The first set 
    contains soft 
    and perturbative particles while in the second set, 
    two Born particles with $p_{t, \hard}=100\GeV$ are present in addition. 
    For reference, the analytic curve from fig.~\ref{fig:analytical-soft-UE} 
    for the pure soft case is also shown. 
    The mean number of soft particles per unit area, $\nu$, was $5$ and 
    their average transverse momentum $\mu=0.4\GeV$, which corresponds to 
    $\rho=2\GeV$. The perturbative emissions
    are distributed between scales $Q_0=1\GeV$ and $Q=50\GeV$.
    % 
    %\comment{in the corresponding text: mention about $p_t$-scheme for
    %  jets}
  }
  \label{fig:toy-combined-UE}
\end{figure}

Fig.~\ref{fig:toy-combined-UE}  summarises the study of the two component model by showing the analytic results from eqs.~(\ref{eq:rho-AM-Gauss+pert-v-R}) and (\ref{eq:nh-over-Atot-num}) together with the numerical results from the median tile-based and median jet-based approaches. 
Two sets of results are presented, corresponding to the events with
and without Born particles. We see the very good agreement between the
analytic and the numerical tile-based approaches and the median
jet-based result (except in the
threshold region, as in section~\ref{sec:toy-low-pt-part}). To achieve
this it was essential to include the term $\sim R^3$ in
eq.~(\ref{eq:rho-AM-Gauss+pert-v-R}).

\begin{figure}
  \centering
  \includegraphics[width=0.6\textwidth]{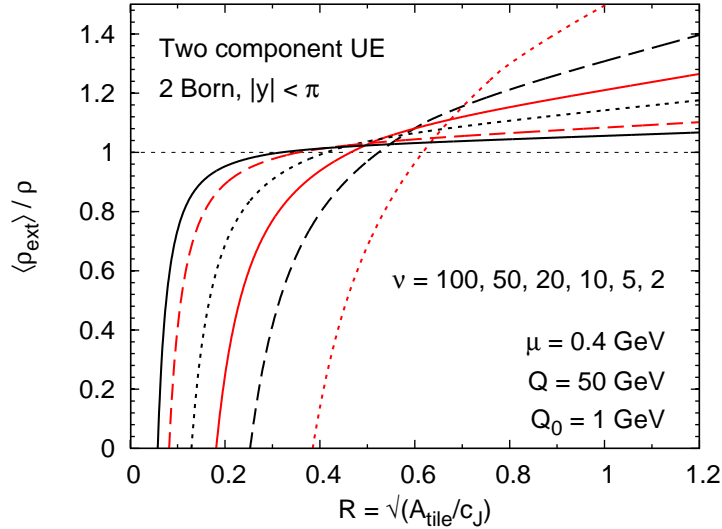}
  \caption{The analytical approximation for $\rho_\ext$ as a function
    of $R$,
    eqs.~(\ref{eq:rho-ext-tiles-approx},\ref{eq:rho-AM-Gauss+pert-v-R}),
    shown for a range of soft-particle densities ($\nu$ values), for
    the gluonic case, $C_i = C_A$.
    The curves from left to right (at small $R$ values) correspond to
    $\nu=100,50,20,10,5,2$ (or equivalently $\rho$ from $40\GeV$ to
    $0.8\GeV$).  }
  \label{fig:many-nu}
\end{figure}

Fig.~\ref{fig:toy-combined-UE} suggests that the region of $R$ with
the least bias for the determination of $\rho$ is $R=0.4-0.6$.
If one requires that the biases in eqs.~(\ref{eq:rho-jet-offset}) and
(\ref{eq:rho-AM-Gauss+pert-v-R}) cancel each other, then one finds that
$R$ should be chosen proportional to $\nu^{1/6}$ or equivalently
proportional to $(\sigma/\rho)^{1/3} \sim R_{\crit}^{1/3}$. 
Ignoring the $n_b/A_\tot$ component and the $R^3$ term in
the equations of this section, one finds
\begin{equation}
  \label{eq:R-zero-bias}
  R_{\text{zero-bias}} \simeq \sqrt{\frac{\pi}{2^{\frac13}
      c_J}}\left(\frac{\sigma}{\rho} \frac{b_0}{C_i L}
  \right)^{\frac13}
  \simeq 0.65 \left(
    \frac{\sigma}{\rho} \frac{C_A}{C_i}\right)^{\frac13}
  \simeq 0.87\,  R_\crit^{\frac13} \left(\frac{C_A}{C_i}\right)^{\frac13}\,,
\end{equation}
where the numerical values have been obtained setting $L=1$ in
eq.~(\ref{eq:nh-over-Atot}). 
% 
% 
% 
% %
The result for $R_{\text{zero-bias}}$ can be seen to be consistent with fig.~\ref{fig:many-nu}, which
shows the analytical approximation for $\rho_\ext$ as a function of
$R$ for a broad range of particle densities $\nu$. 
For a variation in the particle density (and $\rho$) by a factor of
$50$ ($\sigma/\rho$
by a factor of $7$), the  zero-bias $R$ value  changes only moderately and
in close accord with the expectations of
eq.~(\ref{eq:R-zero-bias}).
Fig.~\ref{fig:many-nu} also illustrates that a fixed $R\sim 0.6$ leads
to a $\rho_\ext$ value to within $20\%$ of the correct result for a
whole range of $\nu$,
with the relative impact of the biases steadily decreasing as the
particle density is increased.

%% %

%----------------------------------------------------------------------
\subsection[Fluctuations in the estimation of $\rho$]{Fluctuations in the estimation of $\boldsymbol{\rho}$}
\label{sec:toy-fluctuations}
In the simple model studies discussed here, the same $\rho$ value has been used to generate all events. Nevertheless, extracted values of $\rho$ vary from one event to another. 
This is because any method of measuring UE can use only a limited part
of an event (restricted $A_\tot$, $A_\trans$, according to $y$ and $\phi$ cuts)
and works with a finite number of objects (particles, jets).
The magnitude of the observed event-to-event fluctuations in this case
is an important
characteristic of a method, because it sets a lower limit on the uncertainty
of the event-to-event $\rho$ measurement. These intrinsic fluctuations
also affect the measurement of the true fluctuations and correlations
of realistic UE models.
Therefore, in order to measure properties of the underlying event that can then be used for efficient subtraction or tuning of simulation programs, one is interested in reducing the fluctuations that come with the method itself.

For the soft underlying event from section~\ref{sec:toy-low-pt-part},
ignoring the small systematic biases that we found there in the
determination of $\rho$, one can show that 
the standard deviations $S_d$ of the extracted $\rho$ values read
\begin{subequations}
\label{eq:soft-fluctuatioins}
  \begin{align}
    \label{eq:soft-fluct-med}
    \frac{S_{d,\mathrm{med}}^{(\soft)}}{\rho}  &= \sqrt{\frac{\pi}{\nu A_\tot}}
    = \frac{\sigma}{\rho}\sqrt{\frac\pi{2A_\tot}}\,,\\
    \label{eq:soft-fluct-av}
    \frac{S_{d,\mathrm{Av }}^{(\soft)}}{\rho} &= \sqrt{\frac{1}{\nu A_\trans}}
    = \frac{\sigma}{\rho}\frac1{\sqrt{2 A_\trans}}
    \,, \\
    \label{eq:soft-fluct-minmax}
    \frac{S_{d,\mathrm{Min}}^{(\soft)}}{\rho} = 
    \frac{S_{d,\mathrm{Max}}^{(\soft)}}{\rho} &= 
    \sqrt{\frac{2}{\nu A_\trans}\left(1-\frac{1}{\pi}\right)}
    = \frac{\sigma}{\rho} \sqrt{\frac{\pi - 1}{\pi A_\trans}}
    \,.
  \end{align}
\end{subequations}
The result for the median case
is derived in appendix~\ref{sec:pure-soft-case-fluct}. 
The formula for the Min/Max regions was obtained in the Gaussian approximation, valid in the limit $\nu A_\trans \gg 1$.
Substituting realistic values for the areas, \ie $A_\trans=2\pi/3$, $A_\tot=4\pi$, and the density $\nu=5$, one arrives at the following numerical estimates
\begin{equation}
\label{eq:fluct-num}
S_{d, \mathrm{med}}^{(\soft)}/\rho     =  0.22, \qquad
S_{d, \mathrm{Av}}^{(\soft)}/\rho      =  0.31, \qquad
S_{d, \mathrm{Min/Max}}^{(\soft)}/\rho =  0.36.
\end{equation}
The lower expected fluctuations for the area/median based approach are
a consequence of the larger area used in the UE determination.
The use of a larger area is possible in the first place because the
method's dynamical separation of UE limits the need to cut away
regions from the $y-\phi$ plane to reduce contamination from hard
jets.

% %
 
\begin{figure}[p]
  \centering
  \includegraphics[width=0.4\textwidth, angle=-90]{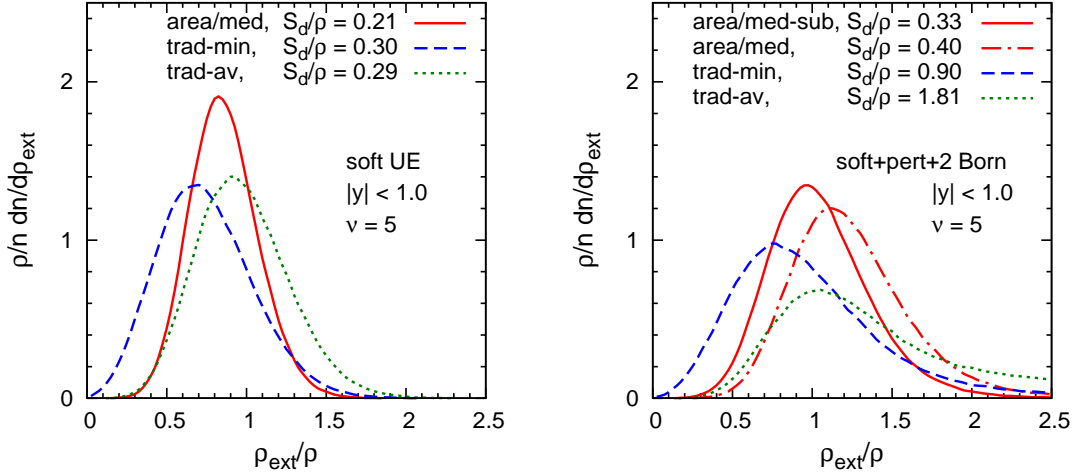}
  \caption{%
  Distributions of $\rho_\ext$  from traditional and area/median based
  approach for the case of pure soft underlying event from
  section~\ref{sec:toy-low-pt-part}, (left) and for the combined model
  from section~\ref{sec:toy-two-comp} with the soft and perturbative
  components and 2~Born particles (right). The average density of soft
  particles is $\nu=5$ and their average transverse momentum
  is $\mu=0.4\GeV$, which corresponds to 
  $\rho=2\GeV$. The perturbative emissions are distributed
  between scales $Q_0=1\GeV$ and  $Q=50\GeV$.
  To guarantee that the two hardest jets are associated with the Born
  particles, their transverse momenta were set to
  $p_t=400\GeV$. 
   The jet clustering was performed with the Cambridge/Aachen
   algorithm with R=0.6.
   %\comment{When we combine UE and PT parts, as well as
   % specifying $\nu$ we should also say what $\mu$ (or $\rho$) actually
   % is -- just to situate the different scales; we should also
   % comment on the other parameters of the model? ($p_{t,Born}$, $Q$)}
   %
 }
  \label{fig:toy-rhohis}
\end{figure}
\begin{table}[p]
  \centering
  \begin{tabular}{lcc}\toprule
                            &  \multicolumn{2}{c}{$S_{d,\ext}/\rho$}\\\cmidrule{2-3}
                            &   pure soft  &  soft\,+\,PT\,+\,2 born\\\midrule
    median/area, all jets   &     0.21     &    0.40            \\
    median/area, all-2 jets &      --      &    0.33            \\
    traditional, TransAv          &     0.30     &    1.81            \\
    traditional, TransMin         &     0.29     &    0.90            \\\bottomrule
  \end{tabular}
  \caption{Results for $S_{d,\ext}/\rho$ for various methods of
    extracting $\rho$ in the toy model
    where the input standard deviation of $\rho$ is $S_d=0$. These values provide an indication of the
    lower bound of  $S_{d}$ values that could be observed in real
    events.} 
  \label{tab:toy-Sd-values}
\end{table}

In fig.~\ref{fig:toy-rhohis}, we show histograms of $\rho$ extracted
in the traditional and area/median approaches for the case of purely soft
underlying event from section~\ref{sec:toy-low-pt-part} (left) and for
the combined model described in section~\ref{sec:toy-two-comp} with
the soft and perturbative components and 2~Born particles (right). 
The corresponding standard deviations of the $\rho_\ext$ values
are given in table~\ref{tab:toy-Sd-values}.
For the case of the purely soft UE, we see that these standard
deviations follow the pattern from eq.~(\ref{eq:fluct-num}), with the
area/median approach performing best. Note however that finite-density
($\nu$) effects do tend to slightly reduce the standard deviation
results as compared to expectations, especially in the TransMin case.

Adding perturbative and Born particles increases fluctuations in all
cases. From the shapes of the curves in fig.~\ref{fig:toy-rhohis}
(right), it seems that the area/median and TransMin results are
both degraded by similar amounts, while the TransAv result suffers
significantly more, with a long tail to large $\rho_{\ext}$ values.
The strong degradation of the TransAv result is an expected
consequence of its sensitivity to perturbative radiation, 
\begin{equation}
  \label{eq:SdTransAvPT}
  S_{d,\Av}^{(\hard)} \simeq \sqrt{\frac{C_i \as}{4A_\trans\, \pi^2}}\, 
    Q \left(1 +
    \order{\sqrt{\as A_\trans}}\right),
\end{equation}
where one observes the dependence on $\sqrt{\as/A_\trans}$ (to be
compared to the bias on the mean which goes as $\as$).

However, if one examines the results for $S_d$ in table
\ref{tab:toy-Sd-values}, one sees that the TransMin standard
deviation is also significantly increased by perturbative radiation. 
The toy-model expectation is 
\begin{equation}
  \label{eq:SdTransMinPT}
  S_{d,\Min}^{(\hard)} \simeq \frac{C_i \as}{\pi^2\sqrt{2}} Q \left(1 +
    \order{\as^2 A_\trans^2}\right),
\end{equation}
and for the particular parameters used in
table~\ref{tab:toy-Sd-values}, the result for $S_d$ turns out to be
as large as $\rho$ itself. Compared to the $\order{\as^2 A}$
suppression for the bias to the $\mean{\rho_{\ext,\Min}}$ result, here the
perturbative radiation bias is much stronger, $\order{\as}$.
The physical explanation is simple: while it is relatively rare for
perturbative radiation to affect the TransMin region (hence the
acceptable peak-region of the $\rho_\ext$ distribution), when it does,
the effect on $\rho_\ext$ is large, contribution significantly to the
final result for $S_d$.\footnote{The hard component also has an impact
  on the soft fluctuations, in analogy with the effect discussed for
  $\mean{\rho_{\ext,\Min}}$ in section~\ref{sec:toy-two-comp}. We
  ignore this complication here since the fluctuations of the hard
  component in any case dominate over those of the soft component. }

The area/median approach is much more robust in this respect, because
the hard emissions' contribution to the standard deviation does not
have significant enhancements compared to the average bias on $\rho$:
\begin{equation}
  \label{eq:SdMedHardMainText}
  S_{d,\med}^{(\hard)} \simeq 
  1.79 \frac{\sigma R}{\sqrt{A_\tot}} \left(\frac{C_i}{C_A}
    + 4.0 \frac{n_b}{A_\tot}\right)^{\frac12} + \cdots\,,
\end{equation}
as derived in appendix~\ref{sec:hard-contamination-fluct}. In
particular, the larger numerical coefficient compared to the
$\order{R}$ term of eq.~(\ref{eq:rho-AM-Gauss+pert-v-R-numbers}) is
compensated by the factor of $1/\sqrt{A_\tot}$.
This good behaviour is visible in fig.~\ref{fig:toy-rhohis}, and also
in the values of table~\ref{tab:toy-Sd-values}, 
which are roughly consistent with the above analytical estimate
(maybe $20\%$ higher).
They also highlight the further improvement to be had with the
hard-jet removal procedure discussed at the end of
section~\ref{sec:area-median-approach} (``all-2 jets'' result), which
benefits not just $S_d$, but also the peak position and height in
fig.~\ref{fig:toy-rhohis}.

Overall, the results of this section suggest that for any measurement
of fluctuations of the UE, it will be preferable to use the
area/median method, with hard-jet removal able to provide some extra
benefit.

% 

%----------------------------------------------------------------------
\subsection[Extraction of $\sigma$]{Extraction of $\boldsymbol\sigma$}

The measurement of intra-event fluctuations, $\sigma$, has only
been discussed so far in the context of the area/median
approach.\footnote{In the traditional approach, one might envision
  calculating the perturbative contributions in
  eqs.~(\ref{eq:pt-trad-Gauss+pert-av},\ref{eq:pt-trad-Gauss+pert-min-max})
  to NLO with a program such as NLOJet++ \cite{NLOJet} and then
  fitting for $\langle \rho \rangle$ and $\langle \sigma \rangle$ in
  those equations after averaging over many events. It seems that it
  would be difficult, however, to extract event-by-event estimates of
  $\sigma$.}
Though we shall not go into full analytical detail, it is easy to
convince oneself that many of the considerations that arise in the
extraction of $\rho$ apply also when determining $\sigma$. 
In particular, for pure soft events one underestimates $\sigma$ when $R$
is too small and the presence of perturbative radiation will bias the
extracted $\sigma$ at larger $R$.

One point to be aware of is that our method for extracting it, cf.\
section~\ref{sec:area-median-approach}, has the characteristic that
$\sigma / \sqrt{A_\jet} $ never exceeds $\rho$.
This translates into an $R$-dependent upper bound on $\sigma_{\ext}/\rho_{\ext}$,
\begin{equation}
  \label{eq:sigma-upper-bound}
  \frac{\sigma_\ext}{\rho_\ext} \lesssim \sqrt{c_J}\, R \simeq 1.43 R\,.
\end{equation}%
\begin{figure}
  \centering
  \includegraphics[width=0.48\textwidth]{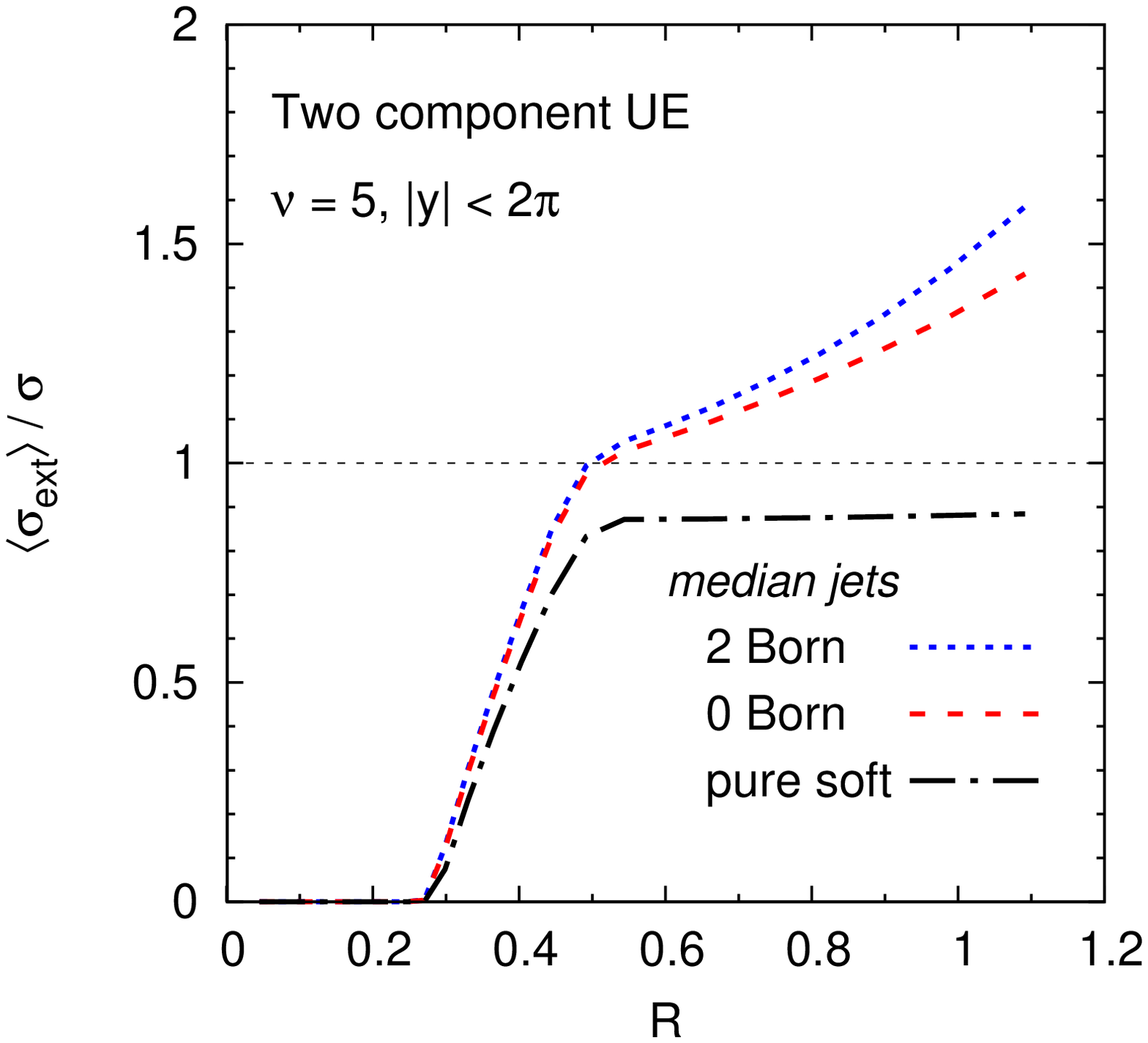}\hfill
  \includegraphics[width=0.48\textwidth]{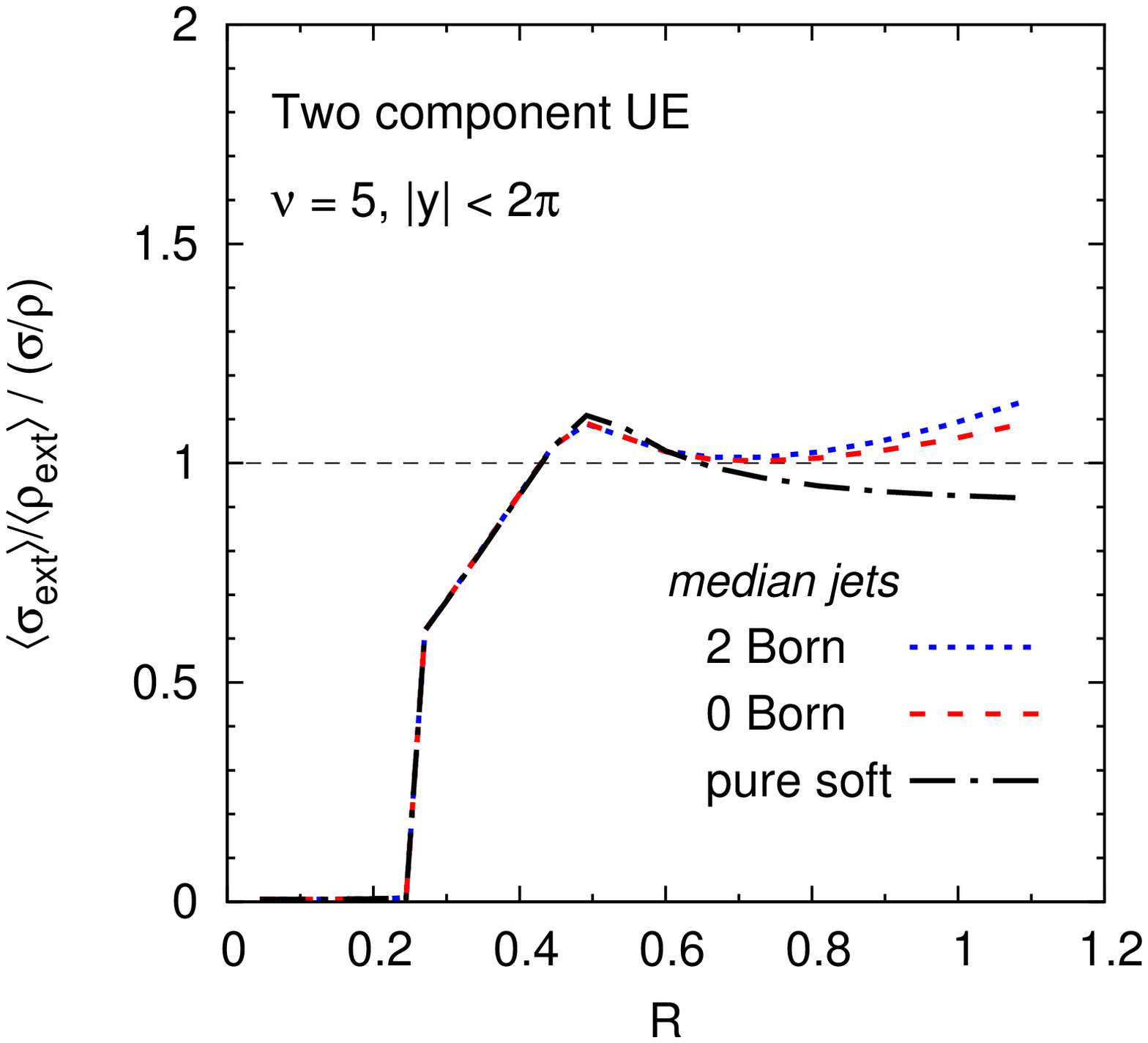}
  \caption{Left: extracted value of $\sigma$ with the area/median
    approach in the pure soft and the two component toy models, shown as a
    function of $R$ and normalised to the
    correct  value for the soft component $\sigma_{\soft} = \rho_\soft \sqrt{2/\nu}$.
    Right: the same set of results, but showing
    $\mean{\sigma_{\ext}}/\mean{\rho_{\ext}}$ normalised to the expected
    $\sigma/\rho$ for the soft component.
    The toy model variant is that corresponding to eq.~(\ref{eq:pt-dist-P1}).
  }
  \label{fig:sigma-ext-toy}
\end{figure}%
Fig.~\ref{fig:sigma-ext-toy} (left) shows the average $\sigma_\ext$ as a
function of $R$ in the toy model ($\nu=5$, with and without the
perturbative radiation and the Born jets), normalised to the correct
result for the soft component.
In the case of just the soft component, one sees a threshold region
followed by a slow approach towards the correct value, much as for
$\rho_\ext$ even if the detailed shapes differ (in part owing to the
one-sided determination of $\sigma$, which causes the residual bias to
be proportional to $1/R$, rather than $1/R^2$ for the bias to $\rho$).
With the inclusion of the perturbative component there is an
additional bias, which grows towards large $R$, again much as happens
for $\rho_\ext$.

To compare the biases on $\sigma_\ext$ and $\rho_\ext$, it is
convenient to examine the ratio $\mean{\sigma_\ext} / \mean{\rho_\ext}$,
fig.~\ref{fig:sigma-ext-toy} (right), normalised to the soft-component
result for $\sigma/\rho$.
The impact of the bound on $\sigma/\rho$,
eq.~(\ref{eq:sigma-upper-bound}), is clearly visible up to $R\simeq
0.5$ (consistent, roughly, with $\sigma/\rho \simeq 0.63$). 
Beyond this point, over a reasonable range of $R$,
$\mean{\sigma_\ext}/\mean{\rho_\ext}$ remains compatible with the true value to
within roughly $10-20\%$, and, as expected, deviations are larger in
the presence of hard radiation.
Overall, in the region of $R$ that is suitable for extracting 
$\rho$, figure~\ref{fig:sigma-ext-toy} suggests that the extraction of
$\sigma$ should also be quite acceptable.

% 
% %
% %

%----------------------------------------------------------------------
\subsection{Summary of main results}
\label{sec:toy-model-summary}

Table~\ref{tab:summary} summarises the main results of this section
for the biases, $\delta\rho$, and the event-to-event fluctuations,
$S_d$, that occur within the toy model in extracting $\rho$, the transverse momentum flow
per unit area, with each of three UE
estimation methods: the traditional approach in its TransAv and
TransMin variants, whose main parameter is the area $A_\trans$ of each
of the two transverse regions; and the area/median approach whose
parameters are the total area under consideration, $A_\tot$ and the
jet radius $R$.
Results are given both for the biases and fluctuations intrinsic to
the soft component and for the additional biases that arise due to the
presence of hard radiation in the event, expressed as
\begin{equation}
  \label{eq:summary-eqn}
  \rho_\ext = \rho + \delta\rho^{(\soft)} + \delta\rho^{(\hard)}
  \pm S_d^{(\soft)} \pm S_d^{(\hard)}\,.
\end{equation}%
The analytical formulae help illustrate the dependence on the
parameters of the measurement methods (the area of the regions used,
the jet radius) and the physical scales present in the events (hard
scale $Q$, and the values for $\rho$ and for the level of intra-event UE
fluctuations, $\sigma$). 
One sees, for example, how the hard biases and event-to-event
fluctuations in the
traditional approach are always proportional to  $Q$ and to some
power of $\as$,
whereas in the area/median approach they are proportional to $\sigma$ (with modest coefficients and weak
additional $\ln \ln Q$ 
dependence contained in the parameter~$L$).

%the levelof UE fluctuations,
\begin{table}[p]
    \centering
    \begin{tabular}{l|cccl}
    \toprule
    quantity  & method & result & numerical value  & equation(s)  \\ \midrule
    & TransAv
        & 0 &   0
        & (\ref{eq:pt-trad-Gauss-av})\\[15pt]
    \multirow{3}{*}[1em]{$\displaystyle \frac{\delta\rho^{(\soft)}}{\rho}$}
    & TransMin
        & $\displaystyle - \frac{\sigma}{\rho} \frac{ \cP}{\sqrt{\pi
            A_\trans}}$ 
        &  $\!\!\!\!-0.09$
        & (\ref{eq:pt-trad-Gauss-min}), (\ref{eq:pt-trad-Gauss+pert-min})\\[15pt]
    & Area/Med$^*$
        %& $ \displaystyle -\frac{\rho}{3 \ln 2}\frac{R_\crit^2}{R^2}$ &   
        & $ \displaystyle -\frac{\sigma^2}{\rho^2} \frac{1}{4c_J R^2}$
        &  $\!\!\!\!-0.14$
        & (\ref{eq:rho-jet-offset})\\[0.7em]\midrule
        %----------------------------------------------------------------------
    & TransAv
         & $\displaystyle \frac{C_i\as}{\pi^2} \frac{Q}{\rho}$ 
         &  $0.99$
         &(\ref{eq:pt-trad-pert-av}), (\ref{eq:pt-trad-Gauss+pert-av}) \\[15pt]
    \multirow{3}{*}[1em]{$\displaystyle \frac{\delta\rho^{(\hard)}}{\rho}$}
    & TransMin
         & $2\displaystyle \left(\frac{C_i\as}{\pi^2}\right)^2
         \frac{A_\trans Q}{\rho}$ 
         &  $0.16$
         & (\ref{eq:pt-trad-pert-min}), (\ref{eq:pt-trad-Gauss+pert-min})\\[15pt]
    & Area/Med
         & $\displaystyle 
                \frac{\sigma R}{\rho}
                \sqrt{\frac{\pi c_J}{2}}
                \left(\frac{n_b}{A_\tot}
                + \frac{C_i}{\pi^2} \frac{L}{2b_0}\right)$
                %\frac{\mean{n_h}}{A_\tot}$ 
              & $0.17$ 
         & (\ref{eq:rho-AM-Gauss+pert-v-R})--(\ref{eq:rho-AM-Gauss+pert-v-R-numbers})\\[0.7em]\midrule
       %----------------------------------------------------------------------
    & TransAv
        & $\displaystyle
        \frac{\sigma}{\rho}\sqrt{\frac{1}{2A_{\trans}}}$ 
        &   $0.31$
        & (\ref{eq:soft-fluct-av})\\[15pt]
    \multirow{3}{*}[1em]{$\displaystyle \frac{S_{d}^{(\soft)}}{\rho}$}
    & TransMin
        & $\displaystyle \frac{\sigma}{\rho}\sqrt{\frac{\pi-1}{\pi A_\trans}}$ 
        & $0.36$
        & (\ref{eq:soft-fluct-minmax})\\[15pt]
    & Area/Med
        %& $ \displaystyle -\frac{\rho}{3 \ln 2}\frac{R_\crit^2}{R^2}$ &   
        & $\displaystyle \frac{\sigma}{\rho}\sqrt{\frac{\pi}{2A_\tot}}$ 
        &  $0.22$
        & (\ref{eq:soft-fluct-med})\\[0.7em]\midrule
       %----------------------------------------------------------------------
    & TransAv
        & $\displaystyle \sqrt{\frac{C_i\as}{4A_\trans\pi^2}}\,\frac{Q}{\rho}$
        &  $1.72$
        & (\ref{eq:SdTransAvPT})\\[15pt]
    \multirow{3}{*}[1em]{$\displaystyle \frac{S_{d}^{(\hard)}}{\rho}$}
    & TransMin
        & $\displaystyle \frac{C_i\as}{\pi^2\sqrt{2}}\,\frac{Q}{\rho}$ 
        & $0.70$
        & (\ref{eq:SdTransMinPT})\\[15pt]
    & Area/Med
        %& $ \displaystyle -\frac{\rho}{3 \ln 2}\frac{R_\crit^2}{R^2}$ &   
        & $\displaystyle \frac{\sigma R}{\rho}\sqrt{\frac{2\pi c_J}{A_\tot}} \left(\frac{n_b}{A_\tot}
                + \frac{C_i}{\pi^2} \frac{L}{2b_0}\right)^{\frac12}$ 
        &  $0.19$
        & (\ref{eq:SdMedHardMainText}), (\ref{eq:SdHardFinal})\\[0.7em]\bottomrule
        %----------------------------------------------------------------------
    \end{tabular}
    %======================================================================
    % %======================================================================
    % \begin{tabular}{c|c|c|l}
    % \hline
    % quantity  &  & numerical value  & references  \\ \hline
    % $ \delta\rho_\soft^\Min$ 
    %     & $\displaystyle -\frac{\sigma \cP}{\sqrt{\pi A}}$ &   
    %     & eqs.~(\ref{eq:pt-trad-pert-av})), (\ref{eq:pt-trad-Gauss+pert-av})\\[15pt]
    % $\delta\rho_\hard^\Min$ 
    %     & $2\displaystyle \left(\frac{C_i\as}{\pi^2}\right)^2 A Q$ &  
    %     & eqs.~(\ref{eq:pt-trad-pert-min}), (\ref{eq:pt-trad-Gauss+pert-min-max})\\[15pt]
    % $\delta\rho_\soft^\med$ 
    %     & $ \displaystyle -\frac{\rho}{3 \ln 2}\frac{R_\crit^2}{R^2}$ &   
    %     & eqs.~(\ref{eq:rho-jet-offset-variant})\\[15pt]
    % $\delta\rho_\hard^\med$ 
    %     & $\displaystyle \sqrt{\frac{\pi c_J}{2}}
    %            \sigma R \frac{\mean{n_h}}{A_\tot}$ &  
    %     & eqs.~(\ref{eq:rho-AM-Gauss+pert-v-R})\\[15pt]
    % \hline
    % \end{tabular}
    \caption{Summary of main biases and sources of fluctuations for
      different UE extraction methods; $Q$ is the hard scale, $Q_0$
      the IR cutoff on perturbative emissions, $L=\ln
      \left[\as(\max(Q_0,\sqrt{c_J\sigma R}))/\as(Q)\right]$, 
      $\cP \simeq \exp\left( 
      -2 A_\trans\, \frac{C_i}{\pi^2} \frac{1}{2 b_0} \ln\left[
      \as(\max(Q_0,\sigma\sqrt{A_\trans}))/\as(Q)\right]\right)$,
      $c_J=2.04$, $C_i$ is the
      colour factor of the incoming partons. Numerical values are
      given for $\rho=2\GeV$, $\sigma/\rho=0.63$, $Q=50\GeV$,
      $Q_0=1\GeV$, $C_i=C_A=3$, $n_b=0$, $R=0.6$, $A_\tot=4\pi$ and
      $A_\trans=2\pi/3$ (the area of a single transverse region),
      corresponding to $L\simeq 1$ and $\cP\simeq 0.35$.
      The result marked with a $^*$ is specific to the form of the soft
      toy model discussed in section~\ref{sec:toy-low-pt-part}.
      \label{tab:summary}
    }
\end{table}

Table~\ref{tab:summary} also gives numerical results for the biases
and $S_d$ values, based on the default set of measurement parameter
choices and physical scales that have been used throughout this
section. This helps illustrate the expected orders of magnitude of
different effects under realistic conditions.
The large results for $S_d^{(\hard)}$ in the traditional method (i.e.\
unreliable event-by-event extraction of $\rho$), together with the
proportionality to $Q$ of the traditional method's biases and
fluctuations and the fact that it offers no easy way of determining
$\sigma$, lead us to prefer the area/median method for the Monte Carlo
UE measurement studies that we will perform below.

Other results of this section that are not summarised in
table~\ref{tab:summary} include: $R_\crit$,
eq.~(\ref{eq:Rcrit}) the $R$-value below which the area/median
approach gives $\rho_\ext=0$; the upper bound on $\sigma_\ext/\rho_\ext$ as
a function of $R$ in the median-area method,
eq.~(\ref{eq:sigma-upper-bound}); and $R_\textrm{zero-bias}$,
eq.~(\ref{eq:R-zero-bias}), the $R$ value for which the soft and hard
biases cancel out in the area/median approach.
Finally, fig.~\ref{fig:toy-combined-UE} shows the characteristic shape
of the $R$-dependence for $\rho_\ext$ in the median/area method, while
fig.~\ref{fig:many-nu} helps illustrate how $R=0.6$ is a reasonable default
choice for a wide range of UE conditions.

%======================================================================
\section{Illustration with Monte Carlo events}
\label{sec:area-based-stuff}

Given the area/median method to determine $\rho$ and $\sigma$ on an event-by-event
basis, let us now explore what kinds of observables we might
construct from them. 
The choices that we shall make are motivated by considerations of how the UE
affects jet measurements at hadron colliders. 
For example the UE leads to an average shift in jet energy;
it's important that one knows that shift as a function of rapidity, and
hence one should determine $\langle \rho(y) \rangle$.
The UE also affects jet energy \emph{resolution}, through a term of
the form $\sigma \sqrt{A_\jet}$. Thus we will also want to look at $\langle
\sigma \rangle $ as a function of rapidity.
A second way in which the UE affects jet energy resolution is that
$\rho$ itself is different event by event, so one might therefore
examine its event-by-event distribution and its standard deviation
$\Sd$.
The different ways in which UE affects jets was summarised in
\cite{Cacciari:2008gn} with the following equation for the variance of
the change in jet $p_t$ due to the underlying event (neglecting
back-reaction):
\begin{equation}
  \label{eq:Delta-pt-fluct}
  \langle \Delta p_{t,j}^2 \rangle - \langle \Delta p_{t,j}
  \rangle^2 \simeq \langle \Sigma_{\JA,R}^2 \rangle \langle
  \rho\rangle^2 + \langle A_{\JA,R} \rangle\, 
  \langle \sigma^2 \rangle + \langle A_{\JA,R} \rangle^2 \Sd^2\,,
\end{equation}
where $\langle A_{\JA,R} \rangle$ is the average jet area and $\langle
\Sigma_{\JA,R}^2 \rangle \sim R^4 $ is the variance of the jet area.
One sees that each of the terms involves a different characteristic of
the UE: $\langle\rho\rangle$, $\langle\sigma\rangle$ and $\Sd^2$.

%% %
%% 

Measurements of UE characteristics, as well as being of direct relevance
to jet measurements, also have the power to constrain UE models.
This has been the motivation for most UE studies to date and we
believe that the range of UE characteristics discussed above would
complement existing types of measurement and so provide an additional
powerful set of constraints.
Furthermore, as well as examining ``local'' quantities, such as
$\langle \rho(y) \rangle $, one can also, for example, ask the
question of whether there are long-range correlations between the
magnitude of the UE in different parts of an event. In (prevalent) UE
models that involve multiple independent $2\to2$ scatterings, one
might expect these correlations to be modest, whereas in a
BFKL-inspired model (as might derive from work like
Ref.~\cite{Avsar:2006jy}), where one or more gluon ladders stretch
across a whole event, one might expect them to be larger.

So far in this article we have  treated the traditional and
area/median based UE measurement approaches on a similar footing. 
The results of section~\ref{sec:systematics} suggest that both provide
reasonably adequate information about average characteristics of the
underlying event: both methods introduced biases and
different sources of biases cancelled partially, limiting their overall
impact (though in the area/median approach the biases had much
weaker dependence on the hard scale of the event).
In contrast, as we saw in section~\ref{sec:toy-fluctuations} and its
fig.~\ref{fig:toy-rhohis}, in the traditional approach the
fluctuations due to hard perturbative emission were likely to dominate
over fluctuations in the soft component of the UE, therefore for
observables that are sensitive to fluctuations, we believe that the
area/median approach is more robust.
For this reason, in what follows, we shall concentrate on this second
approach. 

This section will be structured as follows: after outlining the set of
Monte Carlo simulations that we use (section~\ref{sec:mc-models}) and
the event selection cuts (section~\ref{sec:mc-models}), we
examine to what extent the toy model is qualitatively similar to the
realistic simulations (section~\ref{sec:MC-v-toy}). The motivation is that the toy model guides
our intuition about the measurement procedure, and it is important to
establish that this intuition is well founded.
Having done so we then study (section~\ref{sec:MC-observables}) a selection of observables that are
relevant for eq.~(\ref{eq:Delta-pt-fluct}), including their rapidity
dependence and also their degree of correlation across different parts
of a same event.
This set of observables (together with some of those in section
\ref{sec:MC-v-toy}) would, we believe, be interesting to
examine experimentally, both in terms of the direct information it
provides for understanding the impact of UE on jets (and isolation,
etc.) and in terms of its ability to constrain models.

%----------------------------------------------------------------------
\subsection{Monte Carlo models used}
\label{sec:mc-models}

We shall examine a series of Monte Carlo UE models: 
the UE that comes with the ``old'' virtuality-ordered shower in
Pythia~6.4~\cite{Sjostrand:2006za}, in the DW and DWT tunes by
R.~Field~\cite{Albrow:2006rt}, identical at Tevatron energies, but
with different energy-dependences;
the UE that comes with the ``new'' transverse-momentum-ordered shower in
Pythia 6.4, in the S0A
tune~\cite{Skands:2007zg,Buttar:2006zd,Sjostrand:2004ef} by P.~Skands;
Herwig 6.5's~\cite{Corcella:2000bw,Corcella:2002jc} default ``soft''
UE, which fails to reproduce various aspects of Tevatron
data, but instructive for comparisons between different
types of models;
Herwig 6.5 with Jimmy 4.3~\cite{Butterworth:1996zw} in an ATLAS tune%
\footnote{The non-default parameter setting are: 
%
  %PRSOF=0, JMRAD(73)=1.8, PTJIM=4.9~GeV, JMUEO=1, with CTEQ6L1~\cite{Pumplin:2002vw} PDFs (i.e. just the parameters in \cite{Albrow:2006rt}).
  PRSOF=0, JMRAD(73)=1.8, PTJIM= 2.8$(\sqrt{s}/1.8)^{0.274}$ ,  with
  CTEQ6L1~\cite{Pumplin:2002vw} PDFs (i.e. just the parameters in \cite{Albrow:2006rt}).
  %\comment{was: PRSOF=0, JMRAD(73)=1.8, PTJIM=4.9~GeV, JMUEO=1, with CTEQ6L1~\cite{Pumplin:2002vw} PDFs (i.e. just the parameters in \cite{Albrow:2006rt})}.
}
by Moraes~\cite{Albrow:2006rt}. 
All models are based on multiple interactions except for Herwig's soft
UE.

By default we use a $pp$ centre of mass collision energy of
$\sqrt{s}=10\TeV$, though we will also consider the energy dependence
of some of the observables we study.

A comment is due on the fact that we will carry out our investigations
at particle (hadron) level. 
Experiments may carry out measurements on tracks only (well measured,
though with some low $p_t$ threshold), on calorimeter information
(subject to noise and noise-suppression thresholds) or on some
combination of the the two.
It is beyond the scope of this article to estimate the potentially
substantial impact of detector effects on the results presented here.
Nevertheless, we believe that the \emph{differences} that we will see
between various event-generator tunes should persist even after
detector effects.

%----------------------------------------------------------------------
\subsection{Event selection}
\label{sec:ev-sel}

We consider dijet events, where the leading jet, reconstructed with the
anti-$k_t$ algorithm~\cite{Cacciari:2008gp} with $R=0.6$ has $p_t> 100 \GeV$,
the next hardest jet has $p_t> 80 \GeV$ and both jets are required to lie 
in the rapidity window $|y| < 4$.
Note that since the cross section for jets falls steeply, a cut on jet
$p_t$  introduces a ``trigger-bias'', i.e. favours events where the~UE 
is slightly larger than average.\footnote{
  %
  %% rho measurement done excluding two hardest jets from overall list
  %
  To investigate the impact of the trigger bias, we used the following
  procedure: we first measured the underlying event density $\rho$ in a
  rapidity window $|\delta y| < 1$ around each hard jet, and then
  placed our hard jet cuts on $p_{t,\jet}-\rho A_{\jet}$ rather than on
  $p_{t,\jet}$.
  The result for $\mean{\rho}$ in events that passed these cuts on
  ``subtracted'' jets came out about $10\%$ lower than for events
  where we cut on unsubtracted jets.
  Since this is not too large an effect, and for reasons of
  simplicity, we have however chosen not to apply this procedure to
  our analysis as a whole.  }
The above cut is a default used for the study presented in this
section.  In some places we employ tighter cuts on the rapidities of the
two hardest jets in order to study the impact of these jets on UE
determination. Whenever this is the case, it will be indicated explicitly.
The choice of the anti-$k_t$ algorithm for event-selection here is motivated by the fact
that it will be the first to be supported by ATLAS and CMS
\cite{AtlasCMSantikt} and therefore will have been used by the
experiments for their initial event selection.

For the determination of the properties of the UE within the
jet-area/median approach, the anti-$k_t$ algorithm is not suitable, as
discussed in section~\ref{sec:area-median-approach}. We will therefore
use jets from the C/A algorithm with $R=0.6$ for this task.
All jet finding is performed with {\tt FastJet 2.4.1} \cite{Cacciari:2005hq,FastJet}, and we use the
\texttt{ActiveAreaExplicitGhosts} option to calculate areas, because this
ensures the safest treatment of pure ghost jets. 
The ghost area that is used is 0.01 and ghosts are placed up to
$|y|<9$ (we could, however, have used a smaller upper limit for most
of the plots). 
The other parameters are left at their default values.

%----------------------------------------------------------------------
%
%  %
%  
%
%
%  %
%
%%
%%
%%

%----------------------------------------------------------------------
\subsection{Comparisons of characteristics of MC  and toy model}
\label{sec:MC-v-toy}

\begin{figure}
  \centering
  \includegraphics[width=0.45\textwidth,angle=-90]{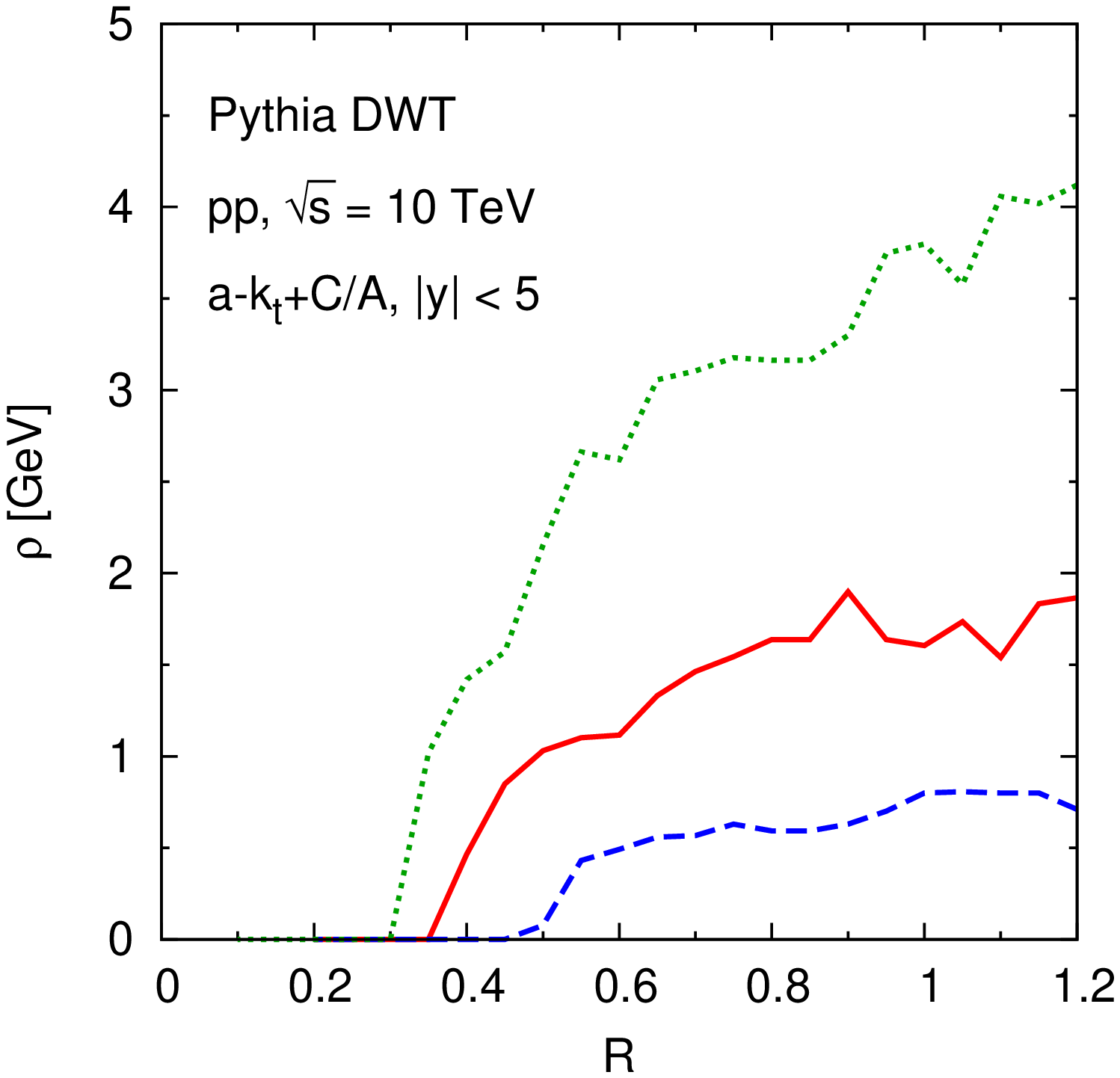}\hfill
  \includegraphics[width=0.45\textwidth,angle=-90]{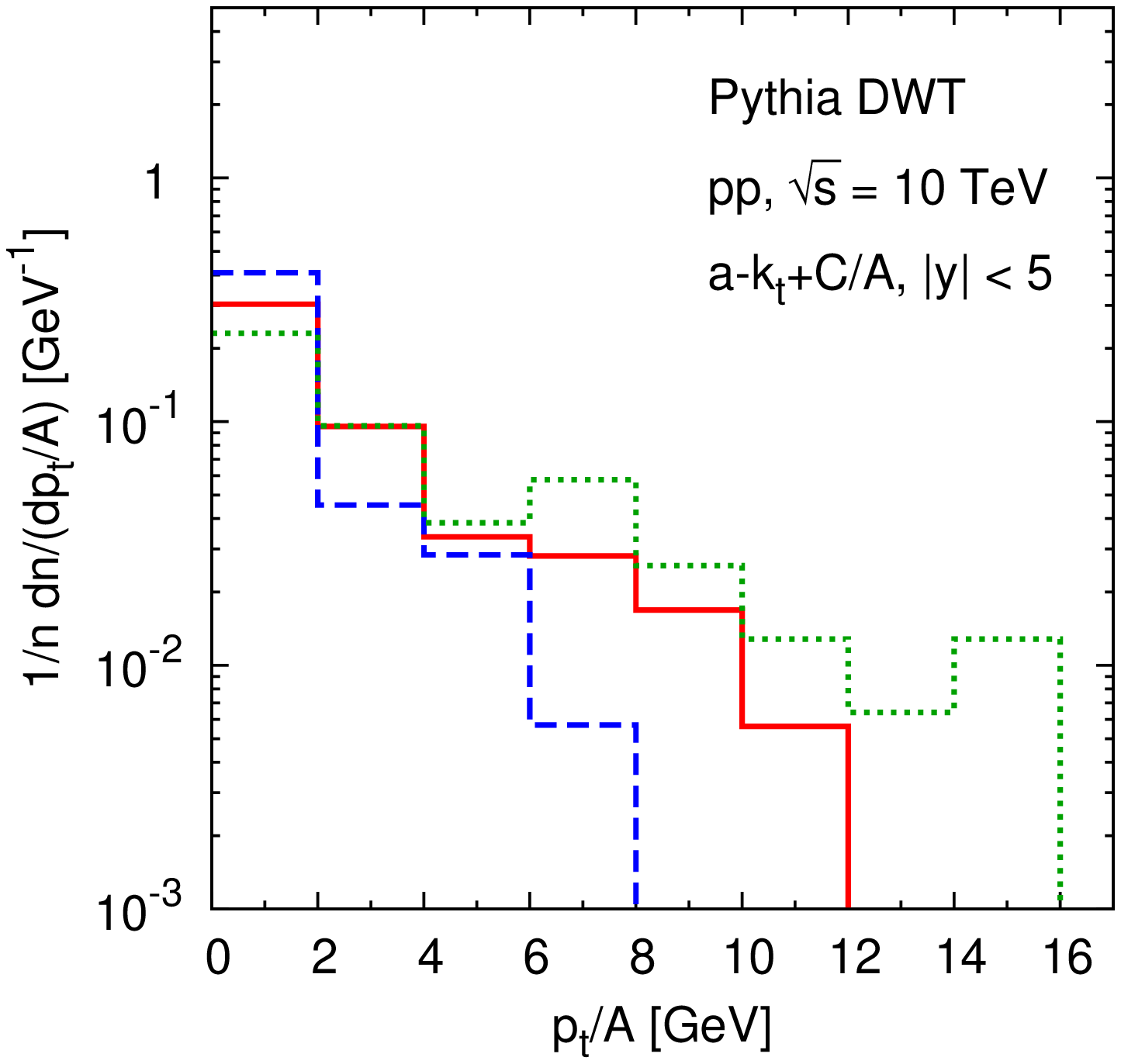}
  \caption{ The left-hand plot shows $\rho(R)$ extracted with C/A jets
    in the area/median approach for 3 representative Pythia (DWT)
    events, which passed the selection cuts of
    section~\ref{sec:ev-sel}. The right-hand plot shows the
    corresponding histograms of $p_t/A$ for the same 3 events.}
  \label{fig:3-DWT-events}
\end{figure}

Given the importance of the toy model in guiding our understanding of
the measurement of the underlying event, let us start by examining
whether realistic Monte Carlo events bear any similarity to toy-model
events.
One way of doing this is to examine the $R$-dependence of the
extracted $\rho$, which, in the toy model, had a characteristic shape,
fig.~\ref{fig:toy-combined-UE}.
Fig.~\ref{fig:3-DWT-events} (left) shows $\rho(R)$ for three
representative Pythia events (DWT tune). In each case one observes the
turn-on at some threshold $R$ value, followed by a roughly linear
slope at larger $R$, precisely as expected.
There is substantial variation in the curves from one event to the
next, and 
one can trace this back to their distributions of $p_t/A$ shown in
fig.~\ref{fig:3-DWT-events} (right): the blue (dashed) line which has
small $\rho$ and little $R$ dependence corresponds to an event in
which there is no activity at high $p_t/A$ values. 
The green (dotted) curve, which has large $\rho$ and strong
$R$-dependence, has a correspondingly broad distribution of $p_t/A$
values, with much activity at intermediate $p_t$ values. In this
event, the toy-model picture of a clean separation between soft and
hard physics is somewhat challenged, though the general pattern of
$\rho(R)$ having a turn-on followed by linear $R$-dependence still holds.

\begin{figure}
  \centering
  \includegraphics[width=0.45\textwidth,angle=-90]{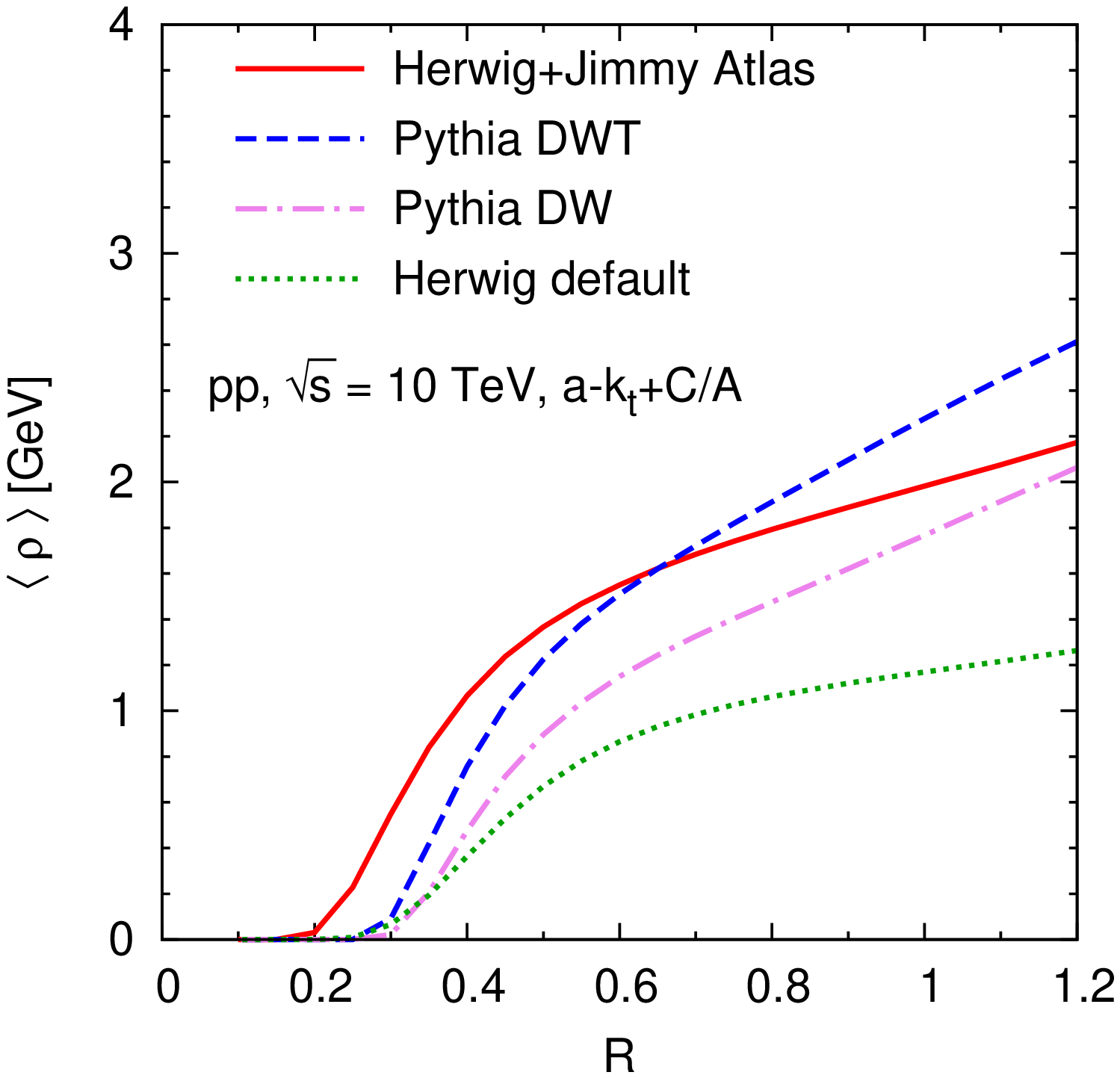}\hfill
  \includegraphics[width=0.45\textwidth,angle=-90]{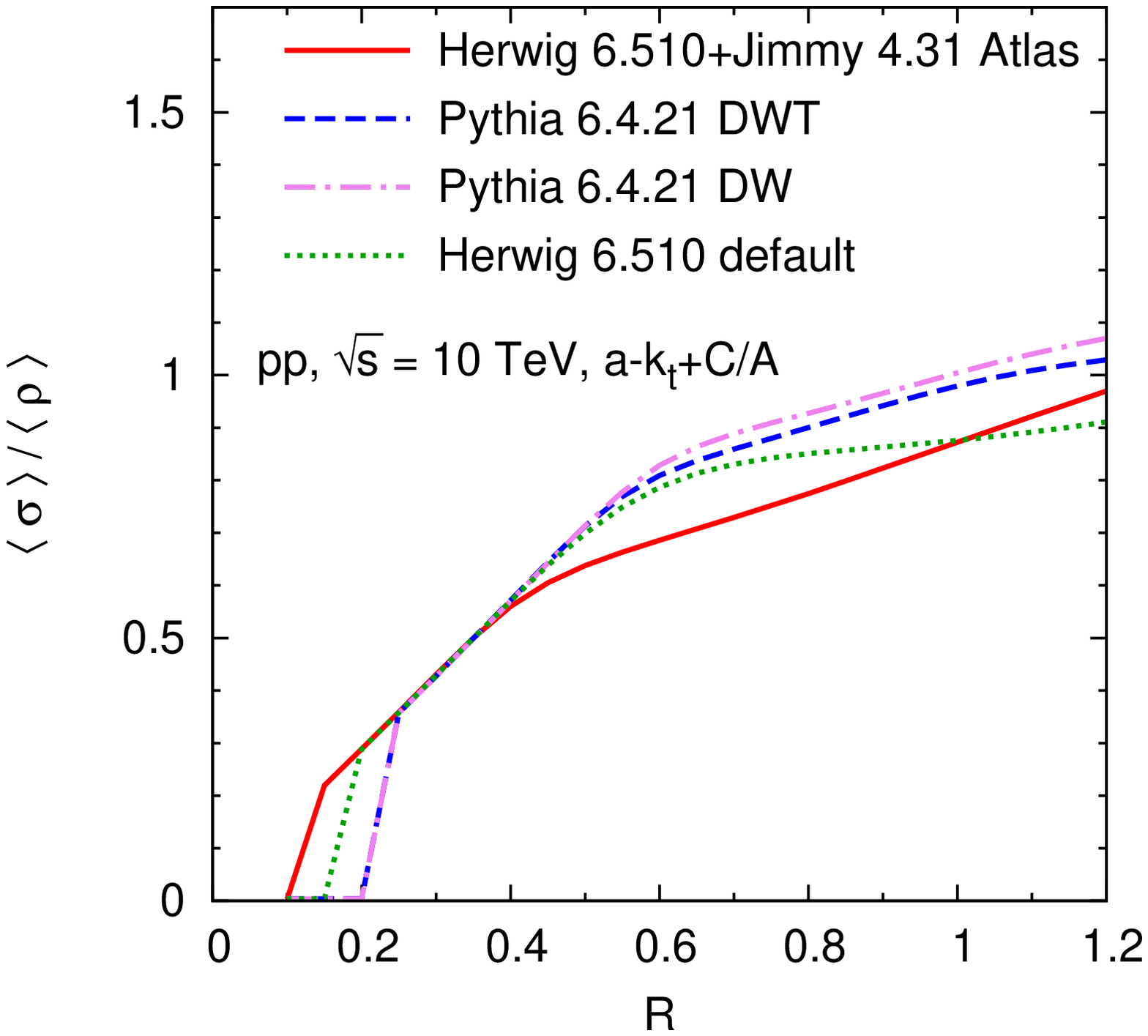}
  \caption{ Average over many events of the extracted $\rho$ (left)
    and $\sigma$ (right), shown as function of $R$ for four
    generators/tunes. }
  \label{fig:avrho-v-R}
\end{figure}

If we average $\rho(R)$ over many events we get
fig.~\ref{fig:avrho-v-R} (left), which shows results for several
generators/tunes.
One observation is that the Pythia tunes have a larger slope: based on
the toy-model calculations, eq.~(\ref{eq:rho-AM-Gauss+pert-v-R}),  this can
be interpreted as meaning that they have a larger value for the
quantity $\sigma n_h$, where $n_h$ is the number of hard jets.
The Herwig default curve, with no MPI, has the smallest slope (and
$\rho$ value).
The average over many events for $\sigma(R)$ (normalised to $\langle
\rho(R) \rangle$) is shown in fig.~\ref{fig:avrho-v-R} (right). 
An interesting feature is the linear rise for $R=0.2-0.4$ (and up to
$R=0.6$ for the Pythia tunes). In this region, one saturates the bound
eq.~(\ref{eq:sigma-upper-bound}), which implies that $R$ is too small
for a proper measurement of $\sigma$. For $R\gtrsim 0.6$ there is a
shallower rise, which we interpret as being due to the presence of
hard jets, as is the case in fig.~\ref{fig:sigma-ext-toy} (though
there is less curvature in the MC events than in the toy model).

To get an idea of the event-to-event variations of $\rho(R)$ we use
the following procedure: given $N$ events, for a given $R$, we sort
the events according to $\rho(R)$. We then define the $10^\mathrm{th}$
percentile result for $\rho(R)$ to be the value of $\rho(R)$ for event
$N/10$, and similarly for other percentiles.
\begin{figure}[t]
  \centering
  \includegraphics[width=0.35\textwidth,angle=-90]{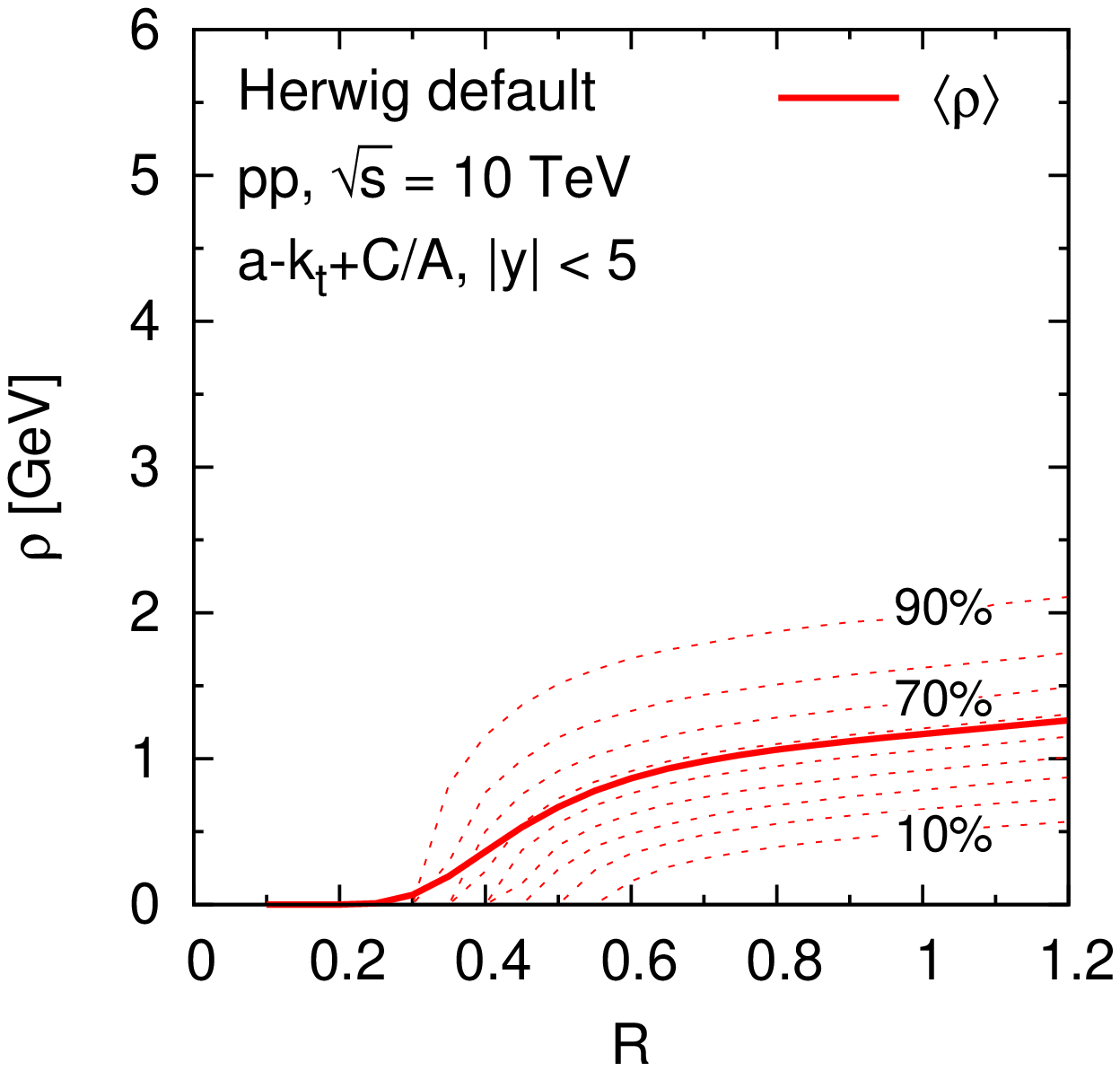}
  \includegraphics[width=0.35\textwidth,angle=-90]{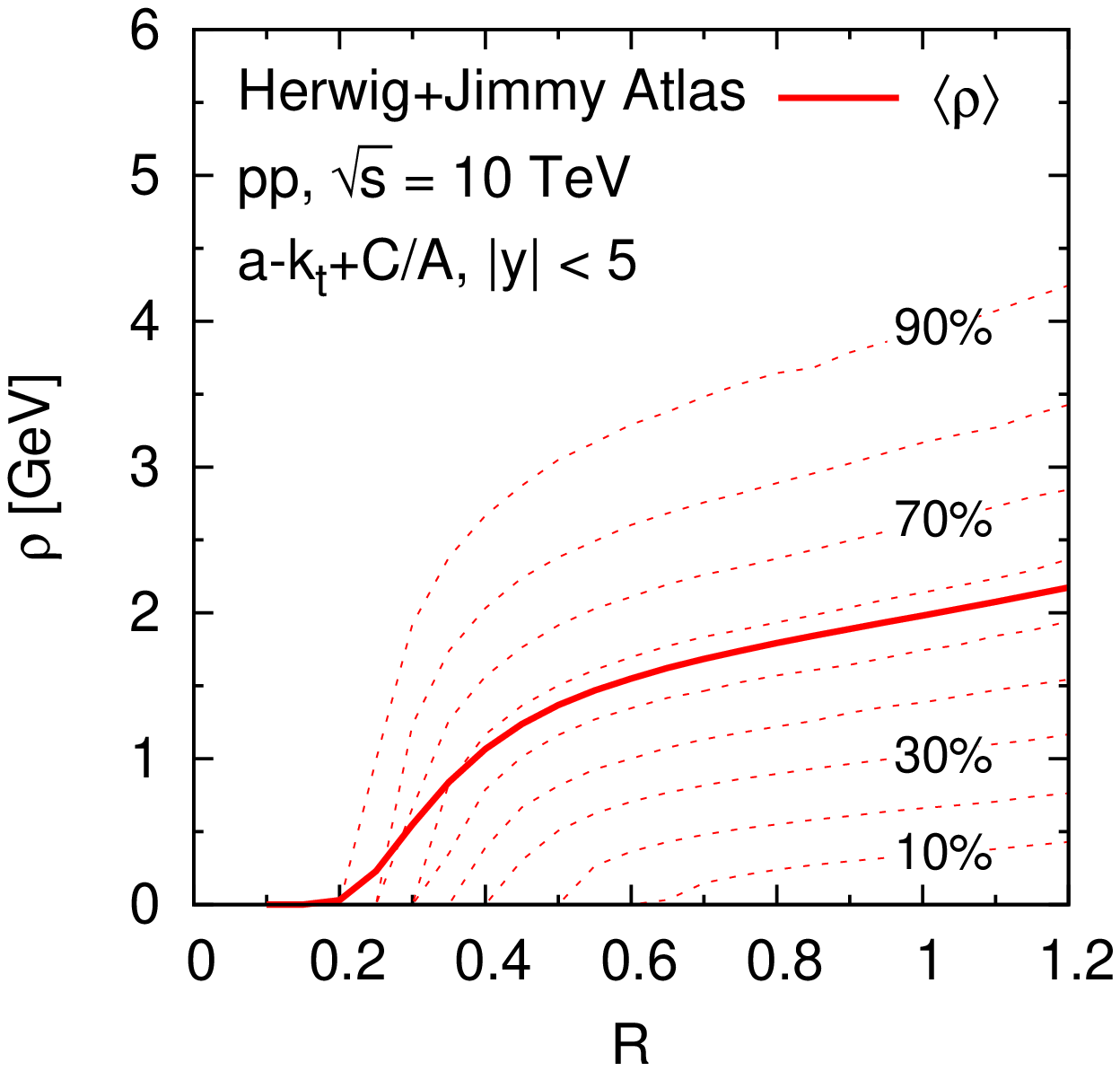}
  \\
  \includegraphics[width=0.35\textwidth,angle=-90]{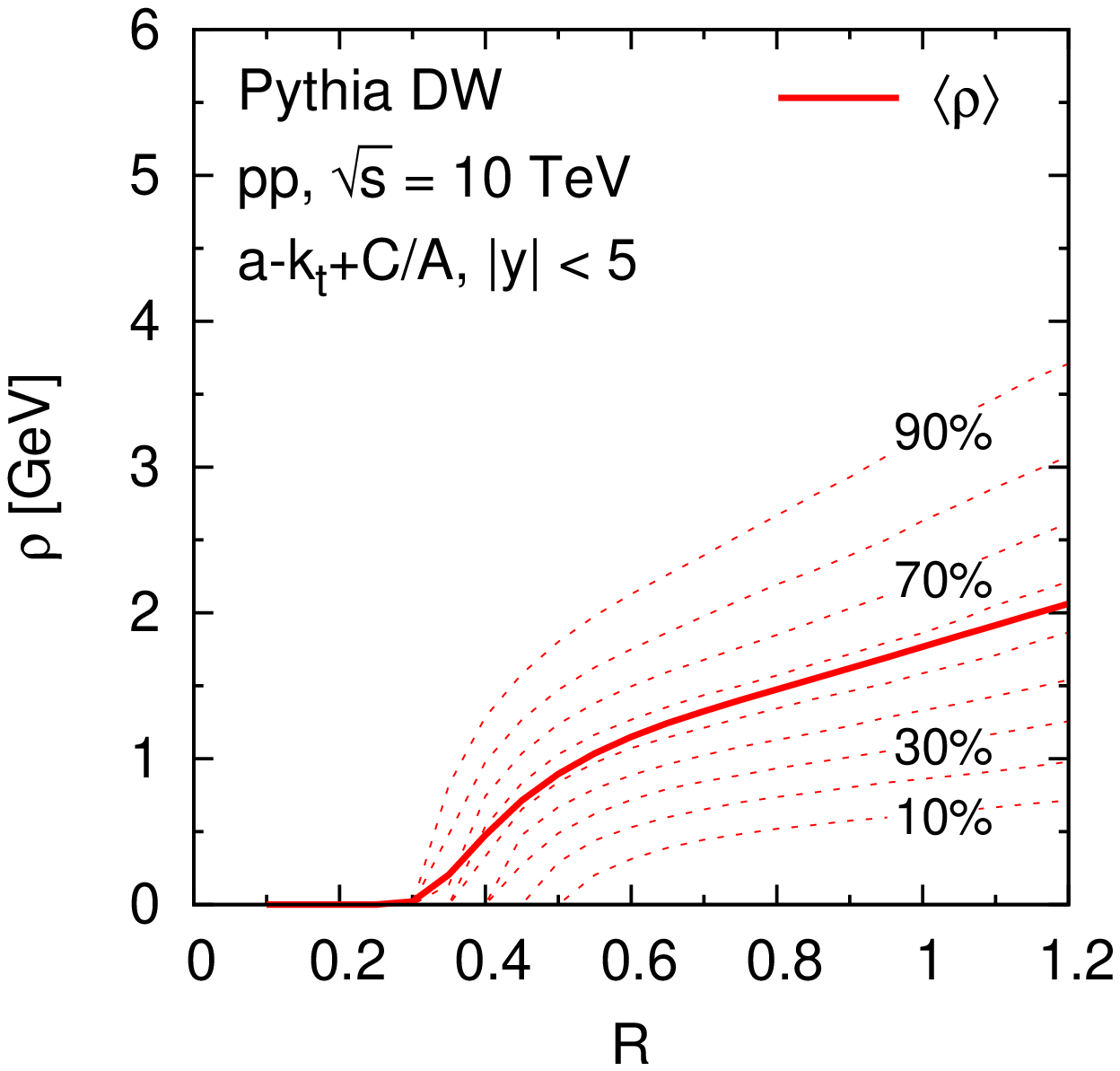}
  \includegraphics[width=0.35\textwidth,angle=-90]{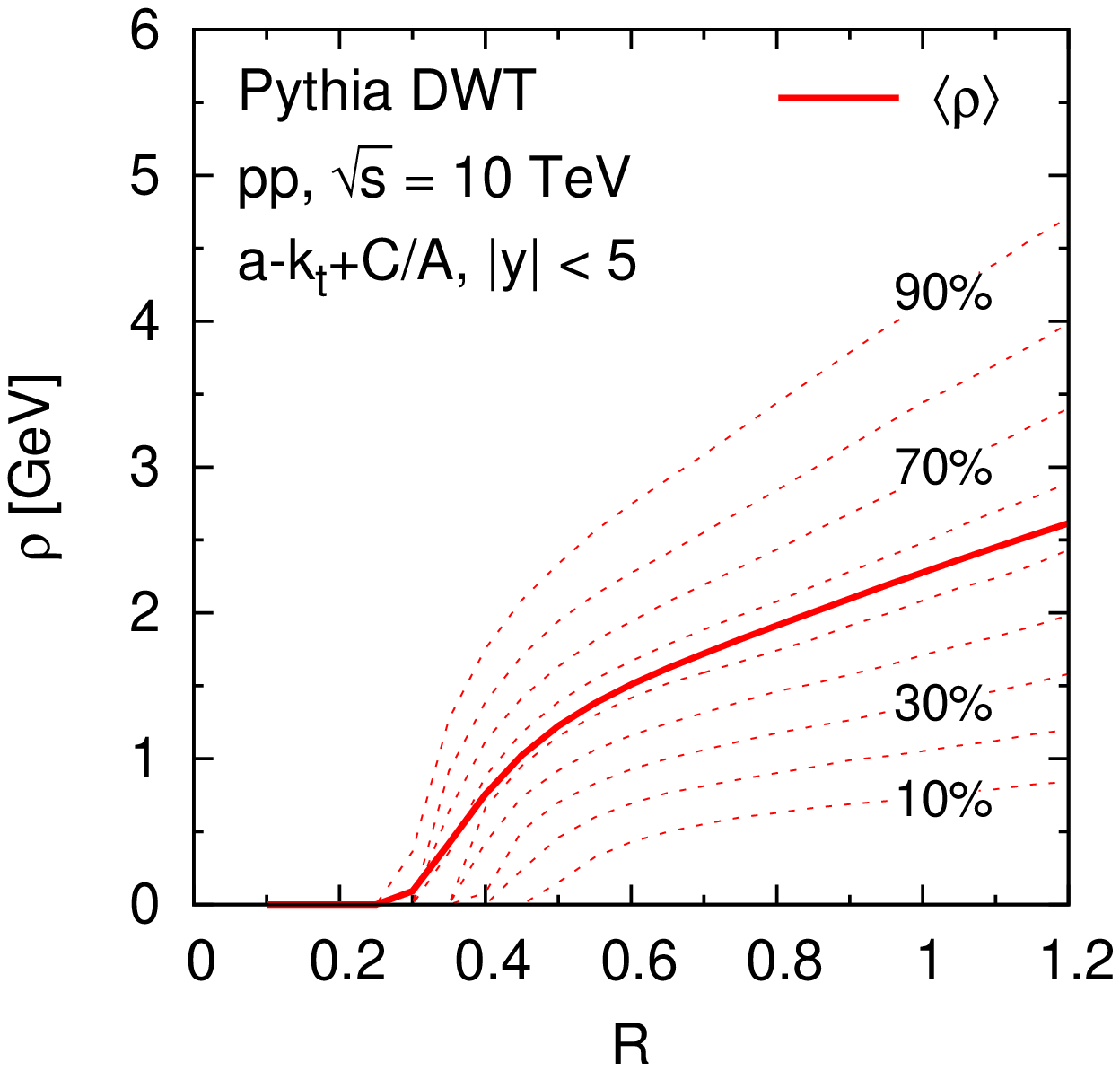}
  \caption{
    $10^\mathrm{th}$, $20^\mathrm{th}$, etc.\ percentiles for
    $\rho(R)$, as a function of $R$ for four generator/tunes.
    Also shown is $\langle\rho(R)\rangle$ in each case.
  }
  \label{fig:rho-v-R-percentiles}
\end{figure}
Fig.~\ref{fig:rho-v-R-percentiles} shows the $10^\mathrm{th}$,
$20^\mathrm{th}$, etc. percentile results for $\rho(R)$, as a function
of $R$ for our 4 generator/tune combinations (together with the
average, for comparison).\footnote{
  A subtle point in the production of
  fig.~\ref{fig:rho-v-R-percentiles} is that the event that provides
  the $10^\mathrm{th}$ percentile for (say) $R=0.6$ is usually not the
  same event that provides it for $R=0.8$. Thus the curves in
  fig.~\ref{fig:rho-v-R-percentiles} are not the curves that would be
  obtained for individual events (these are far less smooth,
  cf. fig.~\ref{fig:3-DWT-events}), but can be thought of as some
  idealisation of these curves in a world free of fluctuations.
}
One observes the sharp turn-on as a function of $R$ (washed out in the
$\langle\rho (R)\rangle$). The smaller the turn-on point, $R_{\crit}$,
the larger than value of $\rho$ (and the larger the slope).
The spread of events is noticeably large, both in the values of $\rho$
and for $R_\crit$, especially, for the latter, in the context of
Herwig+Jimmy. 

Though it is not our intention to highlight figures
\ref{fig:avrho-v-R} and \ref{fig:rho-v-R-percentiles} as main results
of this paper, we do note that it would be possible to make
corresponding experimental measurements, and use them as input to MC
tunes.

In the remaining parts of this section we shall concentrate on results
extracted with $R=0.6$, which, as anticipated in the toy-model section,
seems to offer a reasonable compromise between being sufficiently
large as to be well beyond the turn-on in most events, while not being
too severely affected by the rise at large $R$ that is induced by
semi-hard radiation in the event.

%----------------------------------------------------------------------
\subsection{Study of selected observables}
\label{sec:MC-observables}

%......................................................................
\subsubsection{Mean energy flow}

\begin{figure}[p]
  \centering
  \includegraphics[width=0.45\textwidth,angle=-90]{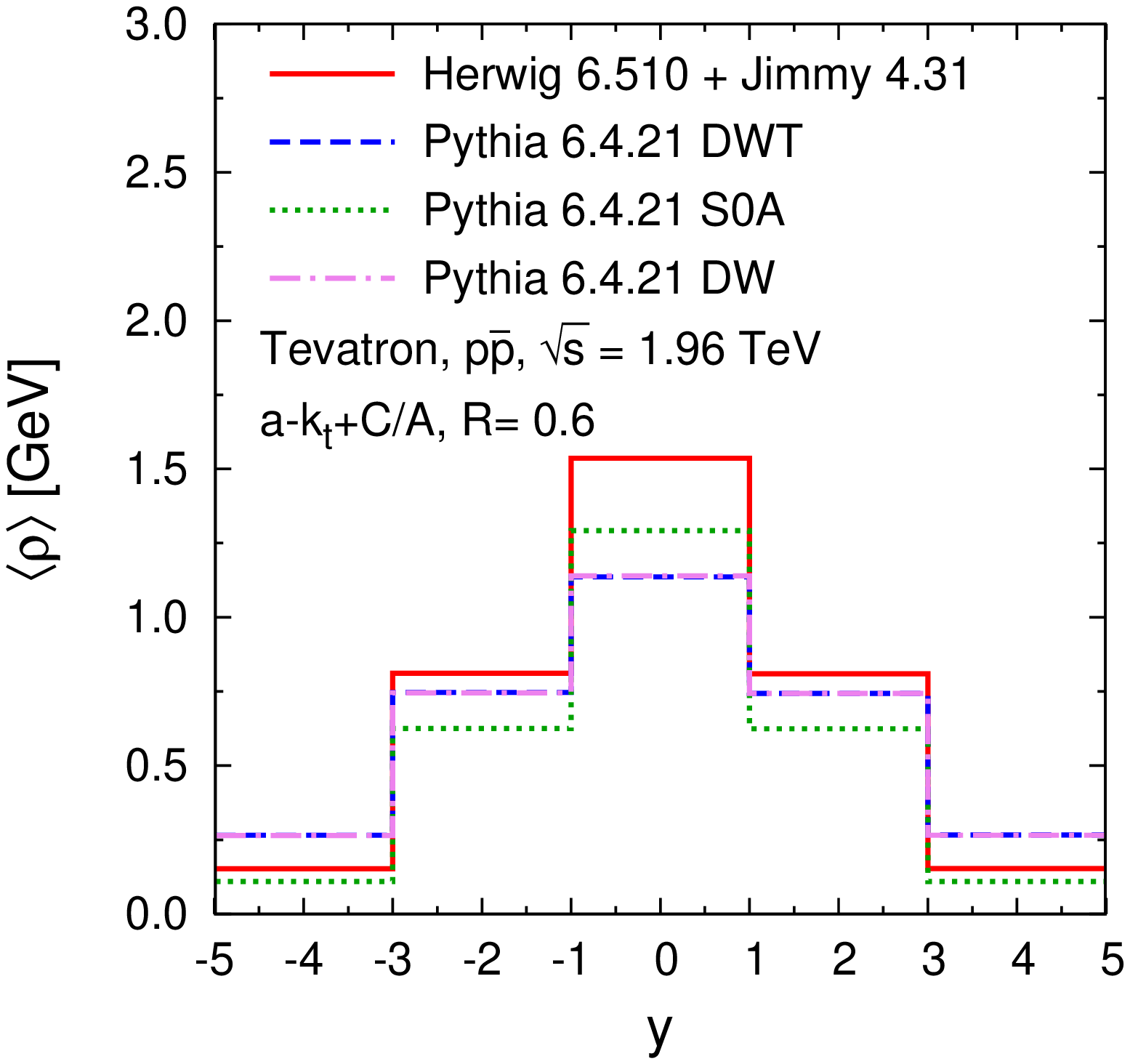}\hfill
  \includegraphics[width=0.45\textwidth,angle=-90]{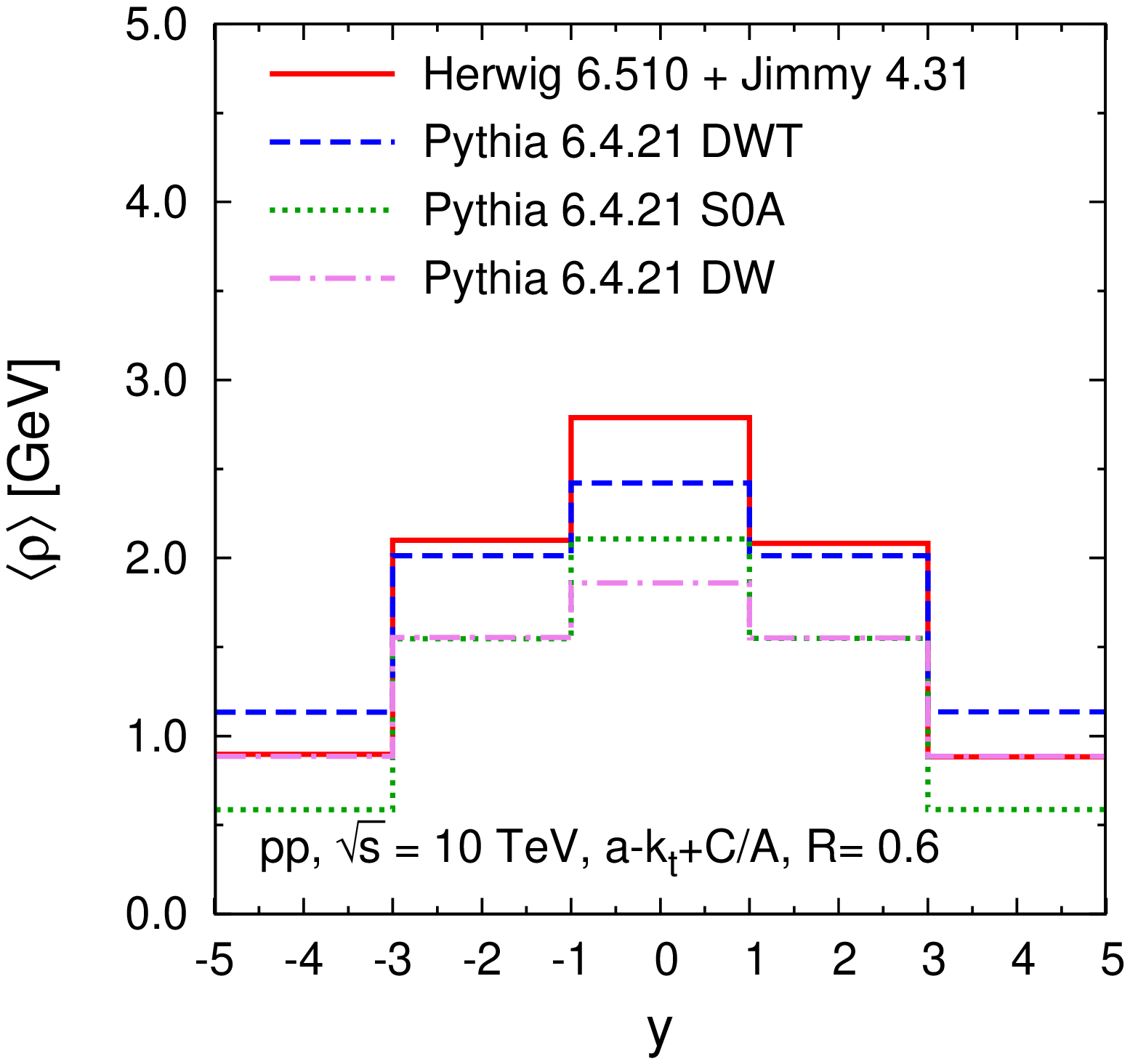}
  \caption{ The rapidity dependence of the average underlying event activity,
    $\mean{\rho(y)}$, shown for $\sqrt{s}=1.96\TeV$ Tevatron ($p\bar
    p$) simulations 
    (left) and $\sqrt{s}=10\TeV$ LHC ($pp$) simulations (right).
  }
  \label{fig:rapav-allgen}
\end{figure}

Let us start by examining $\rho$ and its rapidity dependence,
fig.~\ref{fig:rapav-allgen}, for Tevatron and LHC energies.\footnote{
  To simplify the comparisons, we use the cuts of
  section~\ref{sec:ev-sel} in both cases, though they involve a
  rapidity range that extends beyond the Tevatron's coverage.}
The results for $\mean{\rho}$ in the central rapidity bin for Tevatron
(left) should be
strongly constrained by the standard UE measurements at Tevatron, and
this is reflected in
the small difference between S0A and DW, though there is a somewhat larger
difference with the Herwig+Jimmy tune.
The rapidity dependence is quite strong, with stronger suppression at
forward rapidities for S0A and Herwig+Jimmy than for the DW tune.
One should remember in examining the rapidity dependence that there
are essentially no experimental constraints on the level of the
underlying event at forward rapidities --- it is therefore a
model-dependent extrapolation. 
At LHC energies we see, fig.~\ref{fig:rapav-allgen} (right), 
that differences appear between models also at
central rapidities, reflecting an uncertainty in the extrapolation in
energy. The DWT tune's energy-dependence is disfavoured based on RHIC
\cite{StarUE } and lower-energy Tevatron data~\cite{Acosta:2004wqa},
but we include it to give an idea of the magnitude of possible
differences.

\begin{figure}[p]
  \centering
  \includegraphics[width=0.45\textwidth,angle=-90]{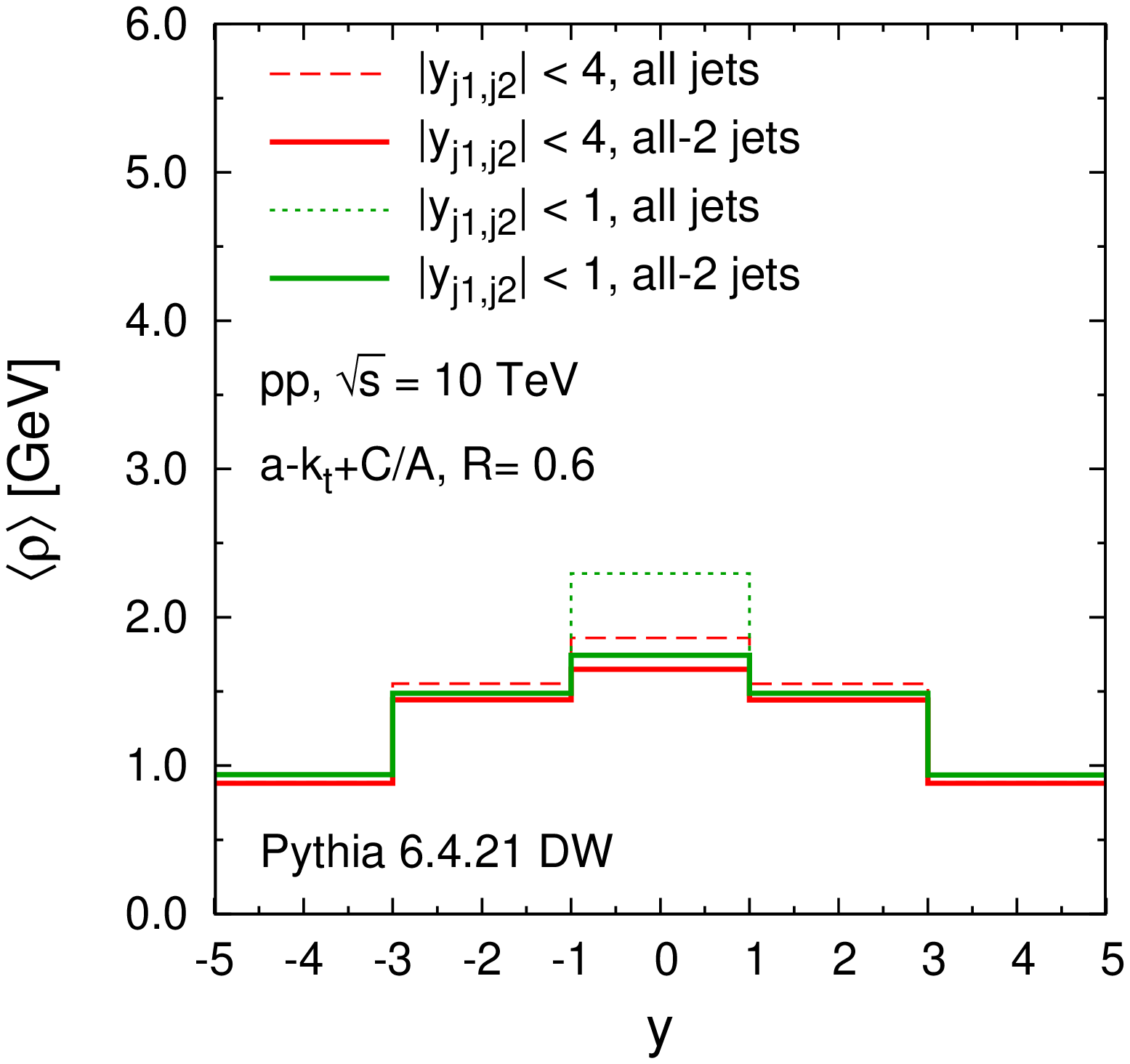}\hfill
  \includegraphics[width=0.45\textwidth,angle=-90]{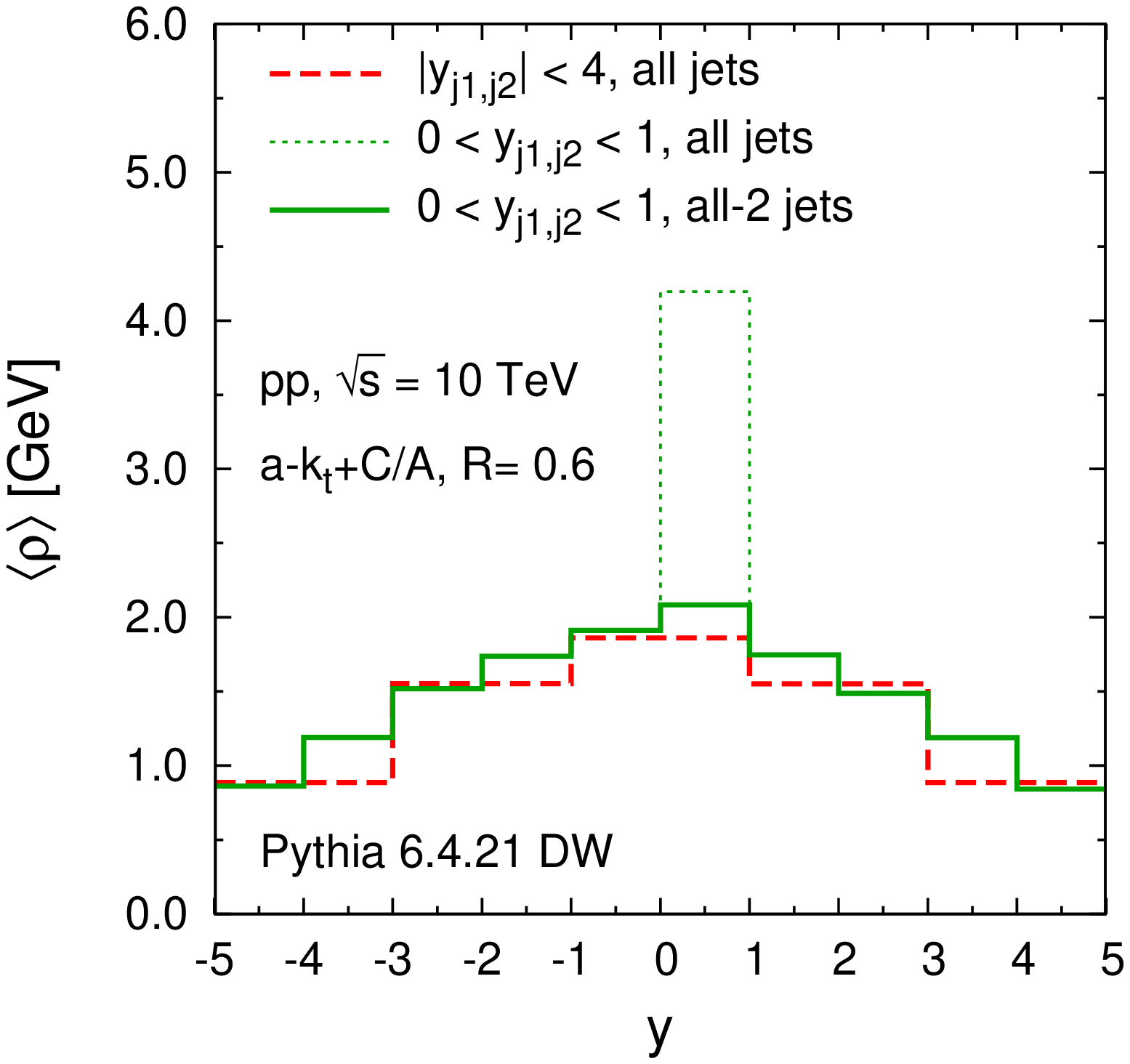}
  \caption{
    Dependence of the $\mean{\rho(y)}$ results on the rapidities of the hard
    jets and on the choice of whether to exclude the two hardest jets
    from the median procedure. The left-hand plot shows this for our
    standard rapidity-bin size, $\delta y = 2$, corresponding to
    $A_\tot=4\pi$, while the right-hand plot shows what happens if we
    choose smaller bins, $\delta y = 1$, i.e.= $A_\tot=2\pi$.
    The results correspond to the Pythia DW tune.
  }
  \label{fig:rapav-with-jety-cuts}
\end{figure}

One question that arises in the measurement of $\rho$ is the possible
bias from hard jets and the relation of this bias with the rapidity
bin size or equivalently the area $A_\tot$ in which one measures $\rho$.
In this context, recall that
eq.~(\ref{eq:rho-AM-Gauss+pert-v-R-numbers}) implies a bias from the
Born jets that scales as $n_b/A_\tot$.
In fig.~\ref{fig:rapav-allgen} the hard jets were free to lie anywhere
within $|y|<4$ and were included in the sample of jets used to obtain
the median.
In fig.~\ref{fig:rapav-with-jety-cuts} (left) the red (dark) dashed curve is
the same as the DW result in fig.~\ref{fig:rapav-allgen} (right). The
red (dark) 
solid curve shows the impact of removing the 2 hardest jets from the
median sample. One sees that this has a rather small effect.
Next we examine an event sample in which the 2 hardest jets are in the central
rapidity bin, so that that bin receives all the bias from the Born
jets.
If those jets are included in the set used to calculate the median
(dotted green histogram), then the impact on $\mean{\rho(y)}$ in that bin
becomes noticeable, $\order{30\%}$.
Removing them from the set for the median (solid green curve) brings
us almost back to the basic ``all-2'' result for the full dijet
sample. That the result is not fully identical is a consequence of the
fact that when the two hard jets are central there is an increased probability
that the 3rd hardest jet will also be central, thus biasing very
slightly the central-$y$ bin.

The right-hand plot of fig.~\ref{fig:rapav-with-jety-cuts} shows what
happens if we reduce the rapidity bin size, causing $\rho$ to be
measured in regions of area $A_\tot=2\pi$ rather than $A_\tot=4\pi$. 
Since the impact of the Born particles
is inversely proportional to $A_\tot$, requiring the Born
particles to be in the central bin has a noticeably larger effect for
the smaller rapidity bin size.\footnote{The effect is definitely
  larger than would be expected based on
  eq.~(\ref{eq:rho-AM-Gauss+pert-v-R}), perhaps a reflection of the
  non-Gaussianity of the distribution of $p_{tj}/A_j$ for the events
  under consideration.}
Discarding the two hard jets brings us back to a result that is
roughly in accord with that for the larger bin size.

The conclusion from fig.~\ref{fig:rapav-with-jety-cuts} is that if
one's event selection does not constrain the hard jets to be in the
same bin as that used for measuring $\rho$ and if the bin area is
sufficiently large, $A_\tot \gtrsim 12$, then biases from the hard jets are
quite small. 
In what follows, we will normally use $A_\tot=4\pi$ and leave in the hard
jets, in order to keep the analysis as simple as possible.

%......................................................................
\subsubsection{Fluctuations}

\begin{figure}[t]
  \centering
  \includegraphics[width=0.45\textwidth,angle=-90]{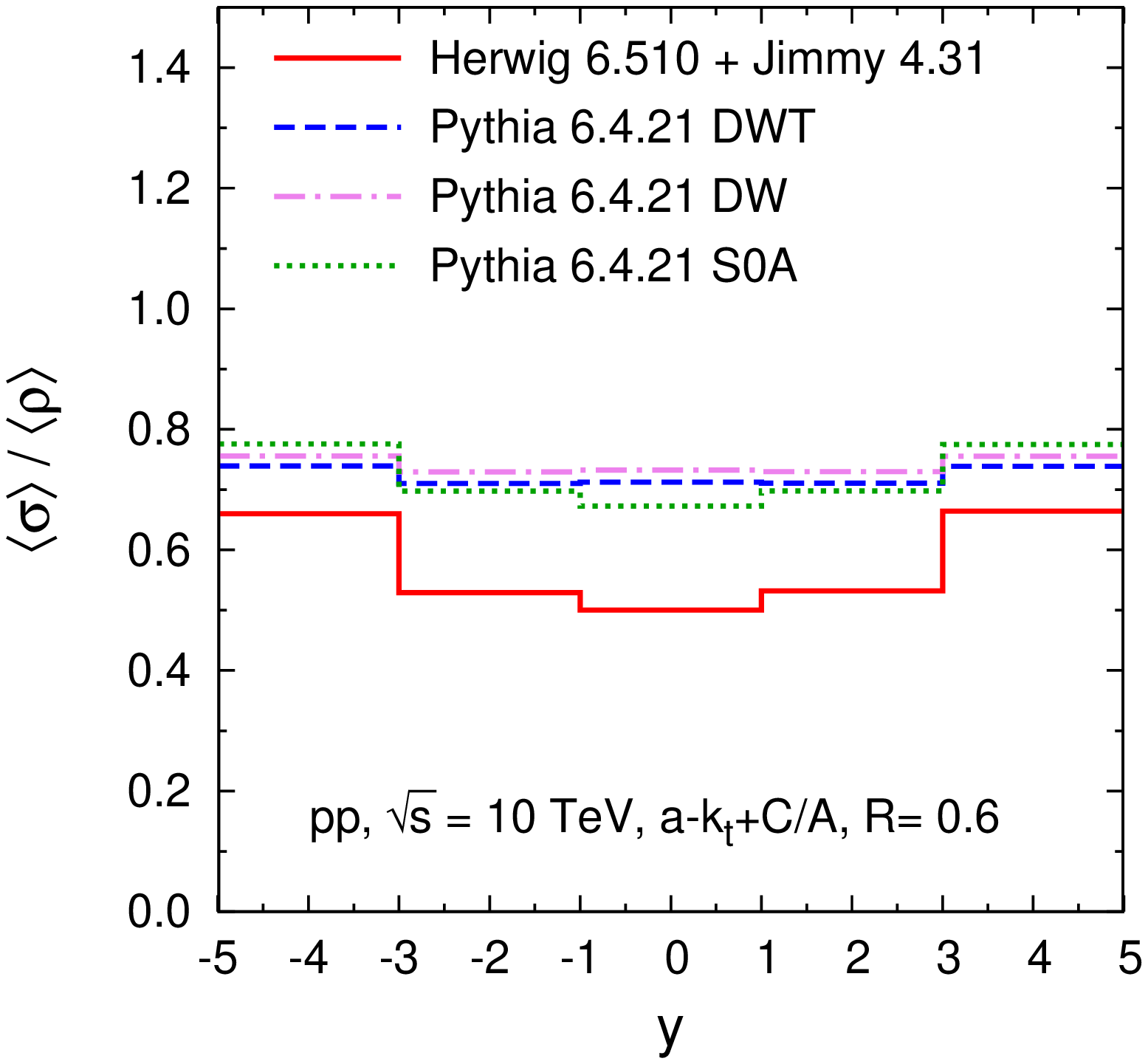}\hfill
  \includegraphics[width=0.45\textwidth,angle=-90]{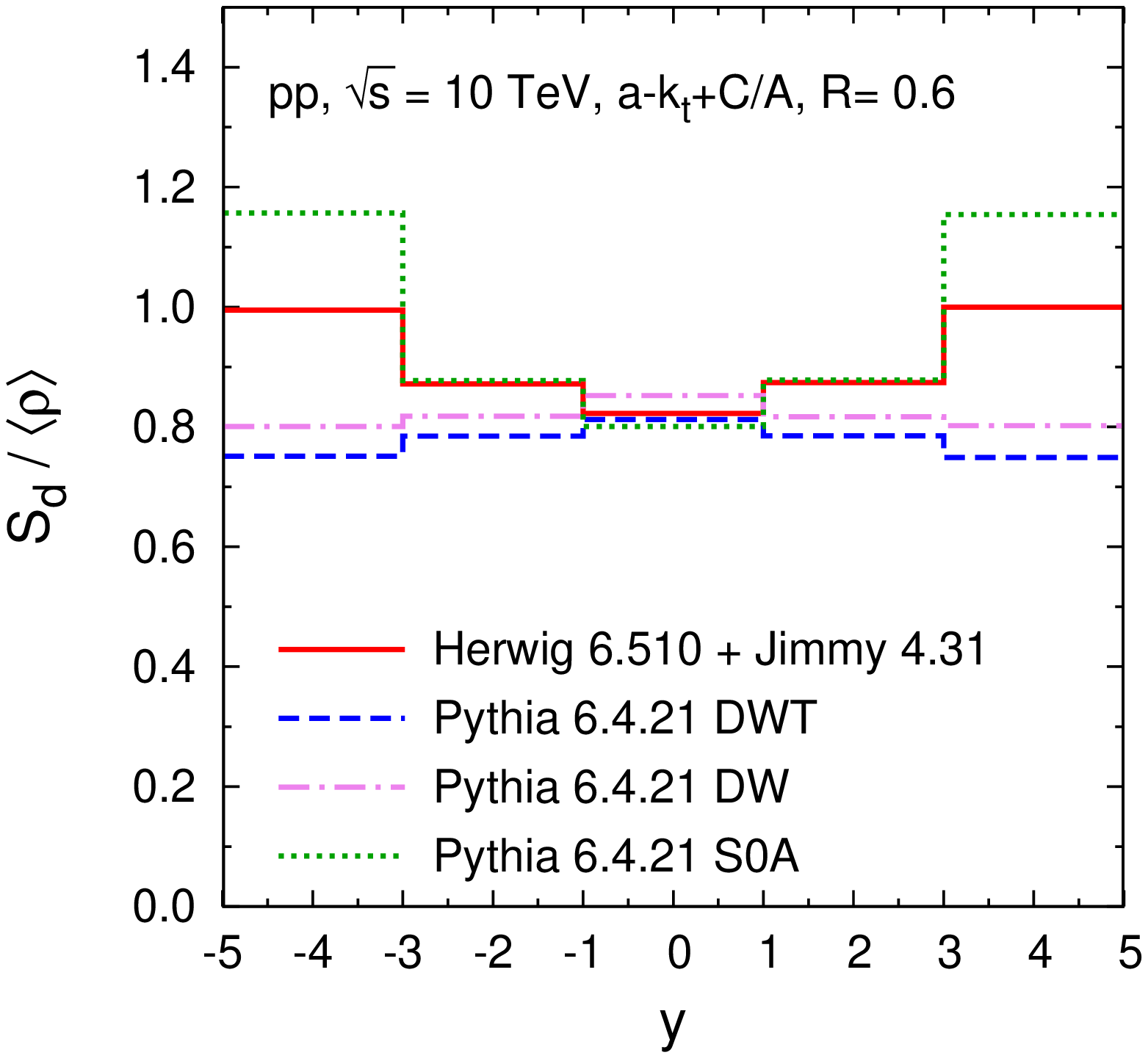}
  \caption{
  Left: $\langle\sigma \rangle /\langle\rho\rangle$ as a function of rapidity.
  Right: $\Sd /\langle\rho\rangle$ as a function of rapidity.
  %% \comment{$\sigma/\rho$ looks like it's saturating at the maximum
  %%   possible value for $R=0.6$ of $0.8$... Do we have it for $R=0.7$? SS: We do not have results of the analysis in rapidity bins but maybe fig.~\ref{fig:avrho-v-R} sheds some light?}
  %% 
}
  \label{fig:rhosdav-allgen}
\end{figure}

In fig.~\ref{fig:rhosdav-allgen} we examine fluctuations, now only
for LHC. The left-hand plot shows the size of intra-event
fluctuations, through the ratio of $\langle\sigma(y)\rangle$ to
$\langle\rho(y)\rangle$, while the right hand plot shows inter-event
fluctuations, through $\Sd(y) / \langle\rho(y)\rangle$.
Fluctuations have not been as directly tuned as energy flow. Despite
this the intra-event fluctuations are very similar across all the
Pythia tunes and almost independent of rapidity when normalised to
$\langle \rho(y)\rangle$.
Herwig+Jimmy's intra-event fluctuations are somewhat smaller, but do have
rapidity dependence.

Concerning inter-event fluctuations, fig.~\ref{fig:rhosdav-allgen}
(right), the two virtuality ordered
(DW/DWT) Pythia models are again flat, whereas the $p_t$-ordered shower
has increasing fluctuations at forward rapidities. Herwig is
intermediate between the two sets of Pythia results.

\begin{figure}[t]
  \centering
  \includegraphics[width=0.45\textwidth, angle=-90]{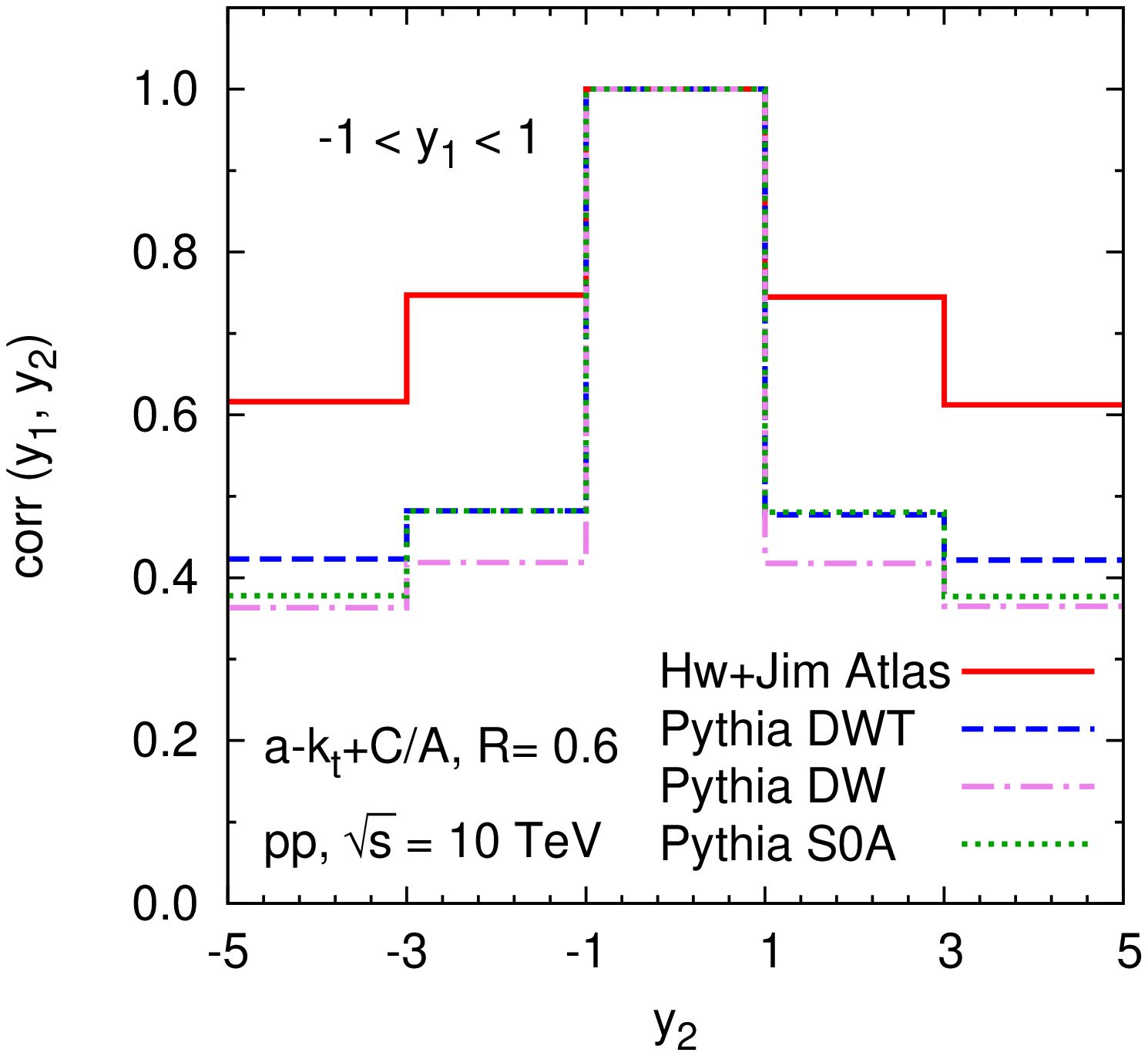}\hfill
  \includegraphics[width=0.45\textwidth, angle=-90]{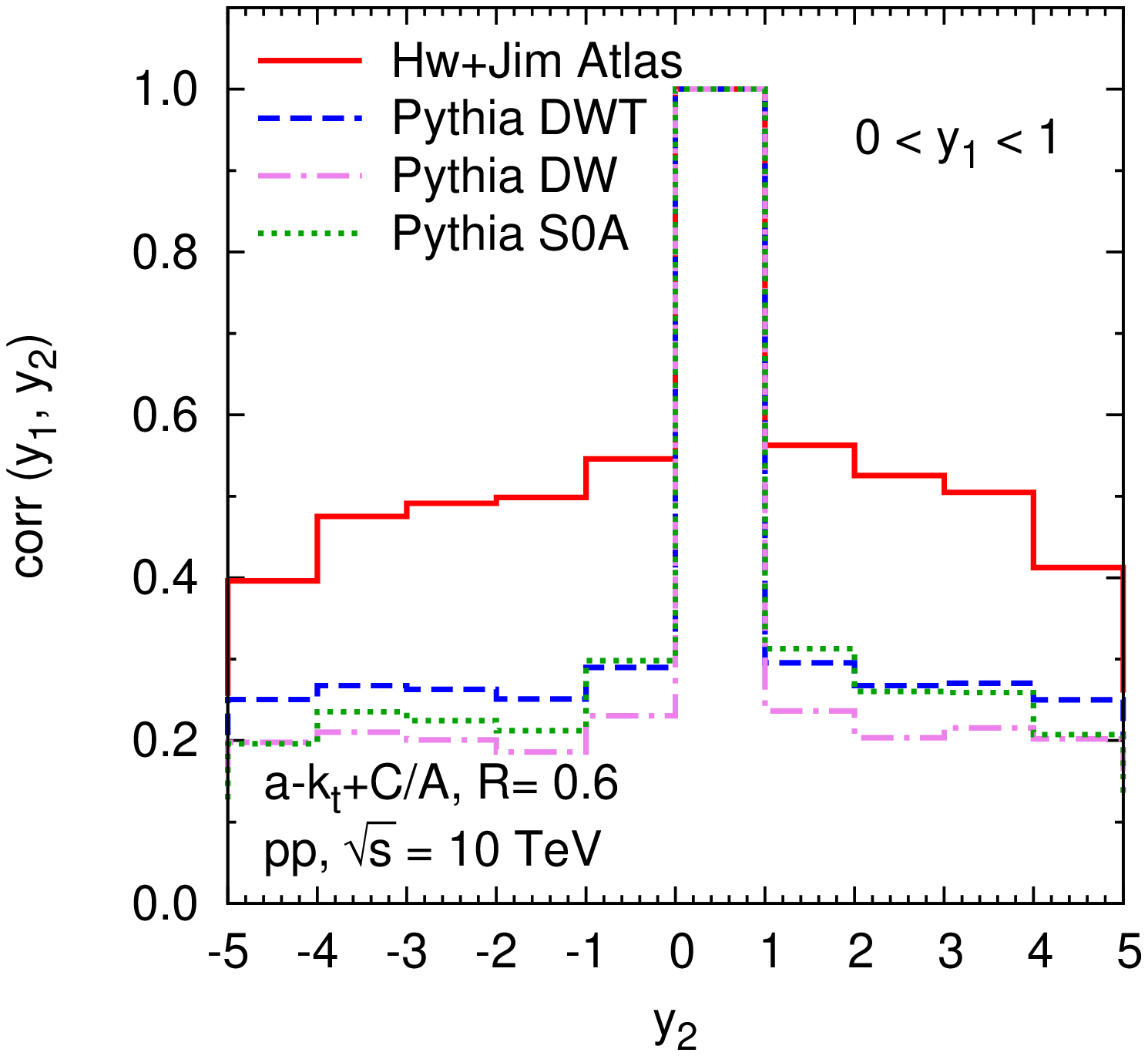}\medskip

  \begin{minipage}{0.5\linewidth}
    \includegraphics[width=0.9\textwidth, angle=-90]{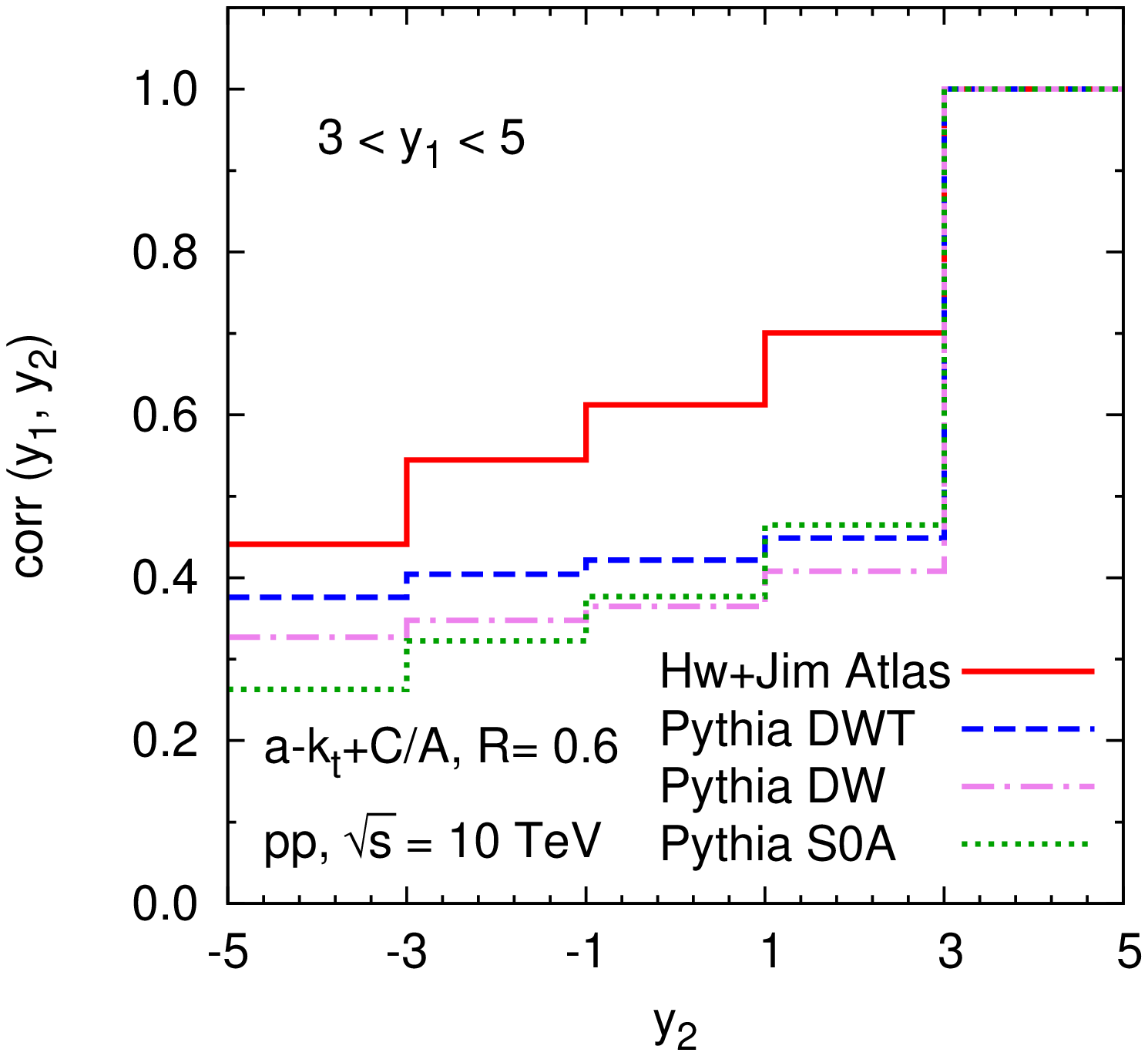}
  \end{minipage}\hfill
  \begin{minipage}{0.45\linewidth}
  \caption{
    \label{fig:rapcor-allgen}
    Correlation of $\rho(y_2)$ with $\rho(y_1)$, shown as a function
    of $y_2$ for $y_1$ in a given rapidity bin: (a) $-1<y<1$, (b)
    $0<y<1$ and (c) $3<y_1<5$.
    In plots (a) and (c) $\rho$ has been determined in bins of size
    $\delta y = 2$, while in (b) it has been determined in bins of
    size $\delta y = 1$.
  }    
  \end{minipage}
\end{figure}
\begin{figure}[t]
  \centering
  \includegraphics[width=0.45\textwidth,angle=-90]{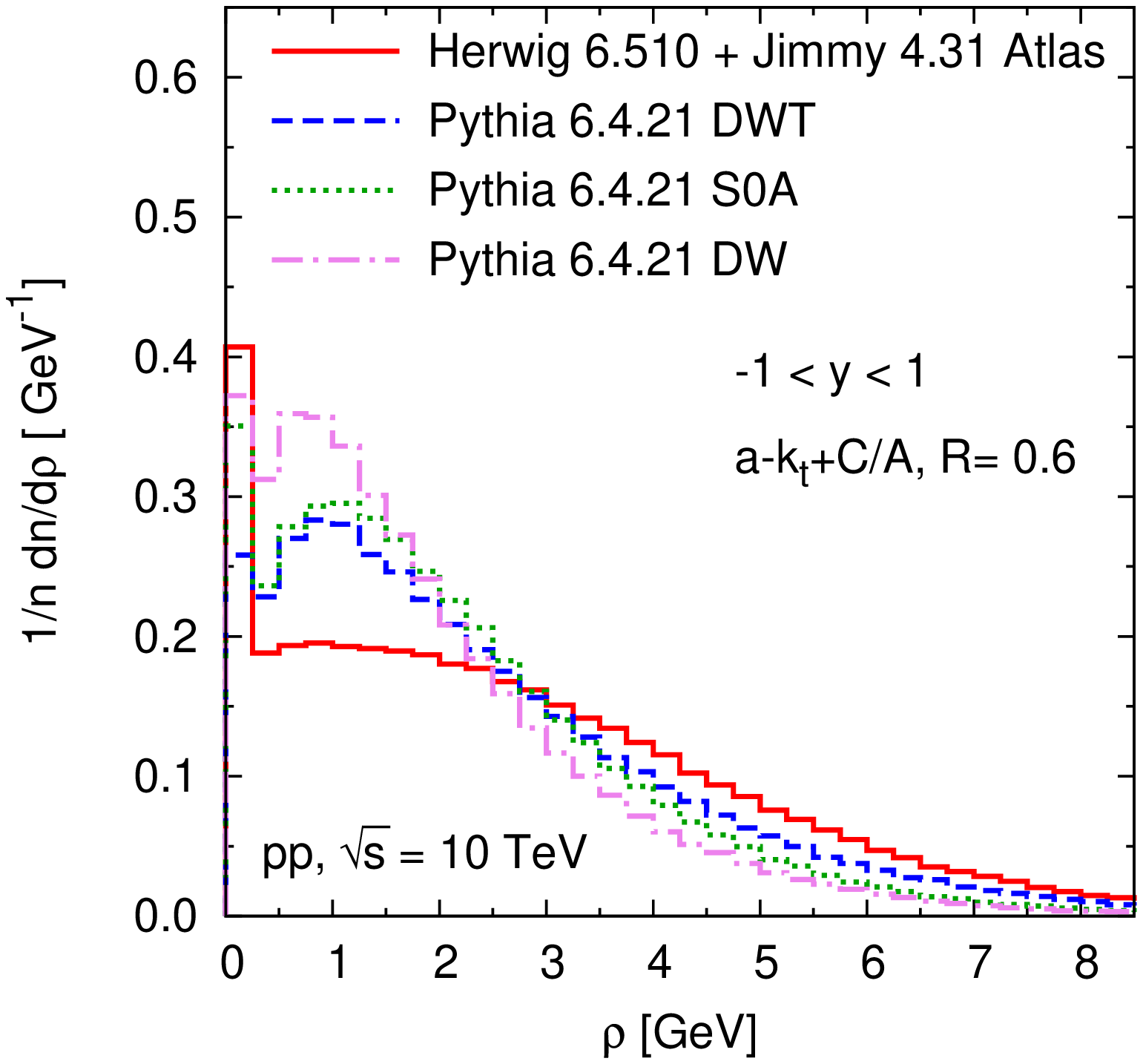}\hfill
  \includegraphics[width=0.45\textwidth,angle=-90]{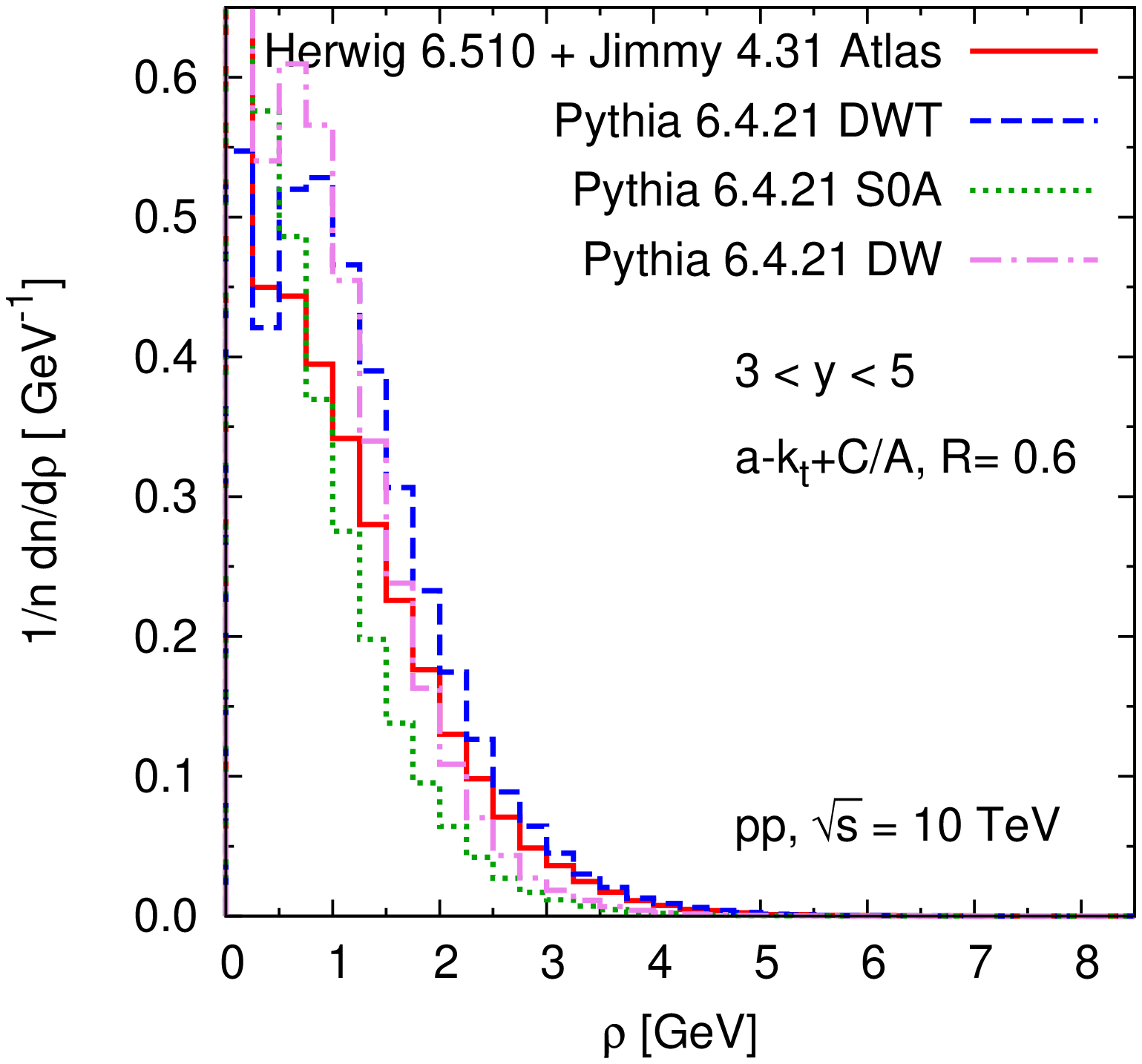}
  \caption{The event-by-event distribution of $\rho$ for a selection
    of generators, in a central rapidity bin, $|y|<1$ (left-hand plot)
    and a forward rapidity bin, $3<y<5$ (right). }
  \label{fig:ev-by-ev-rho-dist}
\end{figure}

One observation is that whereas DW/DWT have almost identical intra and
inter-event fluctuations, Herwig's intra-event fluctuations are nearly
$40\%$ smaller than the inter-event fluctuations.
This is reflected also when we examine correlations between $\rho$ in
different parts of the event, 
\begin{equation}
  \label{eq:correl}
  \mathrm{corr}(y_1,y_2) = \frac{\mean{\rho(y_1) \rho(y_2)} -
    \mean{\rho(y_1)}\mean{\rho(y_2)} }{S_d(y_1) S_d(y_2)}
\end{equation}
as shown in fig.~\ref{fig:rapcor-allgen}. 
The correlations are noticeably larger for Herwig+Jimmy than they are for all the Pythia
tunes. 
In determining the correlations it was important that we used sufficiently
large rapidity bins. Comparing the upper-left plot from fig.~\ref{fig:rapcor-allgen} ($\delta y = 2$) and the upper-right plot  ($\delta y = 1$) one sees
that the smaller rapidity bins lead to  noticeably smaller measured
correlations. 
We interpret this as follows: in small rapidity bins, the
``statistics'' of jets for measuring the $\rho$ value are more
limited. This increases the error on the determination of $\rho$, thus
reducing the maximum amount of correlation that can be observed
between different bins.

The final quantity that we examine is the event-by-event distribution
of $\rho$, fig.~\ref{fig:ev-by-ev-rho-dist}, for a central rapidity
bin (left) and a forward rapidity bin (right). 
Perhaps the most striking characteristic of these plots is the very
broad nature of these distributions, which are far from being Gaussian
distributions of width $\Sd$ centred on $\langle \rho \rangle$. 
The right-hand plot also has a significant bin at $\rho=0$ ---
i.e. there is a substantial number of events for which at least half of the
jets at forward rapidities are pure ghost jets.

%
%
%   }

%......................................................................
\subsubsection{Energy dependence of results}
\begin{figure}[t]
  \centering
  \includegraphics[width=0.45\textwidth,angle=-90]{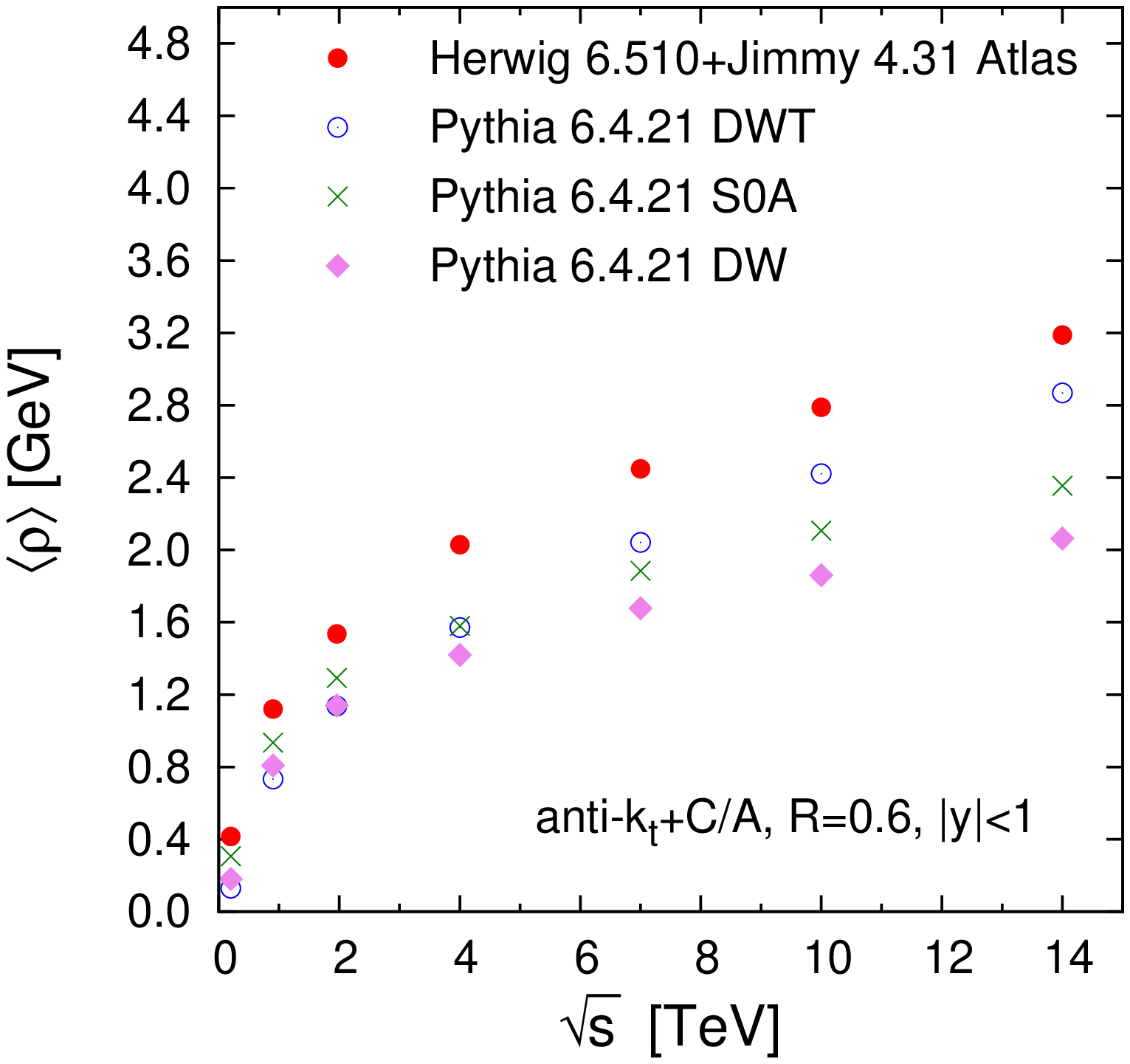}\hfill
  \includegraphics[width=0.45\textwidth,angle=-90]{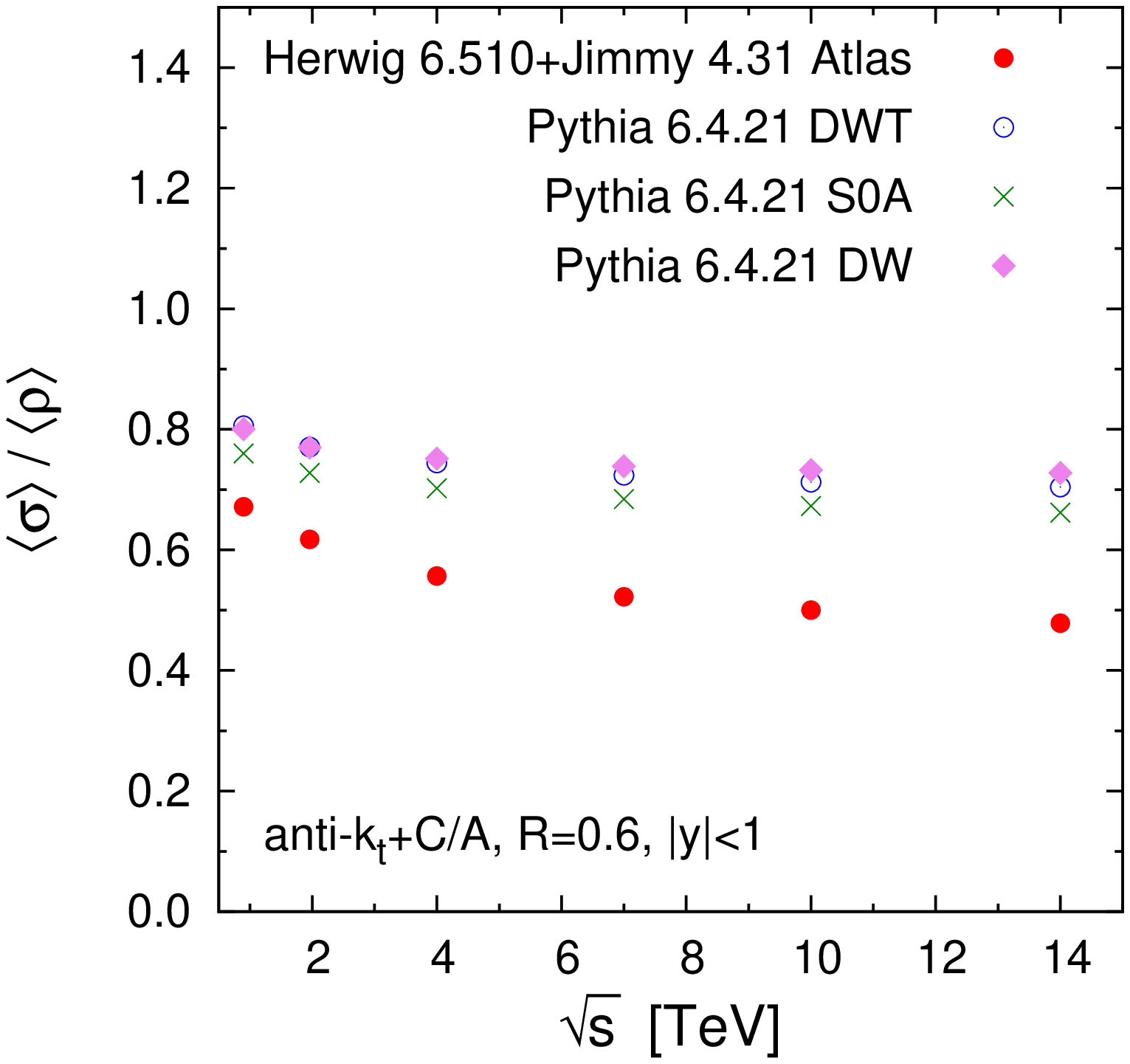}
  \caption{%
  Level of the underlying event $\langle \rho \rangle$ (left) and its
  intra-event fluctuations $\langle \sigma \rangle/\langle \rho
  \rangle$ (right) in central rapidity rapidity bin $|y|<1$ as a
  function of energy of $pp$ collision. The~points correspond to
  the energies of RHIC, Tevatron and a set of possible LHC $\sqrt{s}$
  values. 
  \label{fig:edep-rho-sigma}
  }
\end{figure}

Fig.~\ref{fig:edep-rho-sigma} summarises the energy-dependence of the average energy flow
$\langle \rho \rangle$ (left) and of the intra-event fluctuations
$\langle\sigma \rangle$ normalised to $\langle \rho \rangle$ (right) in the central rapidity bin $|y|<1$.
The features of note are that for $\langle \rho \rangle$, the
Herwig+Jimmy Atlas tune has a significantly steeper energy dependence than the
DW and S0A Pythia tunes, somewhat more like DWT.
For $\langle\sigma \rangle / \langle\rho \rangle$, the Pythia tunes
have almost no energy-dependence, whereas Herwig has substantial
energy-dependence. 
We have not shown the energy dependence of $\Sd/\langle\rho \rangle$
because it is essentially independent of $\sqrt{s}$  for all generators.

%% %======================================================================
%% 
%% %
%% %
%% %
%% %
%% 
%% %
%% %
%% %
%% 
%% 
%%   % 
%%   %
%%   %
%% } 
%% 
%% %
%% %
%% 
%% %
%% 
%% %
%% %
%% %
%% 
%%   %
%% %
%% 

% 
% %%---------------------------------------------------------------------
% 
% %
% 
% %
% %
% 
% 
% %
% %
% 
% 
%   % 
%   % 
%   %
%   %
% } 

% ======================================================================
\section{Conclusions}
\label{sec:conclusions}

The distinction between underlying event and the ``hard'' part of
hadron-collider events is ill-defined in QCD.
Nevertheless, physically, one may picture the UE as being the
component of a hadron-hadron scattering that fills the event fairly
uniformly with low-$p_t$ radiation.
A goal of this article was to investigate how different UE-measurement
strategies fare in separating such a low-$p_t$ component from the hard
part of the event.

To do so, we developed a simple toy-model for events, with two
distinct components, one soft, corresponding to the UE, the other
perturbative and hard.
Within this model it is quite straightforward to establish to what
extent a given UE measurement strategy correctly extracts just the
soft component.

The two UE measurement strategies that we investigated are the
``traditional'' approach, measuring radiation in regions transverse to
a leading jet, and the ``jet-area/median'' approach.
Both involve strategies to help separate out the soft and hard
components: the use of the Av/Min/Max transverse regions in the traditional
approach, the use of median activity rather than average in the
jet-area/median approach.
One result from the toy model is a quantification of how those
strategies fare in the extraction of its soft component. 
A second result is a determination of the nature of the residual
effects due to perturbative (hard) radiation.
These two results could be expressed analytically in terms of the
parameters of the measurement procedures (transverse-region area, jet
radius, providing useful guidance in choosing them) and of
characteristics of the hard scattering (notably the value of the hard
scale and the properties of the soft component).
We also examined the question of event-to-event fluctuations in the
extraction of the characteristics of the soft component. 
Our toy-model results are summarised in
section~\ref{sec:toy-model-summary}. 

Practically, one conclusion from this work is that for determinations
of averaged quantities, for example the mean transverse-momentum
density per unit area,
$\langle \rho \rangle$, both the TransMin and the area/median
measurement methods give a fair determination of the soft component, as long 
as the momentum transfer of the hard scattering is not too large
($\lesssim 100\GeV$ for the LHC; for higher momentum transfers, the
TransMin method is affected by a rapidly increasing hard contribution).
In particular, for the parameter choices used or advocated in the
literature, the two kinds of bias seen in the toy model,
mismeasurement of the soft 
component and contamination from the hard component, tend to partially
cancel each other, giving a limited overall bias, of order $20\%$.
In contrast, for event-by-event measurements and determinations of
fluctuations of the soft component, the traditional approach is
significantly affected by rare ``outliers'',
cf. table~\ref{tab:toy-Sd-values}.

For this reason, in the full Monte Carlo studies of section
\ref{sec:area-based-stuff} we concentrated on the area/median
approach. 
The results included a validation of the main qualitative prediction
from the toy model, namely the structure of the $R$-dependence of the
extracted $\rho$,
section~\ref{sec:MC-v-toy}.
Section~\ref{sec:MC-observables} showed a range of possible observables
whose measure we advocate at LHC: rapidity dependence of the UE,
nature of the event-to-event fluctuations, and intra-event
fluctuations and correlations.
Though existing measurements may indirectly constrain some of these
features of the UE, we believe that they are of sufficient practical
importance that they deserve dedicated measurements, especially as
they differ noticeably between various Monte Carlo models.

%% %
%% %
%% %
%% %

% 
% 
% 
% 
%   %
% 
%   %
% 
% 
% 
% 
% 
% 
% 
% 
% 
% 
% 

\section*{Acknowledgements}
We would like to thank Juan Rojo and Gregory Soyez for collaboration
during the initial stages of this work, and Gregory Soyez for helpful
comments throughout.
We are also grateful to Jon Butterworth and Mike Seymour for
assistance with Jimmy, and Rick Field, Witek Krasny, Torbj\"orn
Sj\"ostrand, Peter Skands and Mark Strikman
 for useful conversations.
This work was supported in part by the French ANR under contract
ANR-09-BLAN-0060.

\appendix

%% 
%% 

%% 

%----------------------------------------------------------------------
\section{Toy model UE calculations}

\label{sec:more-toy-model}

%......................................................................
\subsection{Threshold and asymptotic regions}

With the exponential model for the single-particle $p_t$ distribution,
eq.~(\ref{eq:pt-dist-P1}),
there are two asymptotic limits of interest: the threshold for $\nu A
\to \ln 2$ and the asymptotic $\nu A \to \infty$ region (for brevity
here we write $A \equiv A_{\tile}$).
Writing eq.~(\ref{eq:rho-ext-tiles}) as an expansion around $\nu A =
\ln 2$, one obtains
\begin{equation}
  \label{eq:exp-analytic-work}
  \frac12\left(1 + \ln 2 - \nu A\right) + \frac{\ln 2}2 
  \int_0^{A \mean{\rho_{\ext}}}
    \frac{dp_t}{\mu} e^{-p_t/\mu} + \order{(\nu A - \ln 2)^2} = \frac12\,,
\end{equation}
where we have used eq.~(\ref{eq:pt-dist-PA-sum-ingredients}) for
$dP/dp_t$ and kept only the $\delta$-function term and the first term
of the sum over $n$.
To first order in $\nu A -\ln 2$, this gives us the behaviour in the
turn-on region,
\begin{equation}
  \label{eq:rhoext-exp-near-ln2}
  A \mean{\rho_\ext} = \frac{\nu A - \ln 2}{\ln 2}\mu  + \order{(\nu A - \ln
    2)^2\mu}\, ,
\end{equation}
or, equivalently, using $\mu=\rho/\nu\simeq  \rho A/\ln 2$ for $\nu A
\simeq \ln 2$,
\begin{equation}
  \label{eq:rhoext-exp-near-ln2-alt}
  \frac{\mean{\rho_\ext}}{\rho} = \frac{\nu A - \ln
    2}{\ln^2 2} + \order{(\nu A - \ln 2)^2}\,.
\end{equation}
The approximation eq.~(\ref{eq:rho-ext-tiles-approx}) reproduces the
linear dependence on $\nu A - \ln 2$ at threshold, though its
slope there, $d\!\mean{\rho_\ext}\!/d(\nu A)= 2$, differs slightly from the exact
slope of $1/\ln^2 2$.

At large $\nu A$, by examining the numerical solutions to
eq.~(\ref{eq:rho-ext-tiles}), we have determined the following
relation for the asymptotic behaviour of $\rho_\ext$,
\begin{equation}
  \label{eq:rhoext-exp-large-nu}
  \mean{\rho_\ext} = \nu\mu  - \frac\mu{2A} + \order{\frac\mu{\nu
      A^2}}
  =
  \rho - \frac{\rho}{2\nu A} + \order{\frac{\rho}{\nu^2 A^2}}
  \,.
\end{equation}
This is reproduced by the approximation of eq.~(\ref{eq:rho-ext-tiles-approx}).

An approximation that is closer still to the full result, with the
correct coefficients in both limits, is 
\begin{equation}
  \label{eq:rhoext-exp-betterapprox}
  \rho_{\ext} = \rho \frac{X + X^2}{(\ln 2)^2 + X^2 + \frac32
    X}\,,\qquad\quad
  X = \nu A - \ln 2\,,
\end{equation}
however the difference between this and
eq.~(\ref{eq:rho-ext-tiles-approx}) is irrelevant for all practical
purposes.

%......................................................................
\subsection{Variant of toy model}
We can also consider a model with 
\begin{equation}
  \label{eq:pt-dist-P1-x-exp}
  \frac{1}{P_1} \frac{dP_1}{dp_t} = \frac{4p_t}{\mu^2} e^{-2p_t/\mu}\,.
\end{equation}
and correspondingly
\begin{equation}
  \label{eq:pt-dist-Pn-x-exp}
  \frac{1}{P_n} \frac{dP_n}{dp_t} = \frac{1}{\mu} \frac{2^{2n}}{(2n-1)!} \left(\frac{p_t}{\mu}\right)^{2n-1} e^{-2p_t/\mu}\,.
\end{equation}
This model has the property that $\sigma = \sqrt{3\nu/2}\mu$.
For $\nu$ near $\ln 2$ this leads to 
\begin{equation}
  \label{eq:x-exp-analytic-work}
  \frac12\left(1 - X\right) + \frac{\ln 2}2 \int_0^{A \mean{\rho_{\ext}}}
    dp_t\, \frac{4p_t}{\mu^2} e^{-2p_t} + \order{X^2} = \frac12\,,
\end{equation}
with $X$ defined as in eq.~(\ref{eq:rhoext-exp-betterapprox}),
resulting in
\begin{equation}
  \label{eq:rhoext-A-x-exp-near-ln2}
  A \mean{\rho_\ext} = \mu \sqrt{\frac{\nu A - \ln 2}{2\ln 2}} + \order{(\nu
    A - \ln 2)^{3/2} \mu}
  \,,
\end{equation}
or, equivalently,
\begin{equation}
  \label{eq:rhoext-x-exp-near-ln2}
  \mean{\rho_\ext} = \frac{\rho}{\ln 2} \sqrt{\frac{\nu A - \ln 2}{2\ln 2}} + \order{(\nu
    A - \ln 2)^{3/2} \rho}
  \,.
\end{equation}
At large $\nu$ one finds, again numerically, 
\begin{equation}
  \label{eq:rhoext-x-exp-large-nu}
  \mean{\rho_\ext} = \nu \mu  - \frac\mu{3 A} + \order{\frac\mu{\nu
      A^2}}
  =
  \rho - \frac{\rho}{3\nu A} + \order{\frac{\rho}{\nu^2 A^2}}
  \,.
\end{equation}
The two above equations both imply that $\mean{\rho_{\ext}}$ turns on and
approaches its asymptotic value somewhat faster than in the model with 
$\frac{dP_1}{dp_t} \propto e^{-p_t/\mu}/\mu$.
A reasonable analytic approximation for $\mean{\rho_{\ext}}$ over the
whole domain is
\begin{equation}
  \label{eq:rho-ext-tiles-approx-x-exp}
  \mean{\rho_{\ext}} \simeq \rho\, \sqrt{\frac{\nu A - \ln 2}{\nu A
    - \ln 2 + \frac23}} \,\Theta(\nu A - \ln 2)\,,
\end{equation}
which has the correct large-$\nu$ behaviour, and very nearly the correct
coefficient for the $\sqrt{\nu A - \ln 2}$ turn-on.

%======================================================================
\section{Fluctuations in area/median extraction of $\boldsymbol{\rho}$}

%----------------------------------------------------------------------
\subsection{Pure soft case}
\label{sec:pure-soft-case-fluct}

To determine the event-to-event fluctuations $S_d$ in the area/median
extraction of $\rho$ when the intrinsic event-to-event fluctuations
are zero, it is convenient to work in the limit $\nu A_{\jet} \ll 1 $
such that the probability distribution of $\rho_{\jet}$ is close to a
Gaussian,\footnote{We will ignore the issue of fluctuations in
  $A_\jet$ itself, i.e.\ it is really the tiled case that we consider
  here rather than the full jet case. Nevertheless the good agreement
  that we find with numerical studies in
  section~\ref{sec:toy-fluctuations} suggests that it is not
  illegitimate to ignore jet area fluctuations for our purposes here.}
\begin{equation}
  \label{eq:dist-rho-jet}
  \frac{dP_{\jet}}{d\rho}(\delta\rho_\jet) \equiv \frac1N
  \frac{dN}{d\rho_{\jet}} = 
  \frac{1}{\sigma} \sqrt{\frac{\langle A_{\jet}\rangle}{2\pi}} 
  \exp\left(
    -\frac{\langle A_{\jet}\rangle  }{2\sigma^2 } \delta \rho_\jet^2
    \right)\,,
    \qquad\quad
    \delta \rho_\jet \equiv \rho_{\jet} - \rho\,.
\end{equation}
The corresponding cumulative probability distribution for the jets is
given by 
\begin{equation}
  \label{eq:cumul-dist-rho-jet}
  P_\jet(\delta\rho) = \int_{-\infty}^{\delta \rho+\rho} d\rho'
  \frac1N\frac{dN}{d\rho}=\frac{1}{2}\left(1 + \Erf\left(\frac{\delta\rho}{\sigma}\sqrt{\frac{\mean{A_\jet}}{2}}\right)\right)\,.
\end{equation}
If the number $N$ of jets is odd, $N=2m+1$, then the probability distribution of
the median $\delta \rho$ is obtained from the product of the
probability of having one jet with $\rho_\jet = \rho-\delta\rho$, $m$
jets with $\rho$ smaller than this, and $m$ jets with $\rho$ larger
than this:
\begin{equation}
  \label{eq:dist-median}
  \frac{dP_\med}{d\rho}(\delta \rho) = (2m+1)\frac{(2m)!}{(m!)^2} \,
  \frac{dP_{\jet}}{d\rho}(\delta\rho)\,
  \left[P_\jet(\delta\rho)\right]^m\,
  \left[1-P_\jet(\delta\rho)\right]^m\,.
\end{equation}
Making use of the expansion of the
error function
\begin{equation}
  \label{eq:cumul-p-expansion}
  P_\jet(\delta\rho) = \frac12 +
    \sqrt{\frac{\mean{A_\jet}}{2\pi}} \frac{\delta\rho}{\sigma} + \order{\left(\frac{\delta\rho}{\sigma}\right)^3}\,,
\end{equation}
working in the large $m$ and small $\delta\rho/\sigma$ limit, and
making use also of Stirling's formula $m! \simeq \sqrt{2\pi m}
(m/e)^m$, one can approximate eq.~(\ref{eq:dist-median}) as
\begin{equation}
  \label{eq:P-median-approx}
  \frac{dP_\med}{d\rho}(\delta \rho) \simeq
  %%2m \frac{\sqrt{4\pi m}}{2 \pi m } \frac{(2m/e)^{2m}}{(m/e)^{2m}}
  \frac{\sqrt{2m \langle A_{\jet}\rangle}}{\pi \sigma}
  %\sqrt{\frac{4m}{\pi} \frac{\langle A_{\jet}\rangle}{2\pi}} 
  \left(1 -
    \frac{2\mean{A_\jet}}{\pi}\frac{\delta\rho^2}{\sigma^2}\right)^m
  \simeq 
  \frac{\sqrt{N \langle A_{\jet}\rangle}}{\pi \sigma}
  \exp\left(-
    \frac{N\mean{A_\jet}}{\pi}\frac{\delta\rho^2}{\sigma^2}\right)\,,
\end{equation}
where in the last step we have replaced $m\simeq N/2$.
Using the relation $N\mean{A_\jet} = A_\tot$, we finally obtain the
following result for the standard deviation of the extracted median
$\rho$ values,
\begin{equation}
  \label{eq:Sd-median-result}
  S_{d,\med}^{(\soft)} \simeq \sigma \sqrt{\frac{\pi}{2 A_\tot}}\,.
\end{equation}
This is about $25\%$ larger than the standard
deviation that would be obtained for $\rho$ extracted as an average of
the $\rho_{\jet}$ values over the same total area. 
This moderate enhancement of fluctuations in the pure soft case is
part of the price that one pays in exchange for the median's greater
resilience to hard contamination.

%----------------------------------------------------------------------
\subsection{Hard contamination}
\label{sec:hard-contamination-fluct}

The result eq.~(\ref{eq:rho-AM-Gauss+pert-v-R}) for the average
discrepancy in $\rho_\ext$ due to hard contamination can be obtained
in an alternative manner, which will be more useful for estimating
fluctuations.
In this approach we imagine some distribution of soft jets, and then
add in the hard partons.
Some of the hard partons will enter jets whose $\rho_\jet$ is already above the median
value for $\rho$. These hard partons will not affected $\rho_\ext$. 
The remaining hard partons (a number $k$) will enter jets that were
below the median.
These jets will acquire much larger transverse momenta, taking them
well above the median.
Thus it becomes necessary to recalculate the median, which will be
shifted by $k$ soft jets' worth.
From eq.~(\ref{eq:dist-rho-jet}), and working in the large-$N$ limit
(throughout this section), this translates to an average shift in $\rho_\ext$ of
\begin{equation}
  \mean{\delta\rho} = k\delta_1\,, \quad\qquad \delta_1 \equiv \frac{\sigma}{N}\sqrt{\frac{2\pi}{\mean{A_\jet}}}\,,
\end{equation}
where $\delta_1$ is the 1-jet shift.
Substituting $\mean{k}=\mean{n_h}/2$ gives
\begin{equation}
  \mean{\delta\rho} = 
  \frac{\mean{n_h}}{2}\delta_1
  =\sigma \sqrt{\frac{\pi}{2\mean{A_\jet}}}\frac{\mean{n_h}}{N}
  \,,
\end{equation}
in accord with the first order term in $\mean{n_h}/N$ in
eq.~(\ref{eq:rho-AM-Gauss+pertN}).

We will consider two main sources of fluctuations in this result. 
Let us first imagine that $k=1$. The median will shift up by one jet,
and the distribution of $\delta \rho$ will be simply be given by the
distribution of the difference in $\rho_\jet$ between two neighbouring jets
in the sorted sequence of jets (at position in the sequence that is
near the median).
That distribution is  an exponential distribution with mean $\delta_1$,
\begin{equation}
  \frac{dP}{d\delta\rho}(k=1) = 
  \frac{1}{\delta_1}\exp\left[-\delta\rho/\delta_1 \right]\,.
\end{equation}
The distribution of the shift for $k$ jets is
\begin{equation}
  \frac{dP}{d\delta\rho}(k) = 
  \frac{(\delta\rho)^{k-1}}{k!\,(\delta_1)^k}\exp\left[-\delta\rho/\delta_1 \right]\,,
\end{equation}
with the standard deviation $\sqrt{k}\delta_1$.

The second source of fluctuations comes from the fact that $k$ is not
constant but rather has a Poisson distribution with mean
$\mean{n_h}/2$ and standard deviation $\sqrt{\mean{n_h}/2}$.%
\footnote{The statement about the Poisson distribution holds for the
  perturbatively radiated partons, but not, strictly speaking, for the
  Born partons. Nevertheless, for simplicity, we shall treat the Born
  and perturbatively radiated partons in common here.}
Combining this with the fluctuation on $\rho$ for fixed $k$, leads to
an overall standard deviation of 
\begin{equation}
  S_{d,\med}^{(\hard)} = \sqrt{\mean{n_h}}\delta_1 = \sigma
  \sqrt{\frac{2\pi}{\mean{A_\jet}}}\left(\frac{\sqrt{\mean{n_h}}}{N} +
  \order{\frac{\mean{n_h}^{3/2}}{N^2}}\right)\,,
\end{equation}
In terms of $R$ and $A_\tot$, this becomes
\begin{subequations}
  \label{eq:SdHardFinal}
  \begin{align}
    S_{d,\med}^{(\hard)} &= \sigma
    R \sqrt{2\pi c_J}\left(\frac{\sqrt{\mean{n_h}}}{A_\tot} +
      \order{\frac{\mean{n_h}^{3/2}c_J R^2}{A_\tot^2}}\right)
    \\
    &\simeq 1.79 \frac{\sigma R}{\sqrt{A_\tot}} \left(\frac{C_i}{C_A}L
      + 4.0 \frac{n_b}{A_\tot}\right)^{\frac12} + \cdots
  \end{align}
\end{subequations}
where $L\simeq1$ was defined in eq.~(\ref{eq:nh-over-Atot}).

%   % 
%     +
%     % 
%   \\

%======================================================================

\end{document}